\documentclass[twocolumn]{aastex631}
\usepackage{float}
\usepackage{multirow}
\usepackage[para]{threeparttable}
\usepackage{enumitem}
\usepackage{hyperref}

\begin{document}

\title{Dust production rates in Jupiter-family Comets: A two-year study with ATLAS photometry} 

\author[0000-0003-4094-9408]{A. Fraser Gillan}
\affiliation{Astrophysics Research Centre, School of Mathematics and Physics, Queen’s University Belfast, Belfast, BT7 1NN, UK}

\author[0000-0003-0250-9911]{Alan Fitzsimmons}
\affiliation{Astrophysics Research Centre, School of Mathematics and Physics, Queen’s University Belfast, Belfast, BT7 1NN, UK}

\author[0000-0002-7034-148X]{Larry Denneau}
\affiliation{University of Hawaii Institute for Astronomy, Honolulu, HI 96822, USA}

\author[0000-0001-5016-3359]{Robert J. Siverd}
\affiliation{University of Hawaii Institute for Astronomy, Honolulu, HI 96822, USA}

\author[0000-0001-9535-3199]{Ken W. Smith}
\affiliation{Astrophysics Research Centre, School of Mathematics and Physics, Queen’s University Belfast, Belfast, BT7 1NN, UK}

\author[0000-0003-2858-9657]{John L. Tonry}
\affiliation{University of Hawaii Institute for Astronomy, Honolulu, HI 96822, USA}

\author[0000-0002-1229-2499]{David R. Young}
\affiliation{Astrophysics Research Centre, School of Mathematics and Physics, Queen’s University Belfast, Belfast, BT7 1NN, UK}

\begin{abstract}
Jupiter-family Comets (JFCs) exhibit a wide range of activity levels and mass-loss over their orbits. We analyzed high-cadence observations of 42 active JFCs with the wide-field Asteroid Terrestrial-impact Last Alert System (ATLAS) survey in 2020-2021. We measured dust production rates of the JFCs using the $Af\rho$ parameter and its variation as a function of heliocentric distance. There is a tendency for our JFC sample to exhibit a maximum $Af\rho$ after perihelion, with 254P/McNaught and P/2020 WJ5 (Lemmon) having their maximum $Af\rho$ over a year after perihelion. On average, the rate of change of activity post-perihelion was shallower than that pre-perihelion. We also estimated the mass maximum loss rate for 17 of the JFCs in our sample, finding 4P/Faye to be the most active. We present a subset of comets whose measured $Af\rho$ have been interpolated and extrapolated to a common distance of 2 au pre-perihelion and post-perihelion. From these measurements we found no correlation of intrinsic activity with current perihelion distance. For three of the JFCs in our sample, 6P/d’Arrest, 156P/Russell-LINEAR and 254P/McNaught, there was no visible coma but a constant absolute magnitude which we attributed to a probable detection of the nucleus. We derived upper limits for the nuclear radii of $\leq$ $2.1\pm0.3$ km, $\leq$ $2.0\pm0.2$ km and $\leq$ $4.0\pm0.8$ km respectively. Finally, we found that 
4P/Faye, 108P/Ciffreo, 132P/Helin-Roman-Alu 2, 141P/Machholz 2, and 398P/Boattini
experienced outbursts between 2020 and 2022.
\end{abstract}

\section{Introduction} \label{sec:intro}
Jupiter-family Comets (JFCs) are a dynamical subclass of short-period comets, typically with low-inclination orbits that have an origin that is generally accepted to be in the Kuiper Belt region of the Solar system \citep{{1988ApJ...328L..69D},{2008ssbn.book..397L}}. Since the size distribution in the Kuiper Belt is dominated by collisions for bodies that are $\sim$ 50km or less, JFCs may represent the low-mass fragments from previous collision events in the belt \citep{2007SSRv..128....1G} which can inject the 1 to 10 km bodies into dynamical resonances where they evolve inward and become short-period comets \citep{Farinella1996ShortPeriodCP}. 

JFCs have an orbital period of less than 20 years \citep{1997Icar..127...13L} and have orbits that are heavily perturbed by Jupiter. When a comet nears perihelion, it becomes active and produces a gravitationally unbound coma driven by the outflow of volatiles from the surface of the nucleus or sub-surface \citep{2013Icar..225..475K}. JFCs undergo regular sublimation throughout their lives due to their short orbital period which leads to rapid depletion of their volatile material. 

While there are several hundred known JFCs, only a fraction of them come to perihelion each year. To comprehensively examine a significant number of JFCs and investigate their activity and how they evolve, an effective strategy is to utilize an all-sky survey approach. The large number of observations obtained through current all-sky surveys offer an unparalleled opportunity to study how the nature of JFCs evolve throughout their orbits.

Extensive surveys of ensembles of comets have been conducted in previous studies, such as those by \citet{1995Icar..118..223A} and \citet{2012Icar..218..144C}. These investigations examined comets belonging to both long-period and short-period comet populations, but it is challenging to make accurate cross-comparisons between the comets as they made sparse observations at a low cadence. Other studies of large scale surveys such as \citet{1996ApJ...459..729F} and \citet{2012ApJ...752...15O} were spectroscopic surveys with a focus on the molecular gas emission.

Previous self-contained studies of significant numbers of JFCs have revealed important population characteristics. In a study by \citet{2013Icar..225..475K}, the dust activity of JFCs was examined within the range of 3-7 au, revealing that inactive comets tend to have smaller perihelion distances compared to active comets when measured at larger heliocentric distances. The activity of  low perihelion comets is effectively quenched at larger distances from the Sun. \citet{2013Icar..226.1138F} was one of the largest studies to date on the properties of JFC nuclei. Their findings showed that the current population of JFCs are likely incomplete up to 3 km in radius, but there was no evidence for geometric albedo to depend on size. The authors also identified a potential excess above the nominal cumulative size distribution of JFCs at radii 3-6 km. 

Other studies have focused on individual comets over the years, see \citet{1993PASJ...45L..33Y},  \citet{2004Sci...304.1764B}, \citet{2022P&SS..21805506G}, \citet{2023AJ....165..150K} and \citet{2022Icar..38315042A}. \citet{2016MNRAS.462S.138S} and \citet{2018A&A...615A.154P} also provide examples of high cadence observations of individual comets which reveal the activity over several months. These previous studies showed the evolution of a single comet or focused on particular phenomena such as dust jets and outbursts. However, this does not allow for an understanding of the broader perspective when looking at a single dynamical population of comets and studying the behavior of JFCs as a whole.

Using a high cadence, all-sky survey system facility such at ATLAS allows the monitoring of a large number of JFCs through a significant fraction of their orbit. We observed 42 JFCs that reached their perihelion in 2020-2021 using the ATLAS all-sky survey and investigated the dust production for each JFC.

\section{ATLAS DATA AND TARGET SELECTION} \label{sec:data}
\subsection{ATLAS} \label{subsec:atlas}

ATLAS (Asteroid Terrestrial-impact Last Alert System) is a survey set up in recognition of the hazard posed to the Earth by asteroid impacts \citep{{2011PASP..123...58T}, {2018PASP..130f4505T}} and is designed to detect small (10-140m) asteroids on their trajectory before impacting the Earth \citep{2018AJ....156..241H}. ATLAS is a robotic wide field survey and has been continuously operational since installation in 2016. It is currently comprised of 4 independent units located on Mauna Loa and Haleakala in Hawaii, the South African Astronomical Observatory and El Sauce Observatory in Chile. 

Each ATLAS unit employs a 0.5m Wright-Schmidt telescope and the camera for each telescope (ACAM) provides a field of view covering 5.4\textdegree\ x 5.4\textdegree\ equating to 29 square degrees \citep{{2018PASP..130f4505T}, {2020PASP..132h5002S}}. The $10560\times10560$ STA1600 CCDs deliver a pixel scale of 1.86"/pixel. As this is larger than the typical atmospheric seeing at the sites, our angular resolution is governed by the accuracy of the temperature-adjusted focus, which is $\sim 2$ pixels on most nights. The ATLAS units achieve a limiting magnitude of $\sim$19.7 in dark skies (30 second exposure time) taking 4 exposures of each survey field with a cadence 5-45 minutes during a night of observations. In practise between 800 and 1000 frames are taken routinely per telescope per night. With ATLAS in 4 different locations around the globe, this ensures 24 hour night sky coverage assuming good observing conditions, surveying $\sim$ 80 \% of the sky every night. ATLAS uses two non-standard broadband filters, cyan (\emph{c}) with a wavelength range of 420nm -- 650nm and orange (\emph{o}) that covers a wavelength range of 560 nm -- 820 nm. These filters have effective central wavelengths of $\lambda_{eff}$ = 533 nm for the \emph{c}-filter on al telescopes, and  $\lambda_{eff}$ = 678 -- 679 nm for the \emph{o}-filter depending on the telescope \citep{2018PASP..130f4505T}. These filters are visible region filters that cover most of the \emph{gri} SDSS filter system. 

One of the key features and benefits of using \mbox{ATLAS} is its high cadence and wide field of view, with \mbox{ATLAS} returning to the same area of sky much more frequently than other surveys such as the Pan-STARRS Survey for example, see \citet{2020ApJS..251....6M}. Although while \mbox{ATLAS} covers a larger area of sky than the Pan-STARRS Survey, Pan-STARRS can observe much fainter objects. This high cadence however, distinguishes the ATLAS all-sky survey system from previous studies.

To detect moving objects, each field is observed four times in a given night. This allows us to obtain orbital information on each moving object and distinguish moving objects from stationary transients. As the ATLAS survey is a high cadence survey system, which could observe a given area of sky every two nights during 2020--2021, Solar system bodies can be monitored throughout a significant fraction of their orbit.

The ATLAS images are calibrated using the reduction pipeline which subtracts a static sky image and produces a database of detected sources \citep{2018PASP..130f4505T}. An adapted version of the Pan-STARRS Moving Object Processing System (MOPS) \citep{2013PASP..125..357D} links detections from multiple images as tracklets and then flags them as moving objects. We employed the use of difference images in this study as they offer a rapid and efficient approach to identify variable and moving objects and are also applicable for photometry as the background is already subtracted, removing stars and galaxies. There is no concern that the background subtraction removes any signal from the coma as the sky aperture is larger for the brighter comets to ensure good sky measurements for aperture photometry. This was confirmed by visual inspection for all bright comets. Finally, we note that all of the data has been calibrated using the ATLAS \mbox{Refcat2} \citep{2018ApJ...867..105T} using the survey automatic  pipeline, which also calibrates out the slight differences in bandpasses between the telescopes to provide consistent AB magnitudes and fluxes.

A summary of the ATLAS telescope features is shown in Table~\ref{tab:ATLASTable}.

\begin{table}
\centering
\caption{ATLAS telescope specifications \footnote{\url{https://atlas.fallingstar.com/specifications.php}}}
\begin{tabular}{lc}
\hline
Number of Telescopes & 4   \\
Aperture & 0.5 m  \\
Filters & \emph{orange}, \emph{cyan}   \\  
Pixel scale & 1.86\arcsec \\
Exposure time & 30 s \\
FOV & 29 degrees$^{2}$ \\
\hline
\end{tabular}
\label{tab:ATLASTable}
\end{table}

\subsection{Comet selection} \label{subsec:com_subset}

In this study, we selected all JFCs that reached perihelion in 2020 and 2021. It should be noted that most of the observations of comets in this study were made during the time of 2 operational ATLAS systems (Haleakala and Mauna Loa) with a declination range down to -45\textdegree, but towards the end of 2021 both the Chile and South African units in the southern hemisphere became available. As described in section \ref{subsec:atlas}, ATLAS is capable of observing point sources at $\leq$$\sim$19.7 magnitude in best conditions. To account for faint comets and potential outbursts that might temporarily increase their brightness, we set a predicted lower magnitude limit of 20.5. After searches for faint JFCs near this magnitude limit of $\sim$20.5 however, the JFCs were not visible in the images. The JFCs in this study were selected by using the predicted total magnitude from JPL Horizons. There were 56 JFCs that reached perihelion in 2020 and 2021 with 47 predicted to be brighter than this limiting magnitude. 

From these 47 JFCs, 42 were sufficiently bright to be detected in the \mbox{ATLAS} images during our analysis and are listed in Table~\ref{tab:table3}. Following rejection of outlying magnitudes (see section \ref{subsec:114P}), the resulting dataset used for this study comprised 10,101 individual images. The median airmass of these observations was $X=1.19$, with only $2.6$\% of images having $X> 2.0$. The median fwhm of stars in these data was 2.3 pixels, equivalent to 4.3\arcsec. As described above, ATLAS images are focus and optics limited rather than atmospheric seeing limited. From our detected sample of JFCs, some are visible pre- and post-perihelion, others are detectable to \mbox{ATLAS} only pre-perihelion or post-perihelion. This may be due to a variety of factors such as the comet having zero or little activity and therefore being below the \mbox{ATLAS} detection magnitude, or in other cases the comet may be visible to \mbox{ATLAS} for part of its orbit and then become undetectable due to the comet being in Solar conjunction and no longer observable. 

\begin{table*}
\centering
\caption{Jupiter-family Comets observed by the ATLAS survey in this study. $\dag$ - Only observed on one day}
\begin{tabular}{lcccccc}
\hline
Comet & Number of & Range of observation dates & $e$ & $a$ & $q$ & $i$ \\
 & observations & & & (au) & (au) & (\textdegree) \\
\hline
4P/Faye &                     530 & 2021 May 25 - 2022 May 31 & 0.58 & 3.80 & 1.58 & 8.16 \\                  6P/d'Arrest &                     415 & 2021 Apr 25 - 2022 Jan 25 & 0.61 & 3.50 & 1.35 & 19.51 \\
11P/Tempel-Swift-LINEAR &                     244 & 2020 Aug 08 - 2021 Feb 23 & 0.58 & 3.29 & 1.39 & 14.44 \\             28P/Neujmin 1 &                     604 & 2020 Feb 28 - 2022 Apr 27 & 0.77 & 6.98 & 1.58 & 14.31 \\          58P/Jackson-Neujmin &                     143 & 2020 Jul 09 - 2020 Dec 15 & 0.66 & 4.08 & 1.38 &13.11 \\                 108P/Ciffreo &                     376 & 2021 Jun 23 - 2022 Mar 06 & 0.58 & 3.64 & 1.53 & 13.98 \\           114P/Wiseman-Skiff &                     311 & 2019 Aug 21 - 2020 May 17 & 0.55 & 3.55 & 1.58 & 18.27 \\       132P/Helin-Roman-Alu 2 &                     250 & 2021 Jul 07 - 2022 Mar 13 & 0.56 & 3.89 & 1.69 & 5.38 \\              141P/Machholz 2 &                      93 & 2020 Dec 03 - 2021 Mar 19 & 0.74 & 3.05 & 0.80 & 13.98 \\          156P/Russell-LINEAR &                     348 & 2020 Jun 27 - 2021 Jun 08 & 0.61 & 3.46 & 1.33 &17.26 \\                254P/McNaught &                     688 & 2020 May 31 - 2022 Mar 02 & 0.32 & 4.62 & 3.14 & 32.57 \\                    312P/NEAT &                      95 & 2020 Jul 22 - 2020 Nov 16 & 0.43 & 3.46 & 1.97 & 19.79 \\                  377P/Scotti &                     209 & 2020 Mar 29 - 2021 Aug 03 & 0.25 & 6.73 & 5.04 &  9.02 \\                378P/McNaught &                     152 & 2021 Aug 01 - 2022 Jan 16 & 0.47 & 6.34 & 3.38 & 19.11 \\                   390P/Gibbs &                     366 & 2019 Jul 06 - 2021 Feb 17 & 0.71 & 5.81 & 1.71 & 18.52 \\                  397P/Lemmon &                     289 & 2020 Jul 09 - 2021 Jan 30 & 0.40 & 3.86 & 2.32 & 11.03 \\                398P/Boattini &                     431 & 2020 Jul 18 - 2021 May 13 & 0.58 & 3.13 & 1.31 & 11.02 \\               399P/PANSTARRS &                     332 & 2020 Aug 13 - 2022 May 20 & 0.44 & 3.81 & 2.13 & 13.35 \\               400P/PANSTARRS &                     238 & 2020 Sep 08 - 2021 Mar 26 & 0.41 & 3.56 & 2.10 & 10.93 \\                403P/CATALINA &                     175 & 2020 Nov 25 - 2021 Apr 03 & 0.50 & 5.41 & 2.68 & 12.32 \\                  405P/Lemmon &                     256 & 2020 Sep 03 - 2021 May 19 & 0.69 & 3.61 & 1.12 &  9.37 \\             409P/LONEOS-Hill &                     327 & 2020 Oct 31 - 2021 Jul 14 & 0.71 & 6.05 & 1.75 & 17.14 \\             410P/NEAT-LINEAR &                     198 & 2022 Jan 01 - 2022 Mar 04 & 0.51 & 6.63 & 3.25 & 9.39 \\                  414P/STEREO &                       4 & 2021 Jan 04 \dag  & 0.81 & 2.80 & 0.53 & 23.38 \\            
415P/Tenagra &                     131 & 2020 Nov 05 - 2021 Apr 17 & 0.20 & 4.12 & 3.31 & 31.79 \\                417P/NEOWISE &                     270 & 2021 Jan 14 - 2021 Sep 28 & 0.55 & 3.35 & 1.49 &  8.13 \\                425P/Kowalski &                     248 & 2021 Aug 27 - 2022 Mar 18 & 0.54 & 6.34 & 2.90 & 16.43 \\      449P/Leonard (2020 S6) &                     110 & 2020 Oct 19 - 2021 Feb 07 & 0.48 & 3.60 & 1.87 & 15.46 \\          P/2019 LD2 (ATLAS) &                     695 & 2019 Mar 28 - 2021 Dec 16 & 0.13 & 5.29 & 4.58 & 11.57 \\         P/2020 G1 (Pimentel) &                      87 & 2020 Apr 21 - 2020 Jun 24 & 0.86 & 3.60 & 0.51 &18.47 \\
P/2020 K9  (Lemmon-PANSTARRS) &                     235 & 2020 May 07 - 2021 Jan 10 & 0.32 & 4.20 & 2.85 & 23.20 \\       P/2020 S5 (PANSTARRS) &                      71 & 2020 Sep 17 - 2020 Dec 26 & 0.34 & 4.05 & 2.68 & 12.35 \\        P/2020 T3 (PANSTARRS) &                      88 & 2020 Nov 08 - 2021 Feb 22 & 0.59 & 3.52 & 1.44 & 7.30 \\        P/2020 U2 (PANSTARRS) &                     120 & 2020 Nov 10 - 2021 Mar 30 & 0.51 & 3.78 & 1.85 & 6.42 \\          P/2020 WJ5 (Lemmon) &                     363 & 2021 Jan 06 - 2022 Jul 18 & 0.17 & 6.02 & 4.99 & 22.29 \\           P/2020 X1 (ATLAS) &                      36 & 2020 Dec 01 - 2020 Dec 11 & 0.37 & 4.52 & 2.87 & 31.70 \\          P/2021 L2 (Leonard) &                      64 & 2021 Jul 10 - 2021 Sep 08 & 0.52 & 4.05 & 1.94 & 21.07 \\              P/2021 N1 (ZTF) &                       6 & 2021 Jun 12 - 2021 Jun 18 & 0.68 & 2.98 & 0.96 & 11.51 \\             P/2021 N2 (Fuls) &                     298 & 2021 Jun 02 - 2022 Feb 20 & 0.45 & 6.93 & 3.80 & 13.06 \\   
P/2021 P3 (PANSTARRS) &                      97 & 2021 May 18 - 2021 Nov 08 & 0.34 & 4.42 & 2.92 & 27.16 \\            P/2021 Q5 (ATLAS) &                      46 & 2021 Aug 19 - 2021 Oct 22 & 0.62 & 3.29 & 1.23 & 10.73 \\         P/2021 R6 (Groeller) &                      36 & 2021 Oct 26 - 2021 Nov 27 & 0.59 & 6.27 & 2.55 & 34.93 \\
\hline
\end{tabular}
\label{tab:table3}
\end{table*}

\section{COMETARY ACTIVITY AND MEASUREMENTS} \label{sec:method}

\subsection[Afp]{$Af\rho$} \label{subsec:afrho}
To quantify the activity from an observed comet, we used the $Af\rho$ parameter that was introduced by \citet{1984AJ.....89..579A}. It is the product of the grain albedo, $A$, the filling factor of the grains within the aperture, $f$, and the linear radius of the photometric aperture on the sky at the comet, $\rho$. We use $\rho$ = 10,000 km in our measurements as this is consistent with radii chosen by previous investigators and is the standard scale used for JFC comae, e.g. \citet{2023PASJ..tmp...24L}; \citet{2022Icar..37214752B}. The majority of our measurements were made at geocentric distances $<$ 4 au, giving an aperture radius of $>$ 3.5\arcsec. With the median angular resolution in our data of 4.3\arcsec, approximating the seeing profile by a gaussian implies that using a 10,000 km aperture radius would result in $>$ 94\% of the flux from a stellar source being measured. However, three of the comets we observed (377P/Scotti, P/2019 LD2 (ATLAS) and P/2020 WJ5 (Lemmon)) were detected at distances of $>$ 5 au. For these comets, although the coma was faint, we note that our $A(0^\circ)f\rho$ measurements are possibly lower limits due to aperture losses. $Af\rho$ is measured via Equation~\ref{equation1}

\begin{equation} \label{equation1}
Af\rho = \frac{4 \, R_h^2 \, \Delta^2}{\rho}\frac{F_{com}}{F_\odot}
\end{equation}

where $R_h$ is the distance between the Sun and the comet measured in au and $\Delta$ is the geocentric distance measured in cm. $F_{com}$ is the measured cometary flux and $F_\odot$ is the solar flux at 1 au. The flux and corresponding AB magnitude of the Sun was derived by convolving a standard solar spectrum with the system throughput (atmosphere+telescope+filters) as shown in \cite{2018PASP..130f4505T}. We found the apparent AB solar magnitudes in the ATLAS system to be $m_\odot (c) = -26.696$ and $m_\odot (o) = -26.982$. Note that $Af\rho$ is typically measured in cm.

The $Af\rho$ quantity only depends on the flux from the comet from a given observation which we can directly measure making it a robust quantity for measuring relative dust activity. Due to this, the quantity is often used to compare the dust production rates between different comets e.g. \citet{2021P&SS..20605308G}. We will examine the effects of gas within the bandpasses in Section~\ref{subsec:114P}.

When making $Af\rho$ measurements, we must consider that dust particles in the coma scatter sunlight by different amounts at different angles. The amount of scattering depends on the size and composition of the dust particles as well as the scattering angle. Strictly speaking, we measure an $A(\theta)f\rho$ value where $\theta$ is the phase angle. Most comet observations are obtained at much smaller phase angles than 90\textdegree\ (except for sungrazing comets) and therefore the most obvious normalisation point is at 0\textdegree, giving us a quantity of $A(0^\circ)f\rho$. We corrected $A(\theta)f\rho$ using the \citet{LowellObservatory} dust phase function.

\subsection{114P/Wiseman-Skiff as a representative JFC} \label{subsec:114P}
\begin{figure}[h!]
\epsscale{1.2}
\plotone{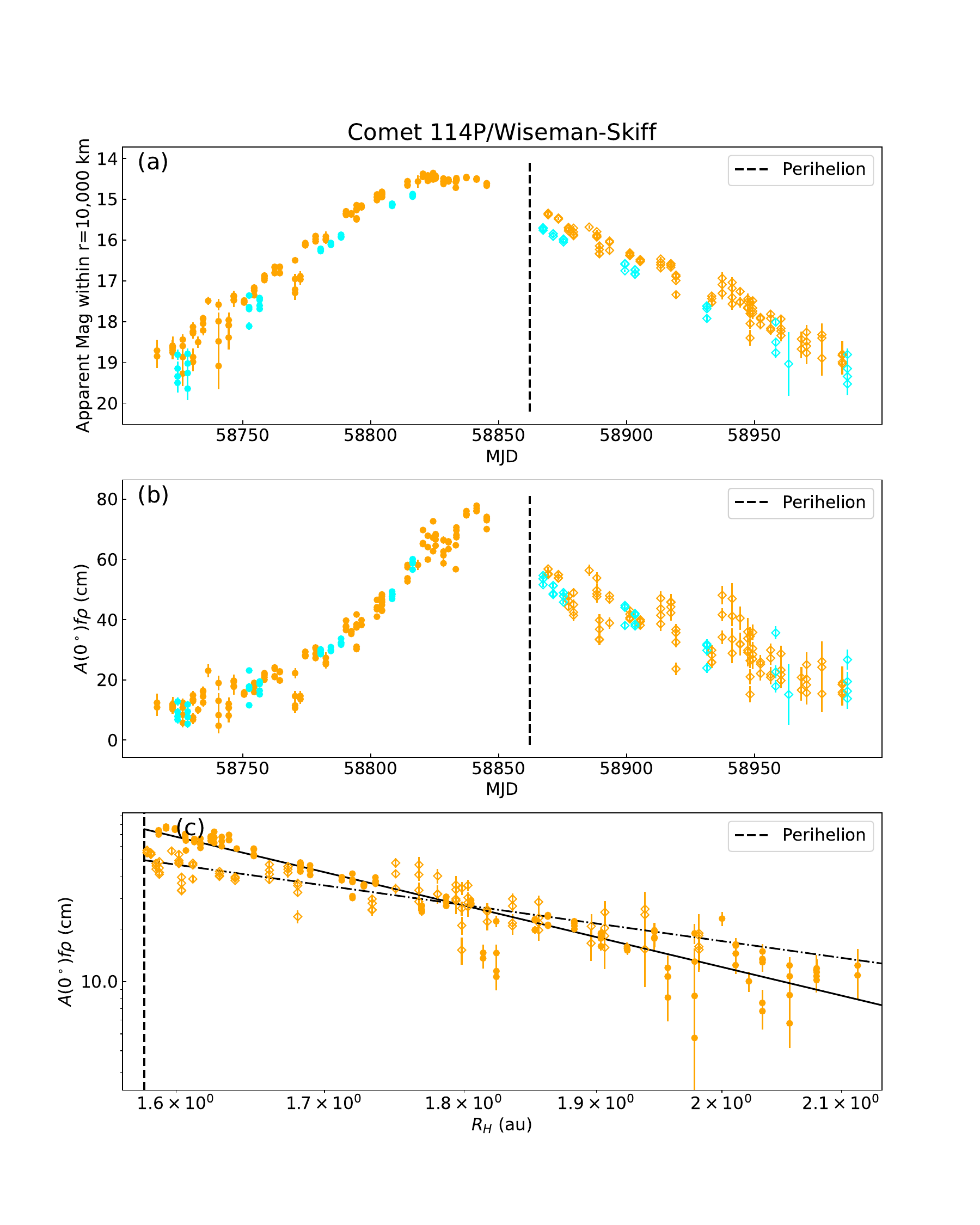}
\caption{The 114P/Wiseman-Skiff light curve is shown in the top panel (a). The dust production parameter $A(0^\circ)f\rho$ vs time is displayed in (b), and (c) shows $A(0^\circ)f\rho$ vs heliocentric distance with the solid line representing the rate of change of activity pre-perihelion and the dot-dashed line showing the rate of change of activity post-perihelion. Orange and cyan points on the plots represent the \emph{o-}filter and \emph{c-}filter respectively. Filled circular points represent pre-perihelion measurements and open diamonds represent post-perihelion. Perihelion is marked with the dashed line in each panel. The complete Figure Set is available in the online journal.}
\label{fig:fig2}
\end{figure}

Each comet in our data set has been measured and analyzed using the same methods that we illustrate using the JFC 114P/Wiseman-Skiff (hereafter 114P) as a representative comet from our sample. Figure \ref{fig:fig2} shows 114P measured over a period of $\sim$ 250 days and over a range of heliocentric distances with 114P being detectable for $\sim$ 0.5 au inward toward perihelion and $\sim$ 0.4 au receding.

The images used in this study were 1000 x 1000 pixel sub-arrays centered on the comet using the object ephemerides from JPL Horizons. To obtain flux measurements for a comet in each image, aperture photometry was performed. As we do not know a-priori the exact center of the JFC in a given image for photometry, we determine this by centroiding to measure the position of the center of the JFC in the frame. If the comet was fainter than 15th magnitude, the outer sky radius used for aperture photometry was $5 \times 10^{4}$ km on the sky, between 14th and 15th magnitude $1 \times 10^{5}$ km was used and brighter than 14th magnitude, $1.5 \times 10^{5}$ km was used. This prevented contamination from the coma in very active comets from interfering with the sky measurement. Figure~\ref{fig:fig2} (a) shows the evolution of the apparent magnitude of the comet with time. For this particular comet, the apparent magnitude within the aperture reached a maximum slightly before perihelion at $\sim$ 14 mag and then decreased in brightness as the comet moved away from perihelion as would be expected. The magnitude then continued to fade until the comet reached the ATLAS limiting magnitude. 

In addition to making the magnitude measurements, $A(0^\circ)f\rho$ was measured for each JFC in the images. We manually inspected each plot to remove any obviously outlying measurements. Outlying measurements were flagged as individual magnitude measurements significantly brighter or fainter than other measurements taking the photometric errors into account. There were also instances where all measurements on a single night were much brighter  fainter than the overall trend in brightness evolution from surrounding nights. These were not outbursts of activity as outbursts normally show a gradual decrease in brightness over subsequent nights as the ejected material moves out of the measurement aperture. All such outliers were then inspected individually. These measurement rejections were due to a number of reasons from cloud cover during observations leading to inaccurate measurements, to the comet being positioned next to another bright object in the frame such as a star or satellite which caused contamination within the aperture. Other reasons for image rejection were problems with the wallpaper subtraction, the JFC being located near the edge of the frame with poor image quality or the JFC having low brightness compared to the surrounding sky. The plots for every JFC in the study can be found in the Figure Set associated online with Figure~\ref{fig:fig2}.

In normal JFCs, the \emph{c}-band could contain prominent cometary gas emissions due to the C$_2 (0-0)$ band and others. The \emph{o}-band should be far less contaminated by gas emission, containing only weaker emissions such as part of the C$_2 (0-1)$ band, NH$_2$ bands and [OI]. To assess the possibility of gas contamination in the \emph{o}-band, we calculated the average $(g'-r')$ and  $(g'-i')$ color for active comets as measured by \citet{2012Icar..218..571S}. Transforming to the Pan-STARRS system via the relationships in \cite{2016ApJ...822...66F}  and then into the ATLAS system using the expressions in \cite{2018PASP..130f4505T}, we find that an average comet with no detectable gas emission should have a \emph{c}-band flux 6\% fainter than in \emph{o}-band. Hence we compensate for this by increasing all \emph{c}-band $A(0^\circ)f\rho$  measurements by this amount. With this assumption of average comet dust color, if there is no significant gas emission in the \emph{c}-band, and therefore in the \emph{o}-band, the transformed \emph{c}-band $A(0^\circ)f\rho$ should follow the same trend as the \emph{o}-band values. If the new \emph{c}-band values are higher, this implies there is excess emission in this filter probably due to gas, and we have to be careful about interpreting the \emph{o}-band $A(0^\circ)f\rho$ measurements. 

Using this procedure, we find there was measurable gas emission in the \emph{c}-band in comets 6P/d'Arrest, 11P/Tempel-Swift-LINEAR and \mbox{141P/Machholz 2}. Hence our \emph{o}-band $A(0^\circ)f\rho$ measurements may be upper limits for these comets and should be treated with caution. This may also be the case for 108P/Ciffreo and \mbox{141P/Machholz 2}, although the data here is sparser. The transformed \emph{c}-band measurements for \mbox{132P/Helin-Roman-Alu 2} and 409P/LONEOS-Hill lie below the \emph{o}-band measurements, implying that the dust in these comets is redder than average. Bluer than average does not necessarily indicate the presence of gas and a probable cause may also be due to bluer than average dust scattering, however we do not have the spectra to definitively determine this. Although as this normally occurs as the comet approaches perihelion, it follows the trend expected for gas contamination of the \emph{c}-filter.

Figure~\ref{fig:fig2} (b) shows the dust production rate  $A(0^\circ)f\rho$  with respect to time (MJD). Clearly shown is the dust production reaching a maximum around (albeit slightly before) perihelion. Figure~\ref{fig:fig2} (c) shows the \emph{o-}band $A(0^\circ)f\rho$ with respect to heliocentric distance. These measurements are discussed in detail in section~\ref{subsec:AI} along with the comparisons with other JFCs in this sample.

\subsection[Detectability]{Detectability $R_h$ range for each comet} \label{subsec:detect_range}

\begin{figure*}[h!]
    \centering
    \includegraphics[width=1\linewidth, height=0.95\linewidth]{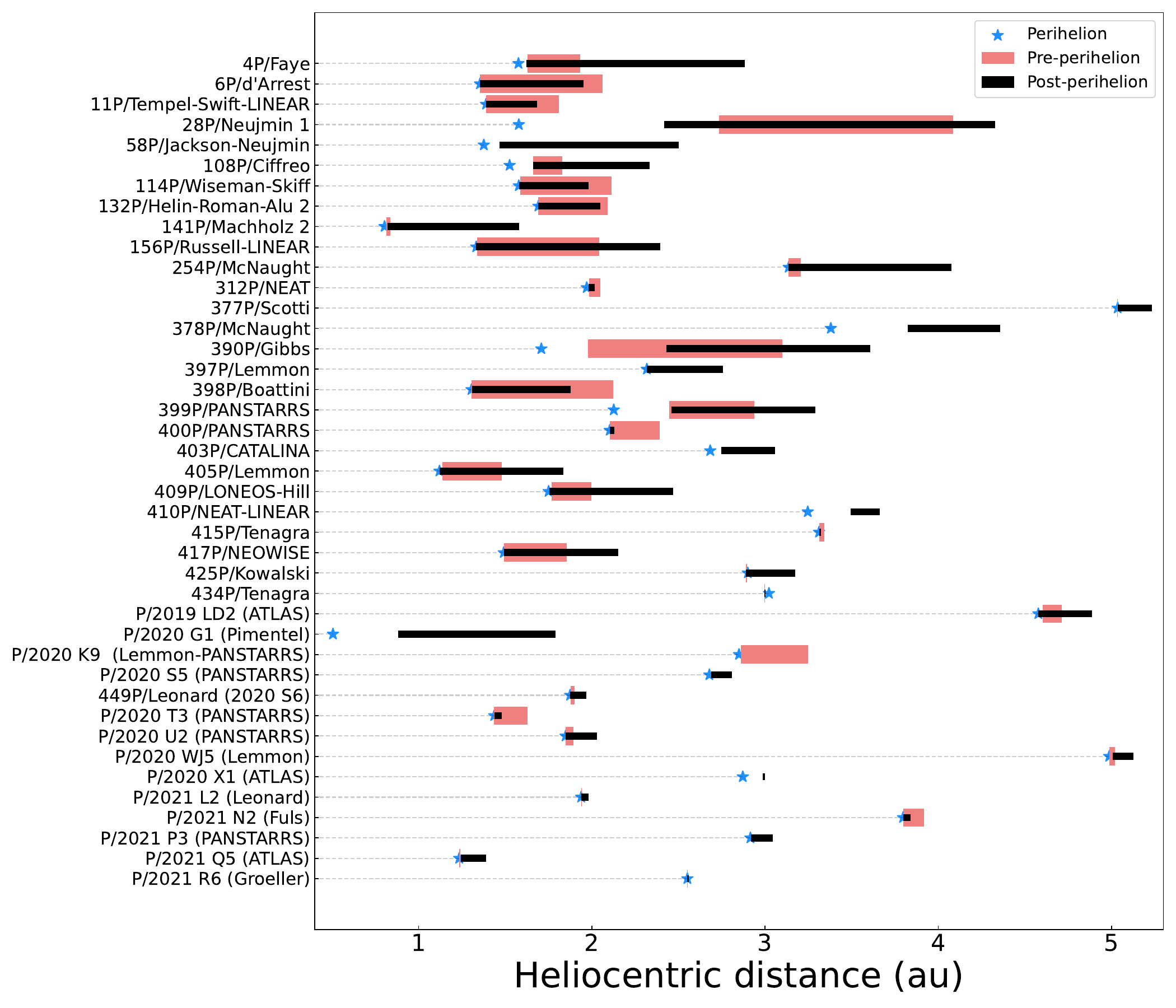}
    \caption{Every JFC reaching perihelion in 2020 and 2021 visible to ATLAS. Blue stars mark the perihelion of each comet, coral/pink represents the pre-perihelion $R_h$ range coverage by ATLAS and black represents the post-perihelion $R_h$ range coverage by ATLAS.}
    \label{fig:fig1}
\end{figure*}

Figure~\ref{fig:fig1} shows every comet in this study that was detected by ATLAS along with the range of heliocentric distances over which we obtained measurements. Figure~\ref{fig:fig1} also shows that numbered comets are detectable over a longer $R_h$ range than recently discovered JFCs with only a provisional designation. This disparity arises from numbered comets being inherently brighter than the rest of the sample and therefore being discovered earlier. 

\subsection[Qd]{Dust production rate $Q_{d}$} \label{subsec:qd}
From $A(0^\circ)f\rho$ it is theoretically possible to estimate the dust production rate $Q_d$, which provides an estimate of the mass- loss-rate in kg/s from particulate matter. $Q_d$ however relies on properties to be known (or assumed) such as dust grain velocity, albedo, density and size distribution. As shown by \cite{Fink2012} and \citet{2018Icar..313....1I} it is a highly problematic and an ambiguous task to derive $Q_d$ from $Af\rho$. Not only is it highly dependent on the assumed albedo and size distribution, the ejection velocity and maximum size ejected will vary as a function of heliocentric distance. Indeed, Ivanova et al. found that for a single epoch of observation of the distant comet C/2012 S1 (ISON), the measured $A(0^\circ)f\rho$ led to estimates of $Q_d$ that varied by two orders of magnitude, depending on the dust composition, density, phase function and terminal velocity.

Nevertheless, $Af\rho$ is less intuitive to interpret than a  mass loss rate in kg/s. Hence although we reiterate that any calculation $Q_d$ is highly uncertain, we have calculated this parameter to provide an approximate order of magnitude estimate. We make simplifying assumptions that all of the dust grains in the coma are ejected at the same velocity, are the same size and have the same density and albedo, the relationship between $Af\rho$ and $Q_d$  is shown below in Equation~\ref{eq2}. 

\begin{equation}\label{eq2}
Q_d = Af\rho \, \frac{4\, \pi \, r_d \, \sigma_d \, v_d}{3\, p_o}
\end{equation}

Here, $\sigma_d$ is the grain density, $v_d$ is the velocity of the dust grains, $r_d$ is the size of the dust grains and ${p_o}$ is the geometric albedo in the o-filter. We assume that the geometric albedo and the grain density are ${p_o}=0.04$ and $\sigma_d$ $\approx$ 1 g/cm$^{-3}$, see \cite{1990ApJ...351..277J}; \citet{2011Icar..211..559I}; \citet{2014ApJ...791..118M}. We also assume that we are dealing with fine, fast moving dust grains and therefore use $v_d=300$ m s$^{-1}$ \citep{2005Natur.437..987K}. Finally, we choose 1 $\mu$m as the particle size. The resulting upper limits of mass loss rates are shown in Table~\ref{tab:table1}.

\section{RESULTS} \label{sec:results}
\subsection[maxafp]{Maximum $Af\rho$ for each JFC in our sample} \label{subsec:afrho_vs_days}

The maximum dust production ($Af\rho$) value for comets is expected to occur at perihelion as this is where the Solar flux will be the highest and therefore the temperature will also be at its maximum. We therefore measured where the maximum $A(0^\circ)f\rho$ occurs for each comet in our sample. This could only happen when $A(0^\circ)f\rho$ reached a clear maximum such as 114P/Wiseman-Skiff shown in Figure~\ref{fig:fig2}, and the activity curve was not `flat' or had a lack of data near perihelion. For each comet we identified the date on which the maximum $A(0^\circ)f\rho$ value was measured by visual inspection of the data. If we could not determine that there was a clear maximum $A(0^\circ)f\rho$, i.e. the previous and following nights were of very similar measurements and the $A(0^\circ)f\rho$ measurements and the variation of $A(0^\circ)f\rho$ looked constant with time, then we did not say that particular JFC had reached a maximum $A(0^\circ)f\rho$. To obtain the maximum $Af\rho$ value for the JFCs, we calculated a weighted mean from the images taken of the comet on the same night where the maximum $A(0^\circ)f\rho$ was measured (3 to 4 images as ATLAS measured in quadrature). Table~\ref{tab:table1} gives the measured $A(0^\circ)f\rho$ and Q$_{d}$(Maximum) for each JFC. The associated uncertainties for each maximum $A(0^\circ)f\rho$ is the standard deviation on the mean, as the formal uncertainties were sometimes not a good representation of the variation between measurements in a single night.

\begin{figure}[h!]
\epsscale{1.3}
\plotone{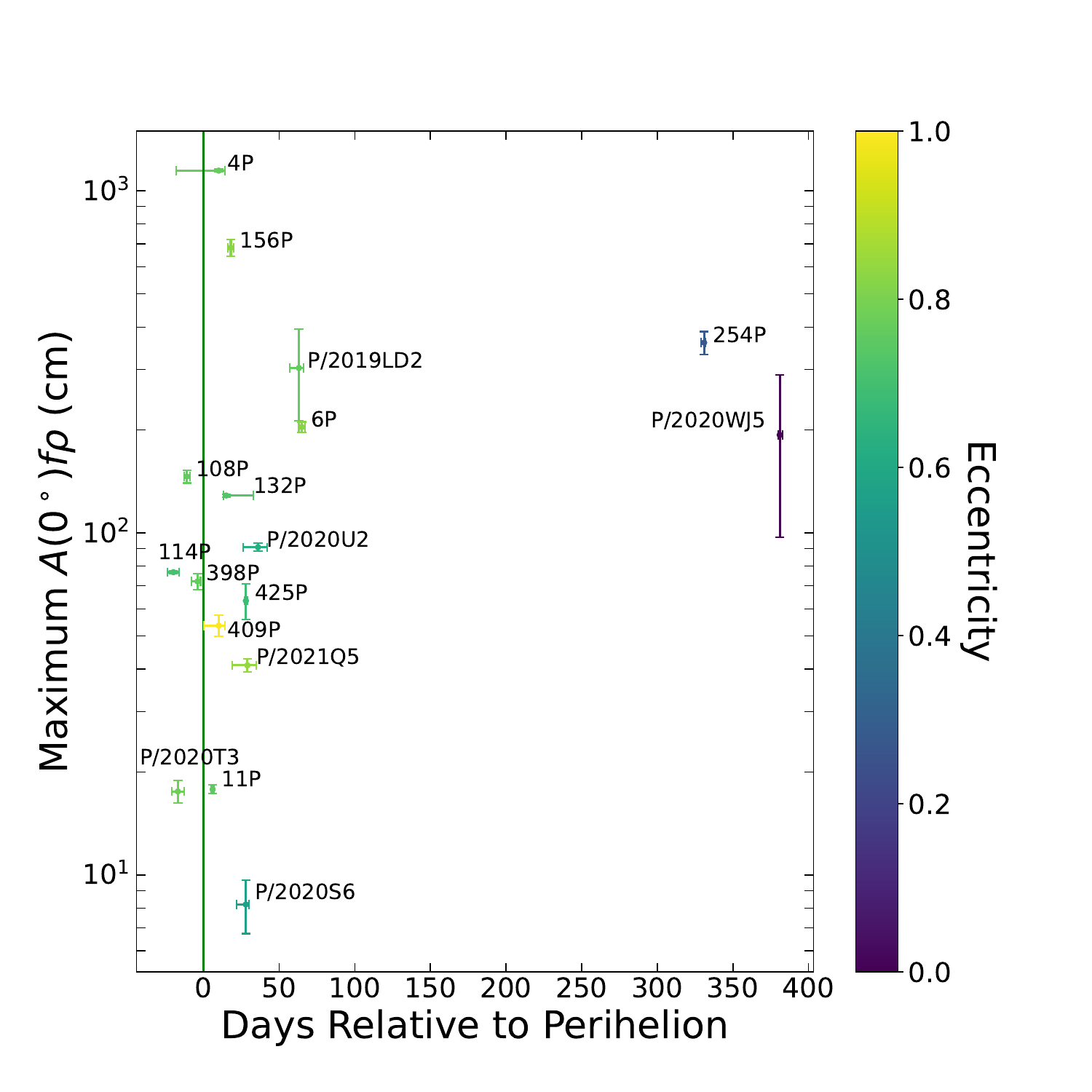}
\caption{Maximum $A(0^\circ)f\rho$ vs days from perihelion for each JFC is shown. The green vertical line represents the perihelion for each individual comet and the plot is color-coded according to orbital eccentricity.}
\label{fig:fig3}
\end{figure}

In our sample, a total of 17 comets exhibited a discernible maximum $A(0^\circ)f\rho$. Shown in Figure~\ref{fig:fig3} is the maximum $A(0^\circ)f\rho$ for each JFC plotted against the date this occurs relative to perihelion. Immediately apparent is the range of $A(0^\circ)f\rho$ values spanning over 2 orders of magnitude. This shows the substantial range in dust production among comets within the same dynamical population and how much it can differ between JFCs. This can be due to a number of factors such as the size of the nucleus, the amount of volatile material available on different areas of the nucleus, the temperature sensitive process of sublimation occurring at different rates inside and outside of thermal shadows \citep{2011ApJ...743...31G} or the heliocentric distance \citep{2018Natur.553..186B}.

The `lag-time' from perihelion to maximum $A(0^\circ)f\rho$ measured in days for each JFC is given in Table~\ref{tab:table1}. The uncertainties in the days from perihelion measurements were determined from the previous and next night of observations of the JFC before and after the image containing the maximum $A(0^\circ)f\rho$.
A small number of JFCs reach their maximum $A(0^\circ)f\rho$ before perihelion. The cause for the maximum activity pre-perihelion may be due to the cometary seasonal effects as described in \citet{2015A&A...583A...1L} and \citet{2015A&A...583A...7B}. This is where there is a volatile-rich surface of the comet illuminated by the Sun pre-perihelion and then a different region of the comet which is less abundant in volatiles becomes exposed after perihelion causing a decrease in activity. An example of this asymmetric activity behavior is shown in \citet{2023A&A...670A.159M} and  \citet{2019AJ....158....7E} for 46P/Wirtanen and 21P/Giacobini-Zinner respectively, both reaching a maximum $A(0^\circ)f\rho$ before perihelion.

The illuminated surface area of the nucleus exposed to the Sun is generally not constant during a given orbit. The shape of the nucleus  plays a crucial role, particularly when we consider real non-elliptical nuclei like 19P/Borrelly or 103P/Hartley 2. In such cases, the dust production rate can vary significantly depending on the orientation of the Sunward pole. This effect becomes more pronounced as the nucleus tends from spherical to elongated. It is important to note however that just because a larger surface area is exposed, it does not necessarily infer that there will be more sublimation, as ice abundance may vary across the surface as well as the thickness of any dust mantle.

Figure~\ref{fig:fig3} shows that the majority of JFCs in our sample reach their maximum $A(0^\circ)f\rho$ after perihelion but within $\sim$ 100 days. This may be due to a number of factors, one of which may be the thermal inertia of the JFC, see \citet{{1950ApJ...111..375W}, {2019SSRv..215...41G}}. 
If comets posses a low thermal inertia, they will lose their perihelion heating much faster than comets with a higher thermal inertia. 67P/Churyumov-Gerasimenko for example, has a notably low thermal inertia \citep{2018A&A...616A.122M}. Thermal inertia measurements of comets have been conducted in many prior investigations
\citep{2009Icar.Groussin, 2009A&A.Licandro, 2013Icar..226.1138F}, 
all of which show relatively low values for measured thermal inertia and infra-red beaming parameter as a proxy. In the absence of direct thermal measurements in our study, we can only infer from these studies that our JFCs probably also posses a low thermal inertia, ruling this out as a prominent contributing factor to the substantial lag in our JFCs reaching maximum $A(0^\circ)f\rho$. It is important to acknowledge that seasonal effects may also occur post-perihelion, however it is difficult to understand why these would not cause an equal number of comets to exhibit a pre-perihelion maximum in activity, assuming random spin-axis orientations.

An alternative reason for this delay in maximum $A(0^\circ)f\rho$ may also be the viewing effect of the larger dust grains which need more time to move out of the aperture. If the largest dust grains move at very low velocities, it is possible we may have measured dust grains that had been released several days before perihelion, depending on the velocity of the dust grains. For a large dust grain of radius 1 mm, the ejection velocity may be 39 ms$^{-1}$ at 1 au for a comet with small 1 km radius \citep{2002MNRAS.337.1081M}. Considering most JFC nuclei are a few km in size, we used this velocity to calculate that the dust would take $\sim$3 days to vacate the 10,000 km aperture. If we increase the distance to 1.5 au, a slightly longer dwell time within the aperture of $\sim 5$ days might be expected due to the lower ejection velocity, still smaller than the tens of days we measured for many comets. However, the amount of time the dust remains in the aperture can also be extended based on observational geometry. If we observe a comet near opposition for example, the dust may spend significantly more time within the aperture due to our line of sight along the direction of the dust tail. 

There are two JFCs in our sample that reach their maximum $A(0^\circ)f\rho$ approximately a year after their perihelion. 254P/McNaught and P/2020 WJ5 (Lemmon) have a much lower orbital eccentricity than the other JFCs shown in Figure~\ref{fig:fig3}. As these comets have a lower eccentricity, this will warm the ices on surface of the nucleus more gradually and take longer for the heat to penetrate through the layers of the nucleus. The low eccentricity may also affect the rate of change of sublimation with respect to time, making it less pronounced. This may also allow the spin-pole orientation and/or any seasonal effect to assume a more prominent role in such cases. Consequently, this suggests that the time of perihelion relative to maximum activity may be of lower significance for JFCs with lower eccentricities. Other explanations are also possible. For example, the average surface area exposed to the Sun by a tri-axial nucleus during rotation may be significantly larger at this point, leading to a higher sublimation and dust-ejection rate. We have no information on the nuclear shapes of 254P/McNaught and P/2020 WJ5 (Lemmon) to test this hypothesis, but an interesting future investigation might be to study nuclear shapes that would dominate the normal heliocentric distance/temperature dependence for specific spin axis orientations.

\begin{table*}
\centering
\caption{JFC maximum $A(0^\circ)f\rho$ measurements. $Q_d$ = upper limit of mass loss rate in kg s$^{-1}$}
\begin{tabular}{lcccc}
\hline
Comet & Max $A(0^\circ)f\rho$ & $Q_d$ & Days from perihelion & $R_h$ at maximum $A(0^\circ)f\rho$ \\
 & (cm) & (kg s$^{-1})$ & & (au) \\
\hline
4P$/$Faye & $1146.0\pm1.2$ &          $\leq$283 & $10_{-28}^{+4}$ &                  1.62 \\
6P$/$d$'$Arrest &  $203.8\pm1.5$ &           $\leq$52 &  $65_{-2}^{+2}$ &                  1.54 \\
11P$/$Tempel-Swift-LINEAR &   $17.8\pm0.3$ &            $\leq$5 &   $6_{-1}^{+1}$ &                  1.39 \\
108P$/$Ciffreo &  $146.2\pm1.5$ &           $\leq$36 & $-11_{-2}^{+2}$ &                  1.67 \\
114P$/$Wiseman-Skiff &   $76.7\pm0.3$ &           $\leq$19 & $-20_{-4}^{+4}$ &                  1.59 \\
132P$/$Helin-Roman-Alu &  $128.6\pm0.5$ &           $\leq$31 & $15_{-2}^{+18}$ &                  1.70 \\
156P$/$Russell-LINEAR &  $681.6\pm0.4$ &          $\leq$184 &  $18_{-2}^{+2}$ &                  1.35 \\
254P$/$McNaught &  $360.0\pm6.5$ &           $\leq$59 & $331_{-2}^{+1}$ &                  3.68 \\
398P$/$Boattini &   $72.1\pm0.1$ &           $\leq$20 &  $-4_{-4}^{+2}$ &                  1.31 \\
409P$/$LONEOS-Hill &   $53.6\pm1.0$ &           $\leq$13 & $10_{-10}^{+4}$ &                  1.75 \\
425P$/$Kowalski &   $63.3\pm3.6$ &           $\leq$12 &  $28_{-1}^{+1}$ &                  2.90 \\
449P$/$Leonard (2020 S6) &    $8.2\pm0.6$ &            $\leq$2 &  $28_{-6}^{+2}$ &                  1.88 \\
P$/$2019 LD2 (ATLAS) & $303.4\pm21.7$ &           $\leq$44 &  $63_{-6}^{+3}$ &                  4.61 \\
P$/$2020 T3 (PANSTARRS) &   $17.5\pm0.5$ &            $\leq$5 & $-17_{-4}^{+4}$ &                  1.47 \\
P$/$2020 U2 (PANSTARRS) &   $90.8\pm0.8$ &           $\leq$21 & $36_{-10}^{+6}$ &                  1.88 \\
P$/$2020 WJ5 (Lemmon) & $193.4\pm16.1$ &           $\leq$27 & $381_{-1}^{+2}$ &                  5.07 \\
P$/$2021 Q5 (ATLAS) &   $41.0\pm0.9$ &           $\leq$11 & $29_{-10}^{+6}$ &                  1.27 \\

\hline
\end{tabular}
\label{tab:table1}
\end{table*}

\subsection{Maximum $Af\rho$ and heliocentric distance} \label{subsec:afp_vs_rh}
\begin{figure}[h!]
\epsscale{1.2}
\plotone{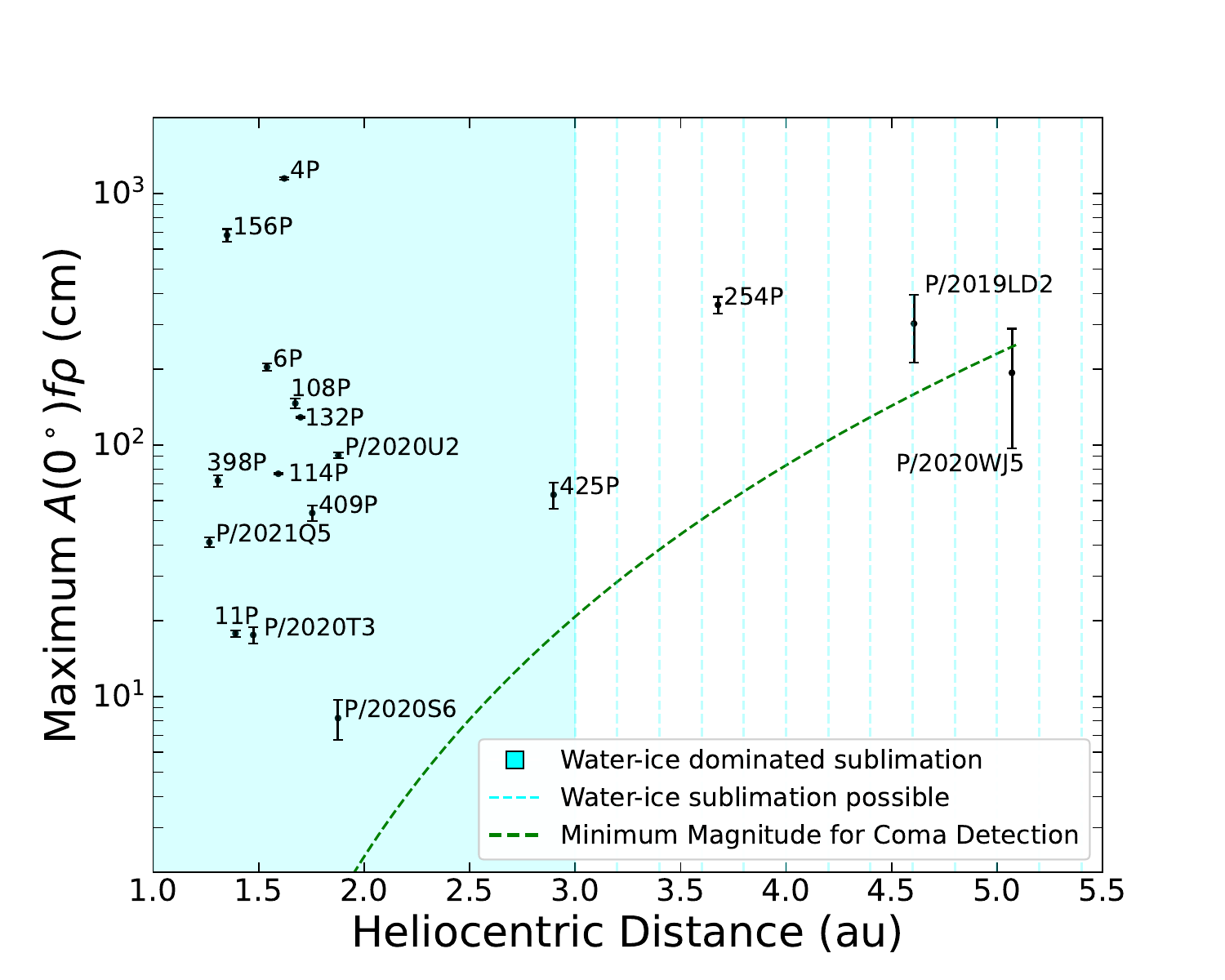}
\caption{JFC maximum activity as a function of heliocentric distance is shown. The minimum $A(0^\circ)f\rho$ for the detection of the coma is displayed by the green dashed line, assuming m(o)=18.5. In cyan, the water-ice sublimation line.}
\label{fig:fig4}
\end{figure}

Given our knowledge of the time between maximum $A(0^\circ)f\rho$ and perihelion, we utilize the same data to determine where this occurs as a function of heliocentric distance. Figure~\ref{fig:fig4} shows the expected tendency for the JFCs to reach their maximum activity near perihelion. Most of the JFCs in this study reach their maximum $A(0^\circ)f\rho$ within 2 au where water ice (H${_2}$O) is typically the primary driver of sublimation, with CO or CO${_2}$ probably dominating the activity outside of 3 au \citep{2015SSRv..197....9C}. 

From Figure~\ref{fig:fig4}, 254P$/$McNaught, P$/$2019 LD2 (ATLAS) and P$/$2020 WJ5 (Lemmon) may have activity driven by CO or CO${_2}$, with 425P$/$Kowalski in the transition region between CO/CO${_2}$ and H${_2}$O dominated sublimation at the time of maximum $A(0^\circ)f\rho$. The JFCs in this sample also display reasonably high levels of activity, explaining why they have been detected at larger heliocentric distances with ATLAS. 

The green dashed line displayed in Figure~\ref{fig:fig4} indicates the minimum $A(0^\circ)f\rho$ for the detection of the coma (and therefore dust production) with ATLAS, assuming the JFC is observed at opposition. The normal stellar limiting magnitude in the \emph{o}-band is 19.5 in dark skies. From experience, we need a magnitude of $\sim$18.5 to ensure detection of the coma. We then convert this magnitude to $A(0^\circ)f\rho$ in cm for varying heliocentric distances at opposition. Any JFC with a total magnitude far below the green dashed line would be too faint to allow an accurate coma measurement in ATLAS data. The lower right area in Figure~\ref{fig:fig4} appears unpopulated as a direct result of the ATLAS limiting magnitude. 

\subsection[AI]{$Af\rho$ activity index} \label{subsec:AI}
We define the activity index, $n$, as the power law exponent of how $A(0^\circ)f\rho$ varies with heliocentric distance, r.

\begin{equation}\label{eq3}
A(0^{\circ})f\rho \propto r^{n}
\end{equation}

As shown Figure~\ref{fig:fig2} (c), the rate of change of $Af\rho$ was measured by performing least squares fits (LSF) to the pre-perihelion and post-perihelion observations. For some comets, the activity index was measured after reaching maximum activity, rather than post-perihelion. As we wish to compare how the dust production in these JFCs varies between pre- and post-perihelion, and we know from Section~\ref{subsec:afrho_vs_days} that there is a lag to reach maximum $A(0^\circ)f\rho$, to make a meaningful comparison, we employ a linear power law fit. However, if we were to perform a power law fit exclusively for the periods before and after perihelion without considering the activity after reaching maximum levels, a linear power law fit would not provide a sensible measurement for all cases. Additionally, some comets could also only have their activity index determined pre- or post-perihelion as this was the only time they were visible to ATLAS as previously discussed. We assume a power law for the variation of activity as a function of heliocentric distance as that has been used by many previous studies \citep{{{1978M&P....18..343W}, {1997EM&P.Kidger}, {2021PSJWomack}}}. It also mirrors a simple model of a comet where the gas sublimation from the nucleus is completely controlled by solar insolation, when if heat conduction into the interior is minimal then the sublimation rate is proportional to $r^{-2}$ within 3 au.

To ensure reliable fitting conditions using the LSF, three specific criteria were established through trial and error and are as follows: 
\begin{itemize} 
    \item The data was used to determine the activity index must have had six or more observations on different nights to ensure adequate sampling.
    \item The heliocentric distance in which the activity index was measured over must have spanned \mbox{$\Delta R_h$ = 0.15 au} or greater.
    \item The final criterion was a measure of the quality of the data for each JFC. We divided each $A(0^\circ)f\rho$ measurement by the uncertainty to obtain the effective signal to noise. We then calculated an average SNR for each comet. After qualitatively assessing the data, we found JFCs with a SNR $>$7 to give acceptable measurements of $n$. 
\end{itemize}

The activity indices for the 20 JFCs that met these criteria are listed in Table~\ref{tab:table2}. Activity indices ranged from -16.1 to 0.4 pre-perihelion and -8.0 to 0.6 post-perihelion. Most of these JFCs possess negative activity indices as we would expect, but there are a few that possess positive values both in the pre- and post-perihelion samples. These can be attributed to the challenge of accurately fitting to relatively constant activity, taking into account the noise in the data.

\begin{table*}
\centering
\caption{JFC $A(0^\circ)f\rho$ and Activity Index pre- and post-perihelion. $A(0^\circ)f\rho$ pre and post perihelion were interpolated and extrapolated to 2au. Results for 28P/Neujmin 1 should be treated with caution as we observed a weak coma in the \emph{o}-band post-perihelion which implies some dust contamination, however our $Af\rho$ measurements for this JFC may be nucleus dominated.}
\begin{tabular}{lcccc}
\hline
Name & $A(0^\circ)f\rho$ pre-perihelion & $A(0^\circ)f\rho$ post-perihelion &  $n$ pre-perihelion & $n$ post-perihelion \\
\hline
4P$/$Faye &               $385.2\pm57.2$ &                $560.8\pm10.1$ &   $-4.6\pm0.3$ &    $-3.2\pm0.03$ \\         6P$/$d'Arrest &                            - &                 $121.3\pm5.1$ &    - &    $-2.2\pm0.1$ \\
11P$/$Tempel-Swift-LINEAR &                  $1.0\pm0.3$ &                   $3.6\pm0.7$ &   $-5.9\pm0.5$ &    $-3.9\pm0.4$ \\ 
28P$/$Neujmin 1 &               $114.1\pm19.3$ &               $774.0\pm188.9$ &    $0.4\pm0.2$ &    $-3.1\pm0.3$ \\    58P$/$Jackson-Neujmin &                            - &                  $31.9\pm3.6$ &              - &    $-4.0\pm0.2$ \\           108P$/$Ciffreo &                  $8.6\pm3.4$ &                  $52.3\pm1.9$ &  $-16.1\pm0.8$ &    $-4.8\pm0.1$ \\     114P$/$Wiseman-Skiff &                 $12.2\pm1.3$ &                  $17.3\pm2.3$ &   $-7.7\pm0.2$ &    $-4.5\pm0.3$ \\ 132P$/$Helin-Roman-Alu 2 &                 $14.0\pm2.0$ &                  $66.5\pm5.7$ &  $-12.2\pm0.3$ &    $-3.8\pm0.2$ \\        141P$/$Machholz 2 &                            - &                   $9.9\pm1.1$ &              - &    $-0.7\pm0.3$ \\    156P$/$Russell-LINEAR &                            - &                 $239.8\pm8.6$ &              - &    $-2.6\pm0.1$ \\             390P$/$Gibbs &               $155.6\pm20.5$ &                 $97.0\pm18.5$ &   $-5.3\pm0.2$ &    $-2.3\pm0.3$ \\            397P$/$Lemmon &                            - &                $197.0\pm33.3$ &              - &    $-8.0\pm0.2$ \\          398P$/$Boattini &                 $13.9\pm0.9$ &                   $6.8\pm0.8$ &   $-2.8\pm0.1$ &    $-5.0\pm0.2$ \\         399P$/$PANSTARRS &                $32.1\pm10.0$ &                $313.5\pm90.7$ &   $-1.6\pm0.4$ &    $-6.2\pm0.4$ \\         400P$/$PANSTARRS &                $48.2\pm16.2$ &                             - &  $-10.6\pm0.5$ &               - \\            405P$/$Lemmon &                  $3.7\pm0.6$ &                   $2.5\pm0.4$ &    $0.4\pm0.3$ &     $0.6\pm0.3$ \\       409P$/$LONEOS-Hill &                 $14.9\pm6.3$ &                  $31.2\pm2.4$ &   $-7.2\pm0.8$ &    $-3.4\pm0.1$ \\           417P$/$NEOWISE &                            - &                   $4.4\pm0.6$ &              - &     $0.6\pm0.3$ \\   P$/$2020 G1 (Pimentel) &                            - &                  $17.7\pm0.9$ &              - &     $0.4\pm0.1$ \\  P$/$2020 T3 (PANSTARRS) &                  $0.2\pm0.2$ &                             - &  $-14.8\pm2.4$ &               - \\
\hline
\end{tabular}
\label{tab:table2}
\end{table*}

\begin{figure}[ht!]
\epsscale{1.3}
\plotone{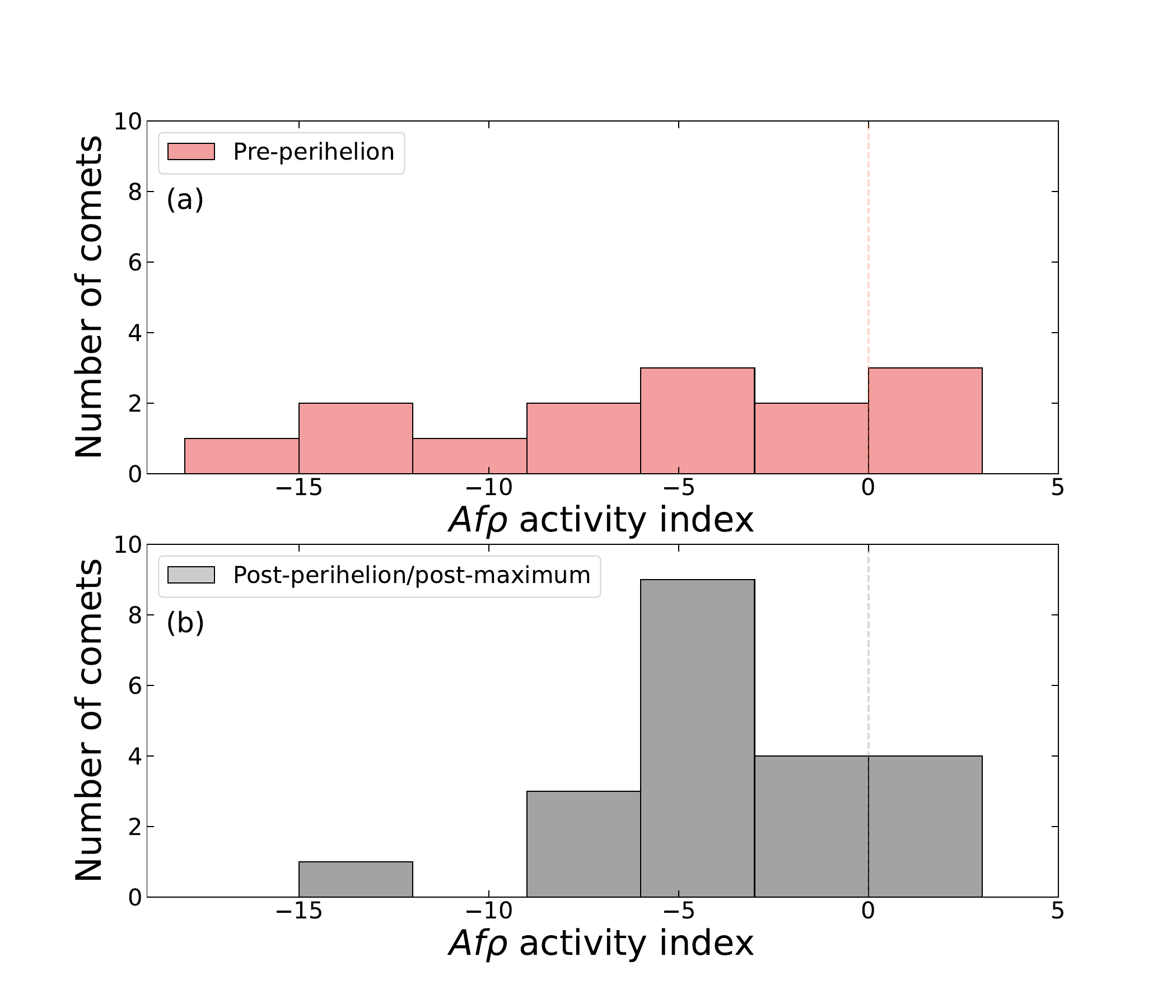}
\caption{The top panel, (a) shows the activity index, $n$, for the JFC subset pre-perihelion. Plot (b) shows $n$ for JFCs post-perihelion/post-max. Binned to $\Delta n = 3$}
\label{fig:fig5}
\end{figure}

The pre-perihelion activity indices in Figure~\ref{fig:fig5} span a larger range than the activity indices post-perihelion and does not show any preference for a particular value of $n$. This may be a result of the difference in the properties of the nuclei as discussed previously in section \ref{subsec:afrho_vs_days}.

From theoretical expectations, we would expect any gas sublimation driven activity to predict an activity index $n=-2$ as $Af\rho$ depends on distance and falls off as the inverse square. The activity index $n=-2$ will only hold true however for nuclei at small heliocentric distances where the absorbed solar energy goes into sublimation and there is no transmission of heat into the sub-surface. As the heliocentric distance increases, $n$ depends on the relative volatility of the grains and nucleus as well as the initial size of the grains \citep{1980M&P....23...41A}. We would also expect the overall amount of sublimation to increase as the comets get closer to the Sun as the increase in temperature causes a larger gas pressure gradient to be created, allowing larger dust particles to be lifted from the nucleus. If we also consider effects such as dust fragmentation, see \citet{1990A&A...231..543B} for possible dust fragmentation scenarios, we would expect a slightly steeper activity index than $n = -2$. Another potential explanation could be attributed to crust insulation. If we consider that the nucleus may be coated with a dust mantle that is partly lost as the comet approaches perihelion, this could potentially result in a more pronounced pre-perihelion activity index. Similarly, net dust fallback onto the surface as observed by the Rosetta Mission at comet 67P/Churyumov-Gerasimenko \citep{2015A&A...583A..17T} could potentially produce a steeper dust activity index. Indeed, dust fallback was found to be more important at larger heliocentric distances for 67P by \citet{2020FrP.....8..227M}.

The average activity index for all JFCs pre-perihelion is $n = -6.2 \pm 5.5$ and $n = -3.6 \pm 3.4$ for all activity indices post-perihelion. This shows that in our sample the overall rate of change of activity as a function of heliocentric distance pre-perihelion is steeper than post-perihelion. We are however dealing with small number statistics making it challenging to draw accurate conclusions. It is therefore crucial to use caution when interpreting these results.

To confirm whether the distribution of pre- and post-perihelion activity indices are the same or different, we employed the two-sample Kolmogorov-Smirnov test \citep{doi:10.1080/01621459.1951.10500769}. We obtained values of d = 0.38 and p = 0.15. Although we expected the two samples to be different, our p-value exceeded the chosen significance level of 0.05 and we therefore could not reject the null hypothesis that the two samples were drawn from the same distribution and are the same. In other words, based on our sample, we lack statistical evidence to confirm that the two samples differ significantly. The small sample size is the key factor that is attributed to the finding from the Kolmogorov-Smirnov test.

Although we are dealing with small number statistics, the conclusion from our study finds support in prior investigations, such as the work by \citet{1978M&P....18..343W} who concentrated on the comparative values of activity index pre- and post-perihelion for different classes of comet. This study also found that for JFCs the pre-perihelion activity index tends to be steeper than post-perihelion.

\subsection{Dust production of JFCs at 2 au} \label{subsec:2au}
To compensate for orbital parameters and using the activity indices from Section~\ref{subsec:AI}, we have interpolated and extrapolated the $A(0^\circ)f\rho$ to a common distance of 2 au to determine the true intrinsic variation in $A(0^\circ)f\rho$. Using 2 au allows us to assume all dust activity at this distance should be water-ice driven so we are comparing under similar sublimation conditions between JFCs. Most of the JFCs in our sample are also observed at $<$ 3 au.

To project $A(0^\circ)f\rho$ to 2 au we employed a Monte Carlo Simulation technique which relies on iteratively sampling from our dataset. We used the relationship calculated from each LSF and considered the uncertainties in the gradient, $m$, and the y-intercept, $c$. To determine the $A(0^\circ)f\rho$ at 2 au, we used random sampling from a uniform distribution between the uncertainties for $m\pm \sigma m$, and $c\pm \sigma c$, which calculated a distinct activity index, $n$, for each iteration. This process was repeated 10,000 times and $A(0^\circ)f\rho$ at 2 au was calculated for each iteration. The mean and standard deviation from the 10,000 runs produced an $A(0^\circ)f\rho$ and the associated uncertainty at 2 au for all of the JFCs in the sample. 

\begin{figure}[h!]
\epsscale{1.2}
\plotone{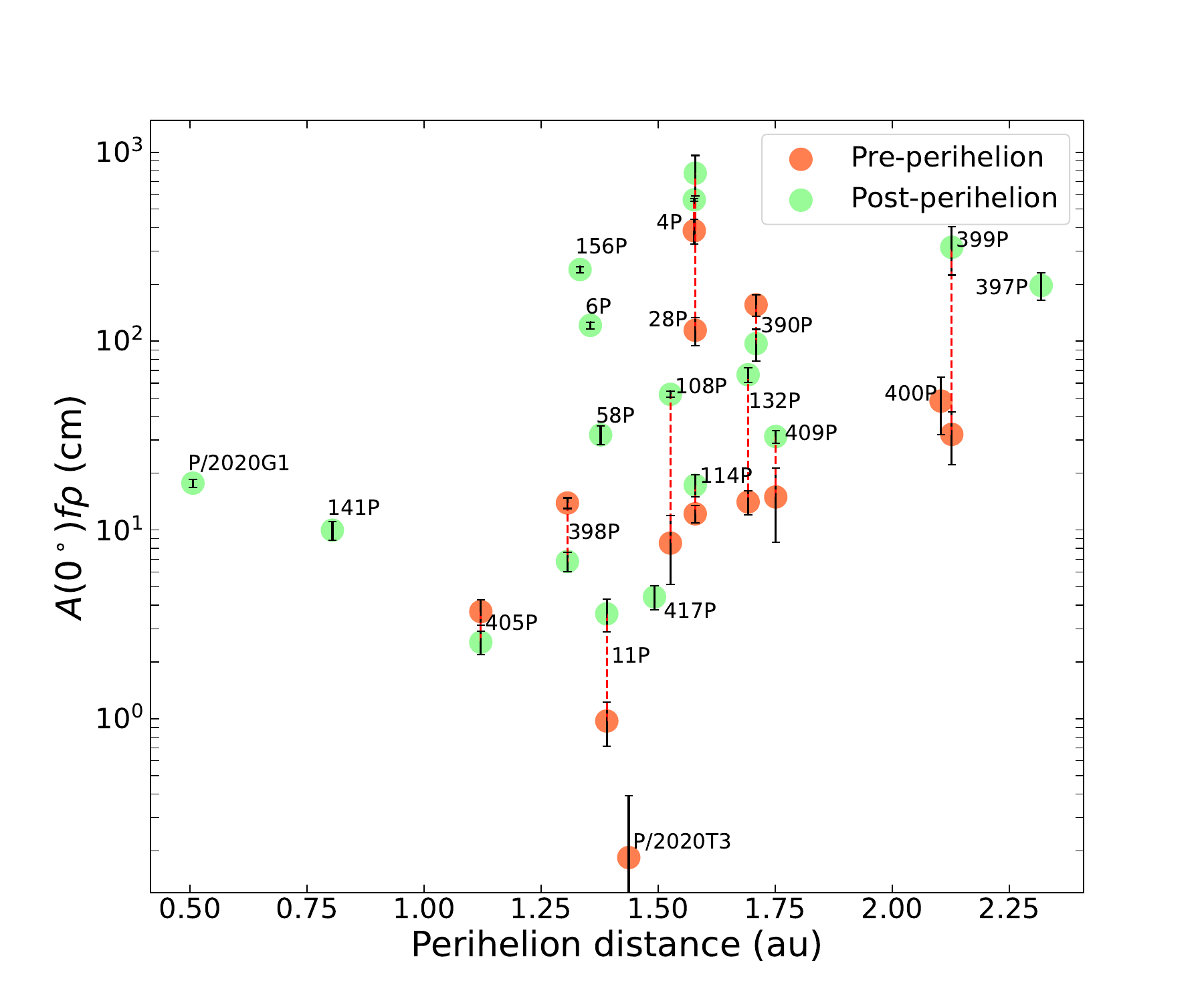}
\caption{$A(0^\circ)f\rho$ projected to 2 au vs perihelion distance for each JFC. In orange, the JFCs projected $A(0^\circ)f\rho$ at 2 au and green shows the projected $A(0^\circ)f\rho$ at 2 au post-perihelion. The red dashed line links the observations of the same JFC pre-perihelion to post-perihelion.}
\label{fig:fig6}
\end{figure}

We then tested the potential influence of perihelion distance on intrinsic activity. Figure~\ref{fig:fig6} shows the JFCs that were projected to 2 au as a function of perihelion distance. It should be noted that some of the JFCs in this sample lacked pre- or post-perihelion measurements of $A(0^\circ)f\rho$. Some JFCs in Figure~\ref{fig:fig6} have a larger $A(0^\circ)f\rho$ at 2 au pre-perihelion and others have a larger $A(0^\circ)f\rho$ measurement post-perihelion. This may be attributed to the same underlying factors discussed in Section~\ref{subsec:afrho_vs_days}.

We found that there was no discernible connection between the intrinsic activity of the JFCs and their perihelion distance. This lack of correlation may be due to the most influential factor being the nucleus itself. JFC nuclei may exhibit substantial variations in shape and size, have different fractional active areas, different regions of the nucleus containing volatile material for sublimation and have different ice to dust ratios. The presence of these numerous variables introduces significant complexity, each of which could contribute to the activity of the JFC by differing amounts. Consequently, isolating a single causal factor from the nucleus becomes a challenging task. Figure~\ref{fig:fig6} is consistent with this as we detect no trend appearing in $A(0^\circ)f\rho$ as a function of perihelion distance.

\section{Discussion}
\subsection{Nuclear radii upper limit estimates} \label{subsec:n_detects}
\begin{figure}[h!]
\epsscale{1.2}
\plotone{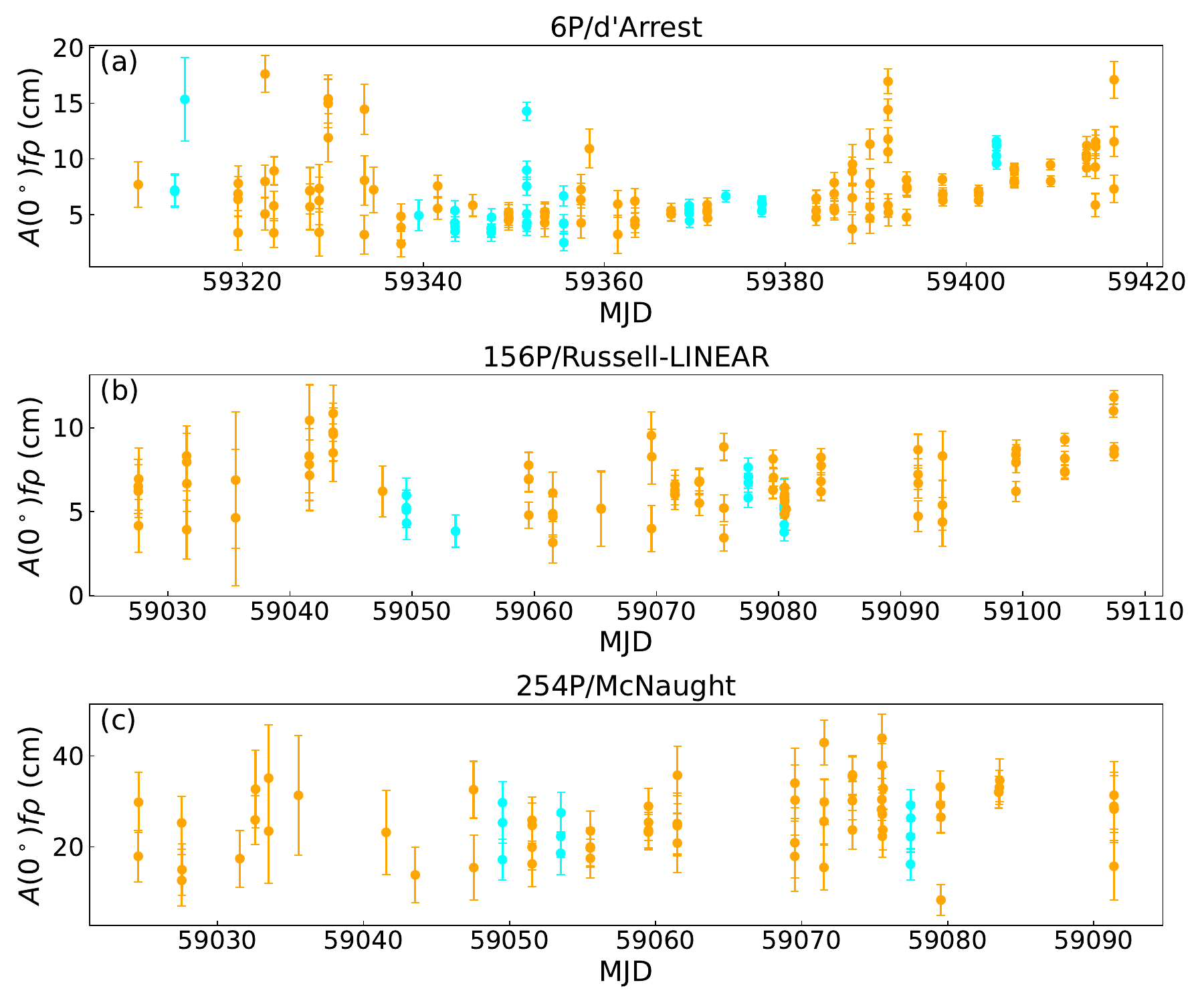}
\caption{$A(0^\circ)f\rho$ for comet 6P$/$d'Arrest during pre-perihelion dates when it exhibited constant values within the measurement uncertainties. Plots (b) and (c) show the same as for 6P$/$d'Arrest but for 156P$/$Russell-LINEAR and 254P$/$McNaught respectively. Orange and cyan points on the plots represent the \emph{o-}band and \emph{c-}band filters.}
\label{fig:Rn}
\end{figure}

For three of the JFCs in our sample, 6P/d'Arrest, 156P/Russell-LINEAR and 254P/McNaught, there was a comet detection but no visible coma in any of the images for a range of detection dates. Figure~\ref{fig:Rn} shows the corresponding pre-perihelion $A(0^\circ)f\rho$ measurements where there is no appreciable change in $A(0^\circ)f\rho$. This is consistent with photometry dominated by a direct detection of the nucleus.

As low level activity cannot be ruled out, we calculated the upper limit to the nuclear radii using the mean $A(0^\circ)f\rho$ measurement over their respective range of dates, assuming a geometric albedo of $p_o$ = 0.04 and using $\rho$ = 10,000 km as previously stated. If we considered geometric albedos from 0.02 to 0.06 for example, the $R_n$ estimates could increase by \mbox{$\sim$ 40\%} and decrease by \mbox{$\sim$ 20\%} respectively depending on the geometric albedo used. Opting for the geometric albedo normally adopted in similar studies of 0.04 \citep{1990ApJ...351..277J} however allows for the cross comparisons of JFCs between other studies to be readily made. We also assume that the dust phase function for the microscopic dust particles is the same for the macroscopic nucleus. We derive a relationship for an estimate of the nuclear radii upper limit by considering the ratio of the aperture area to the JFC cross-section in Equation~\ref{eq4}

\begin{equation}\label{eq4}
R_n = \sqrt{\frac{(Af\rho)\rho}{4 p_o}}
\end{equation}

We have calculated 6P/d'Arrest to have a nuclear radius upper limit of $R_n$ $\leq$ $2.1\pm0.3$ km, which is consistent with \citet{2013Icar..226.1138F} who used thermal measurements to find $R_n$ = $2.23_{-0.15}^{+0.13}$ km. \citet{2003Icar..164..492L} provide a lower limit of $R_n$ = $\geq 1.53\pm0.06$ km and \citet{2001AA...365..204L} provide an upper limit of $R_n$ $\leq 2.1$ km which is also consistent with the measurement reported here. \citet{2004Icar..170..463M} however measured the radii to be smaller at $R_n$ = 1.52–1.70 km. This could be due to unresolved dust coma contamination in our detections which lead to a larger $R_n$ estimate. Taking the largest and smallest estimates above could imply a nuclear elongation of a/b~1.5 assuming that these observations occurred at the minimum and maximum of the lightcurve. This is plausible given the previously measured range of nuclear elongations \citep{2017MNRAS.471.2974K}. From our measurements, 156P/Russell-LINEAR and 254P/McNaught were found to have upper limits of $R_n$ = $2.0\pm0.2$ km and $R_n$ = $4.0\pm0.8$ km respectively. These JFCs have no previous nuclear radii reported in the literature to the best of our knowledge. Overall, the $R_n$ of all three JFCs calculated in this study are consistent with the sizes of JFCs found in previous works, see \citet{2011MNRAS.414..458S}. 

\subsection[Comparison]{Comparison of JFC $Af\rho$ values with previous studies}
For each JFC in this study we produced light curves and activity plots similar to Figure~\ref{fig:fig2} which can be found in the associated, online Figure Set. In this section, we compare the $Af\rho$ values found in this study to literature and through this comparative analysis we also seek to identify any notable trends, variations, or unique features that have emerged over time as the JFCs evolve. For our sample of 42 JFCs, only a small number have had their $Af\rho$ measurements previously reported.

\subsubsection[4P]{4P$/$Faye}
\citet{2009A&A...508.1045L} used HST snapshot observations in the F675W filter to obtain an $A(0^\circ)f\rho$ value for 4P/Faye of 163 $\pm$ 8 cm at $R_h$ = 2.96 au outbound in February 2000. We obtained $A(0^\circ)f\rho$ = 119.1 $\pm$ 15.5 cm at $R_h$ = 2.96 outbound on 2022 Jun 12. Our measurements of $A(0^\circ)f\rho$ are slightly lower than the comparison study so it is possible that 4P/Faye is less active on this orbit. 4P/Faye was still very active at this distance however and was also the strongest dust emitter in our sample (Table~\ref{tab:table1}).
\citet{1996Icar..119..370L} measured the dust activity of 4P/Faye as viewed through perihelion with the HST Planetary Camera and measured $A(0^\circ)f\rho$ = 1126 cm at 1.60 au to $A(0^\circ)f\rho$ = 1256 cm at 1.59 au between October and November 1991. We cannot compare our measurements to these however, as the comet experienced an outburst at this distance in our data.

\subsubsection[6P]{6P$/$d'Arrest}
\citet{2009Icar..201..311F} used CCD spectroscopy to measure $Af\rho$ values for a large number of comets over $\sim$20 years. The $Af\rho$ values calculated in the study were all corrected to a phase angle of 40\textdegree\ using the dust scattering phase function from \citet{1981ESASP.174...47D}. We have corrected these measurements to a phase angle of 0\textdegree\ using the \citet{LowellObservatory} phase function to allow comparison to our measurements. For 6P/d'Arrest, \citet{2009Icar..201..311F} measured $A(0^\circ)f\rho$ = 39.6 cm at $R_h$ = 1.40 au on 1995 Jun 25, increasing to $A(0^\circ)f\rho$ = 335.7 cm at $R_h$ = 1.54 on 1995 Sep 30. In our ATLAS images we measure an increase of $A(0^\circ)f\rho$ = 149.2 cm to $A(0^\circ)f\rho$ = 204.0 cm over the same $R_h$ range from 2021 Oct 18 to 2021 Nov 21. Comparing the measurements between the two studies, the absolute increase in $A(0^\circ)f\rho$ in 6P/d'Arrest is larger in the 1995 apparition than in our observations from 2021. Our measurements of 6P/d'Arrest show significant gas emission.

\citet{2001A&A...365..204L} determined an $A(0^\circ)f\rho$ upper limit of $\leq$ 20.7 cm using the 4.2 m William Herschel Telescope in December 1998 at an $R_h$=5.63 au. We do not have measurements of 6P/d'Arrest beyond 2.1 au in this study so comparison is not possible.

\subsubsection[58P]{58P/Jackson-Neujmin}
Using the same study and phase function corrections as 6P/d'Arrest, \citet{2009Icar..201..311F} measured $A(0^\circ)f\rho$ = 48.4 cm at $R_h$ = 1.39 au on 1995 Sep 27 and $A(0^\circ)f\rho$ = 115.3 cm at $R_h$ = 1.54 au on 1995 Nov 28. We only have observations in this range on a single night, 2020 Jul 09 at $R_h$ = 1.47 au with $A(0^\circ)f\rho$ = 184.4 $\pm$ 0.8 cm. Our measurements show that 58P/Jackson-Neujmin was slightly more active in 2020 than in 1995. Due to Solar conjunction, we have not measured the activity of this comet near perihelion.

\subsubsection[108P]{108P$/$Ciffreo}
We made 376 observations of 108P/Ciffreo from 2021 Jun 23 to 2022 Mar 26 in the ATLAS images. 108P/Ciffreo was observed in the NEOWISE data in 2014 Aug 24 with $A(0^\circ)f\rho$ = 127.2 $\pm$ 29 cm at $R_h$ = 1.79 au by \citet{2023PSJ.....4....3G}. The authors advise their results should be used with caution due to the possible contribution of the nucleus at $R_h$ $<$ 2 au.  We measured $A(0^\circ)f\rho$ = 77.12 $\pm$ 0.59 cm at $R_h$ = 1.80 au which is slightly lower and does not agree with the previous study to within 1$\sigma$. This may be due to a combination of reasons such as the contribution of the nucleus in their measurements as they explicitly mention in their paper. However, NEOWISE makes use of the \emph{W}-band in the infrared, making comparisons between the two data sets difficult. This filter can detect different grain sizes and therefore different coma structures to the ATLAS \emph{o}-band.

\subsubsection[114P]{114P/Wiseman-Skiff} 
\citet{2009A&A...508.1045L} observed 114P/Wiseman-Skiff using Hubble Space Telescope snapshot observations in the F675W filter to obtain an $A(0^\circ)f\rho$ value of 109 $\pm$ 5 cm at 1.57 au in January 2000. The F675W filter was used as it is very close to the \emph{R}-band which is often free of major gas emission. We measured an $Af\rho$ = 50.3 $\pm$ 1.0 cm at 1.58 au in January 2020. The values are not consistent between the two measurements, considering the two measurements were made using similar filters with comparable central wavelengths. 114P/Wiseman-Skiff appears to have been less active on this orbit.

Using Isaac Newton Telescope Wide Field Camera, \citet{2008MNRAS.385..737S} observed 114P/Wiseman-Skiff in July 2005 using eight snapshot images in the \emph{r'} sloan filter at 3.75 au. The comet was unresolved and inactive at the time of observations and an upper limit of $A(0^\circ)f\rho$ $\leq$0.36 cm was calculated. We did not have any observations of this comet beyond $\sim$2.1 au however, so comparison was not possible.

\subsubsection[141P]{141P/Machholz 2} 
Using the same method as above, \citet{2009Icar..201..311F} measured $A(0^\circ)f\rho$ = 400.4 cm on 1995 Sep 05 at $R_h$ = 0.79 au and $A(0^\circ)f\rho$ = 171.4 cm on Sep 06 at $R_h$ = 0.78 au. Our measurements range from $A(0^\circ)f\rho$ = 3.2 $\pm$ 0.4 cm at $R_h$ = 0.84 au pre-perihelion to $A(0^\circ)f\rho$ = 12.1 $\pm$ 0.2 cm at $R_h$ = 0.82 au post-perihelion. Overall the two data set measurements are not consistent. The disparities in these results may be due to the \citet{2009Icar..201..311F} data being obtained spectroscopically. $Af\rho$ in circular apertures assume a $\rho^{-1}$ brightness distribution measured within a circular aperture. The transformation derived by \citet{1984ApJ...282L..43A} allows one to derive the equivalent aperture radius, $\rho$ for a rectangular aperture (slit) that would have the same filling factor, $f$ and also assumes the same $\rho^{-1}$ brightness distribution. These measurements often differ from measurements using circular apertures as shown in \citet{2019MNRAS.484.1347H}, presumably because of non-spherical symmetry in the coma and discrete structures such as jets and shells, which combine to give a brightness profile that deviates from $\rho^{-1}$. However, the primary factor contributing to the observed disparity in results for 141P/Machholz 2 is likely attributable to Fink et al.'s observation of this JFC during a fragmentation event, causing an anomalously large $A(0^\circ)f\rho$.

\subsubsection[156P]{156P/Russell-LINEAR}
\citet{2020ATel14101....1J} noted a bright and compact inner coma on 2020 Oct 14 from TRAPPIST-North at $R_h$ = 1.4 au. They measured an $A(0^\circ)f\rho$ = 23 $\pm$ 4 cm in the red dust continuum filter. We observed 156P/Russell-LINEAR on 2020 Oct 15 with ATLAS and measure an $A(0^\circ)f\rho$ = 92 $\pm$ 0.3 cm. There is clearly a large discrepancy between the two measurements with our measurements being much larger. Our measurements showed no large variation in $A(0^\circ)f\rho$ and there was no large gas contribution in our measurements so the cause of the discrepancy remains unclear.

\subsubsection[405P]{405P$/$Lemmon}
We observed 405P/Lemmon from 2020 Sep 03 - 2021 May 19 in the ATLAS images. 405P/Lemmon was observed in the NEOWISE data in May 2014 with $A(0^\circ)f\rho$ = 5.9 $\pm$ 1.8 cm at $R_h$ = 1.63 au by \citet{2023PSJ.....4....3G}. At $R_h$ = 1.63 au, we measured $A(0^\circ)f\rho$ = 3.90 $\pm$ 1.08 cm pre-perihelion. The two studies show that the comet is weakly active and the $A(0^\circ)f\rho$ values from both studies are consistent. 

\subsubsection{P/2019 LD2 (ATLAS)}
P/2019 LD2 (ATLAS) is a transitioning object between the Centaur and JFC populations and in 2063 will become a JFC \citep{2021Icar..35414019H}. \citet{2021AJ....161..116B} observed P/2019 LD2 (ATLAS) from multiple observatories in \emph{g} and \emph{r} filters and calculated $A(0^\circ)f\rho$ to increase from 85 to 204 cm between 4.7 au on 2019 April 09 to 4.6 au on 2019 November 08. We observed the comet on 2019 April 09 to 2019 October 28 and measured an increase from 91 $\pm$ 6 cm to 156 $\pm$ 37 cm. This is slightly less that that found in the previous study but may be expected as the variation in $Af\rho$ from night to night for P/2019 LD2 (ATLAS) was found to be substantial in this study and by \citet{2021AJ....161..116B}.

\citet{2022Icar..37214752B} found $Af\rho$ to vary between 230 cm and 272 cm between 2020 August 06 to August 13 at 4.59 au in \emph{R} and \emph{V} filters with an average of $Af\rho$ = 256 $\pm$ 15 cm in five observations. Using the data we obtained from ATLAS from 2020 August 06 to August 13, we calculate the $A(0^\circ)f\rho$ = 220 $\pm$ 19 cm. Our measurements agree within 2 $\sigma$.

\subsection{Outbursts}
An outburst is a sudden, unexpected increase in brightness within a few hours to several days \citep{2014AN....335..124G}. 
Examples of the potential causes of an outburst may be due to different factors or a combination of factors. Some examples of the triggering mechanisms behind outburst may be fragmentation of the nucleus, landslides on cliffs that expose fresh water-ice as see on the JFC 67P/Churyumov-Gerasimenko \citep{2016MNRAS.462S.220G} and crystallisation of amorphus ices which release trapped sub-surface gases \citep{2022arXiv220905907P} (See \citet{2021PSJ.....2..131K} and references therein for further discussion on the causes of outbursts). Over the course of our ATLAS observations, we noted five JFCs in our sample that experienced outbursts some of which have been reported previously, see: \citet{2021ATel14967....1L} and [CBET 4940, 2021 March 7] for 4P$/$Faye and 141P$/$Machholz 2 respectively. The outbursts were identified by manual inspection of the lightcurves and ATLAS images for each JFC. Table~\ref{tab:outbursts} shows the duration of the outbursts, the start date they occurred in our data, the visibility duration of the outbursts, the $R_h$ range the outburst spanned and the change in magnitude from quiescent level of activity to the maximum magnitude and the orbital phase. During a typical cometary outburst in the Jupiter family, a change of 1 to 5 magnitudes should be expected \citep{2022Icar..37514847W}. The JFCs in our sample had magnitude increases of 0.5 to 2.8 and all within 2 au of the Sun, with the outburst occurrence showing no preference to pre- or post-perihelion.

\begin{table*}
\centering
\caption{JFC outburst date range and the magnitude change. $\Delta$Magnitude is measured from before the outburst to the maximum magnitude achieved. I = inbound (pre-perihelion), O = outbound (post-perihelion).}
\begin{tabular}{lcccc}
\hline
Comet & Outburst dates start & Outburst duration  & $R_h$ & $\Delta$Magnitude\\
 & & (days) & (au) &\\
\hline
4P$/$Faye & 2021 Aug 27  &  17 & 1.6$^{I}$ & +1\\
108P$/$Ciffreo & 2021 Dec 09  &  6 & 1.9$^{O}$ & +0.5\\
132P$/$Helin-Roman-Alu 2 & 2022 Jan 21  & 24 & 1.8 - 1.9$^{O}$ & +1.8\\
141P$/$Machholz 2 & 2021 Mar 03  & 5 & 1.4 - 1.5$^{O}$ & +2.8\\
398P$/$Boattini & 2020 Sep 28  & 4 & 1.6$^{I}$ & +1.4\\ 
\hline
\end{tabular}
\label{tab:outbursts}
\end{table*}

\subsection{Undetected JFCs}
There were 14 JFCs that reached their perihelion in 2020 and 2021 that do not appear in Table~\ref{tab:table3} as they were undetected in any ATLAS images or had a predicted visible magnitude from JPL Horizons of $<$ 20.5.\\

\emph{Undetected JFCs:} 258P/PANSTARRS, 423P/Lemmon, 434P/Tenagra, P/2020 R5 (PANSTARRS), P/2020 S7 (PANSTARRS), P/2020 W1 (Rankin), P/2021 HS (PANSTARRS), P/2021 R4 (Wierzchos), P/2021 U3 (Attard-Maury) \& P/2022 C1 (PANSTARRS)\\

\emph{JPL predicted magnitude cut-off:} 323P/SOHO, P/2020 O3 (PANSTARRS), P/2020 S1 (PANSTARRS) \& P/2021 R3 (PANSTARRS)

\section{Conclusions} \label{sec:conclusion}

We have measured $A(0^\circ)f\rho$ for 42 Jupiter-family Comets reaching perihelion in 2020 and 2021 using the ATLAS survey data. The results are as follows: \\

\begin{enumerate}[nolistsep]
    \item We investigated the relationship between the maximum $A(0^\circ)f\rho$ relative to their perihelion. We found that most comets in our sample show a preference to reach maximum $A(0^\circ)f\rho$ after perihelion, usually within $\sim$3 months. We rule out thermal inertia as the prominent contributing factor to the substantial lag in our JFCs reaching maximum $A(0^\circ)f\rho$. The lag may be explained by considering the larger dust grains taking longer to move out of the aperture due to their lower ejection velocity and therefore we may be measuring dust that has been released before the date of the maximum $A(0^\circ)f\rho$. Two JFCs in our sample show a maximum $A(0^\circ)f\rho$ over a year after perihelion and this is explained by the low orbital eccentricity and much more gradual heating of the nucleus and ices. The activity in this sample spans $\sim$ 2 orders of magnitude, showing the variation in dust production among comets from the JFC population.
    
    \item The activity index, $n$, was calculated for each JFC. The distribution of $n$ pre-perihelion to post-perihelion shows that the pre-perihelion $n$ spans a larger range than post-perihelion with no preference for any value of $n$. The post-perihelion activity index shows a larger number of JFCs with a shallower $n$ implying the rate of change of activity post-perihelion is overall less than that of pre-perihelion and the dust remains in the aperture for longer after perihelion. Values for the average activity index were calculated as $n = -6.2 \pm 5.5$ pre-perihelion and the average activity index post-perihelion was $n = -3.6 \pm 3.4$. It may however be challenging to draw conclusions from the small sample size. A two-sample K-S test produced values of d = 0.38 and p = 0.15. Based on our sample, we lack statistical evidence to confirm that the two samples differ significantly.
    
    \item We have projected $A(0^\circ)f\rho$ to 2 au pre- and post-perihelion and find no trend in the $A(0^\circ)f\rho$ at 2 au with respect to perihelion distance. 
    
    \item Nuclear radii upper limit estimates were calculated using $A(0^\circ)f\rho$ measurements for three JFCs in our sample, 6P/d'Arrest, 156P/Russell-LINEAR and 254P/McNaught assuming a geometric albedo $p_o$ = 0.04 and using $\rho$ = 10,000 km. Radii of $\leq$ $2.1\pm0.3$ km, $\leq$ $2.0\pm0.2$ km and $\leq$ $4.0\pm0.8$ km respectively were measured. Our $R_n$ estimates for 6P/d'Arrest were consistent with most previous studies. 

    \item We detected five outbursts in our sample of JFCs. 4P/Faye, 108P/Ciffreo, 132P/Helin-Roman-Alu 2, 141P/Machholz 2 and 398P/Boattini all experienced outbursts between 2020 and 2022 with the largest magnitude change occurring in 141P$/$Machholz 2 where it brightened by 2.8 mag.\\
\end{enumerate}

Future all sky surveys such as the Legacy Survey of Space and Time (LSST; \citet{2019ApJ...873..111I}; \citet{2022ApJS..258....1B}) will be able to observe many JFCs throughout their whole orbit, including at aphelion. The use of facilities like ATLAS however will still be vital as its high cadence and wide field of view as previously described can be used in conjunction with larger surveys. For example there will be regions of the northern sky that the LSST will not cover. The LSST may also not have the same cadence in a single filter as other surveys such as ATLAS, Pan-STARRS and the Catalina Sky Survey which will still provide valuable measurements in this area.

\begin{acknowledgments}
We thank the two anonymous reviewers for useful comments and feedback, both of which have improved this paper. We thank Meg Schwamb and Colin Snodgrass for useful discussions. This work has made use of data from the Asteroid Terrestrial-impact Last Alert System (ATLAS) project. ATLAS is primarily funded to search for near earth asteroids through NASA grants NN12AR55G, 80NSSC18K0284, and 80NSSC18K1575; byproducts of the NEO search include images and catalogs from the survey area. The ATLAS science products have been made possible through the contributions of the University of Hawaii Institute for Astronomy, the Queen's University Belfast, the Space Telescope Science Institute, the South African Astronomical Observatory (SAAO), and the Millennium Institute of Astrophysics (MAS), Chile. A.F.G. acknowledges support from the Department for the Economy (DfE) Northern Ireland postgraduate studentship scheme. A.F. acknowledges support from STFC award ST/T00021X/1.
\end{acknowledgments}
\emph{Facility:} ATLAS (Chile, Haleakala, Mauna Loa and South Africa telescopes) 

\emph{Software:} Astropy \citep{{Astropy2013}, {price2018astropy}, {2022ApJ...935..167A}}, Jupyter Notebook \citep{Kluyver2016jupyter}, Matplotlib \citep{Hunter:2007}, Numpy \citep{harris2020array}, Pandas \citep{mckinney2010data}, Photutils \citep{larry_bradley_2022_6825092}, Python (https://www.python.org), SciPy \citep{2020SciPy-NMeth}.

\section*{Data Availability}
The photometric data used in this project is available on Zenodo under an open-source Creative Commons Attribution license: \dataset[10.5281/zenodo.10003622]{https://doi.org/10.5281/zenodo.10003622} 

\bibliography{Two_Years_of_Photometry_of_Active_Jupiter_Family_Comets_with_ATLAS}{}
\bibliographystyle{aasjournal}


\figsetstart \label{JFC_figset}
\figsetnum{1}
\figsettitle{(\textbf{Including outburst}) 4P/Faye light curve is shown in the top panel, (a). The dust production parameter $A(0^\circ)f\rho$ vs time is displayed in plot (b), and (c) shows $A(0^\circ)f\rho$ vs heliocentric distance with the solid line representing the rate of change of activity pre-perihelion and the dot-dashed line showing the rate of change of activity post-perihelion. Orange and cyan points on the plots represent the \emph{orange} filter and \emph{cyan} filter respectively. Filled circular points represent pre-perihelion measurements and open diamonds represent post-perihelion. Perihelion is marked with the dashed line in each panel. The complete figure set (47 images) is available in the online journal}

\figsetgrpstart

\figsetgrpnum{1.1}
\figsetgrptitle{4P}
\figsetplot{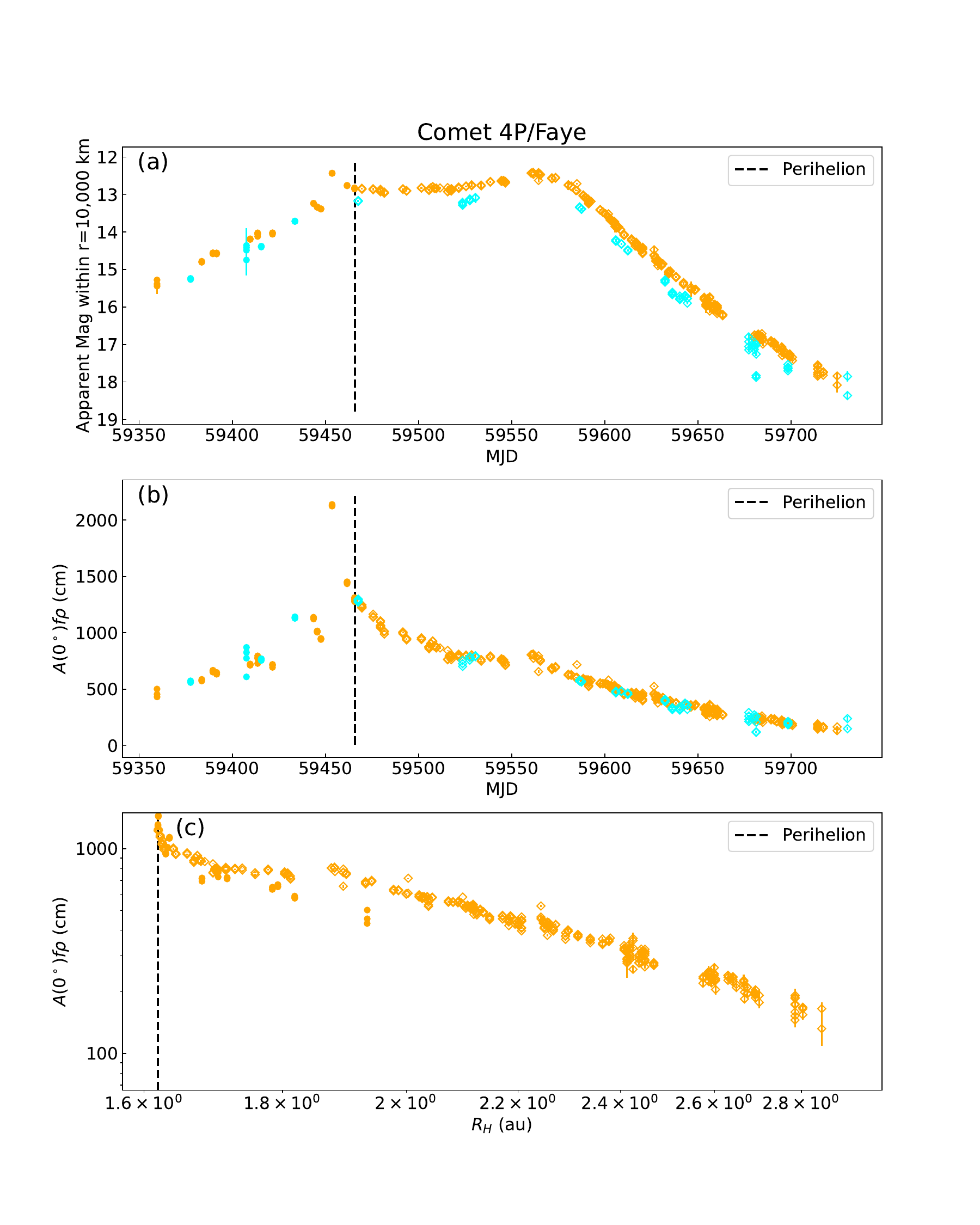}
\figsetgrpnote{(\textbf{Including outburst}) 4P/Faye light curve is shown in the top panel, (a). The dust production parameter $A(0^\circ)f\rho$ vs time is displayed in plot (b), and (c) shows $A(0^\circ)f\rho$ vs heliocentric distance with the solid line representing the rate of change of activity pre-perihelion and the dot-dashed line showing the rate of change of activity post-perihelion. Orange and cyan points on the plots represent the \emph{orange} filter and \emph{cyan} filter respectively. Filled circular points represent pre-perihelion measurements and open diamonds represent post-perihelion. Perihelion is marked with dashed line in each panel.}

\figsetgrpnum{1.2}
\figsetgrptitle{4P}
\figsetplot{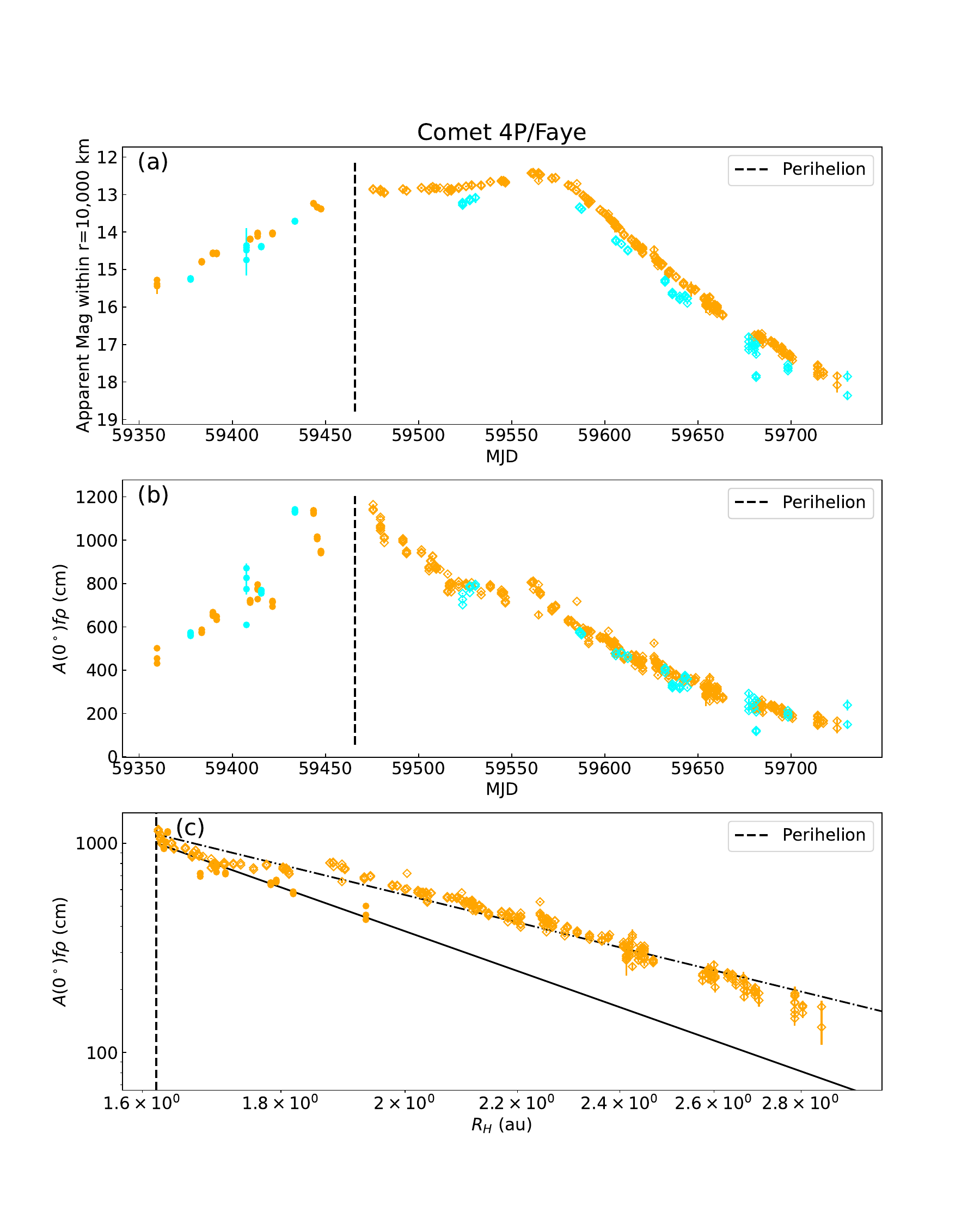}
\figsetgrpnote{\textbf{(Excluding outburst)} 4P/Faye light curve is shown in the top panel, (a). The dust production parameter $A(0^\circ)f\rho$ vs time is displayed in plot (b), and (c) shows $A(0^\circ)f\rho$ vs heliocentric distance with the solid line representing the rate of change of activity pre-perihelion and the dot-dashed line showing the rate of change of activity post-perihelion. Orange and cyan points on the plots represent the \emph{orange} filter and \emph{cyan} filter respectively. Filled circular points represent pre-perihelion measurements and open diamonds represent post-perihelion. Perihelion is marked with dashed line in each panel.}

\figsetgrpnum{1.3}
\figsetgrptitle{6P}
\figsetplot{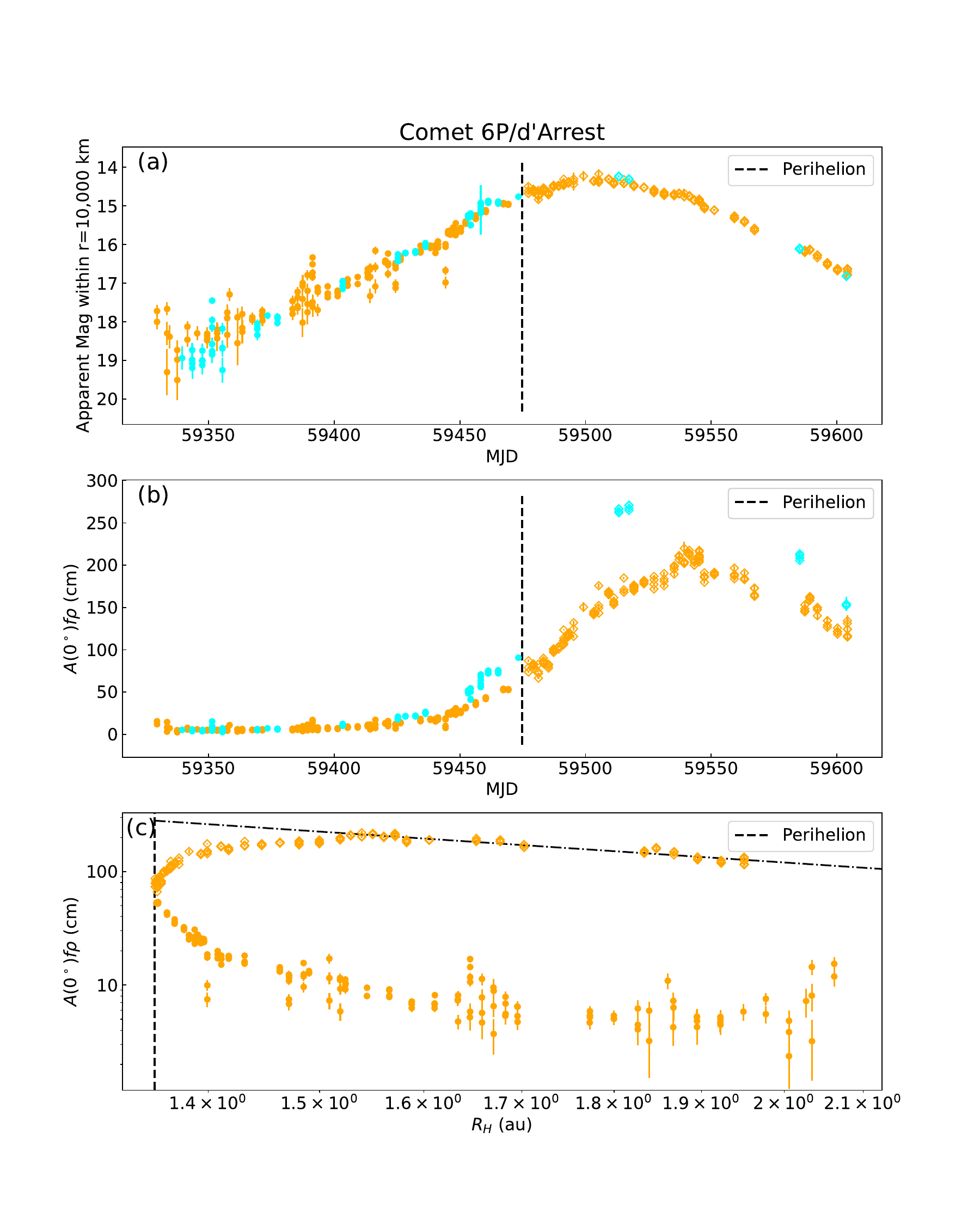}
\figsetgrpnote{6P/d'Arrest light curve is shown in the top panel, (a). The dust production parameter $A(0^\circ)f\rho$ vs time is displayed in plot (b), and (c) shows $A(0^\circ)f\rho$ vs heliocentric distance with the solid line representing the rate of change of activity pre-perihelion and the dot-dashed line showing the rate of change of activity post-perihelion. Orange and cyan points on the plots represent the \emph{orange} filter and \emph{cyan} filter respectively. Filled circular points represent pre-perihelion measurements and open diamonds represent post-perihelion. Perihelion is marked with dashed line in each panel.}

\figsetgrpnum{1.4}
\figsetgrptitle{11P}
\figsetplot{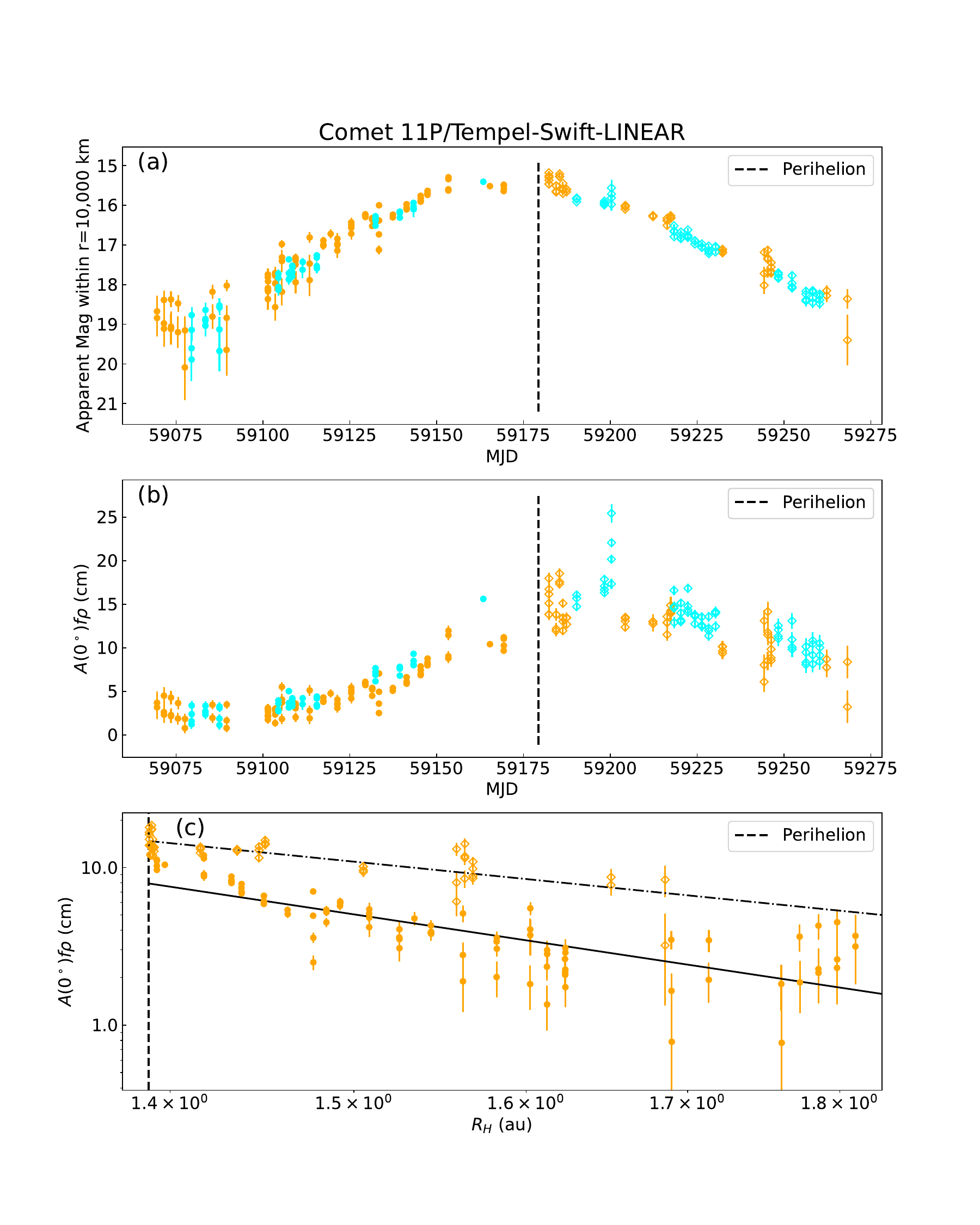}
\figsetgrpnote{11P/Tempel-Swift-LINEAR light curve is shown in the top panel, (a). The dust production parameter $A(0^\circ)f\rho$ vs time is displayed in plot (b), and (c) shows $A(0^\circ)f\rho$ vs heliocentric distance with the solid line representing the rate of change of activity pre-perihelion and the dot-dashed line showing the rate of change of activity post-perihelion. Orange and cyan points on the plots represent the \emph{orange} filter and \emph{cyan} filter respectively. Filled circular points represent pre-perihelion measurements and open diamonds represent post-perihelion. Perihelion is marked with dashed line in each panel.}

\figsetgrpnum{1.5}
\figsetgrptitle{28P}
\figsetplot{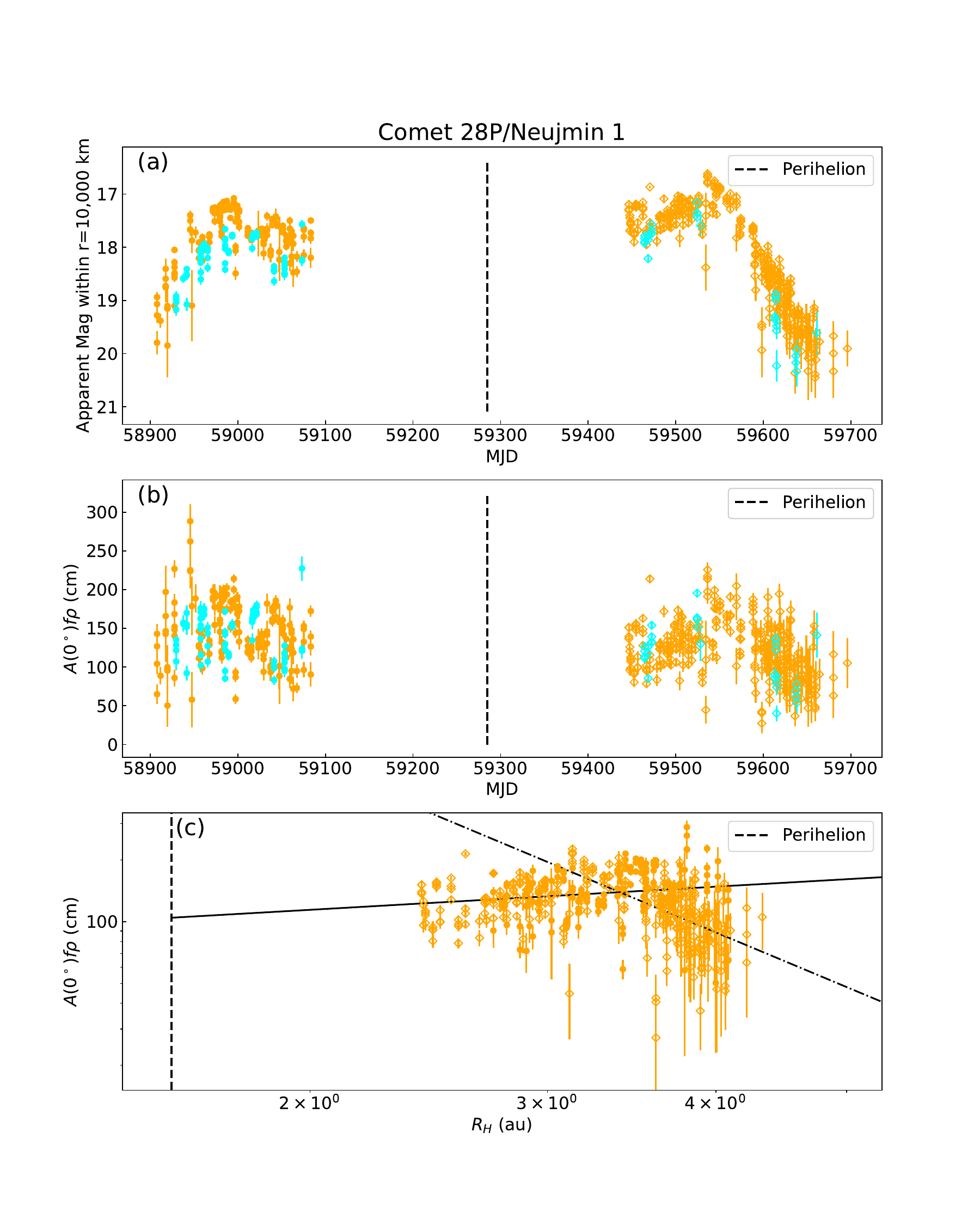}
\figsetgrpnote{28P/Neujmin 1 light curve is shown in the top panel, (a). The dust production parameter $A(0^\circ)f\rho$ vs time is displayed in plot (b), and (c) shows $A(0^\circ)f\rho$ vs heliocentric distance with the solid line representing the rate of change of activity pre-perihelion and the dot-dashed line showing the rate of change of activity post-perihelion. Orange and cyan points on the plots represent the \emph{orange} filter and \emph{cyan} filter respectively. Filled circular points represent pre-perihelion measurements and open diamonds represent post-perihelion. Perihelion is marked with dashed line in each panel.}

\figsetgrpnum{1.6}
\figsetgrptitle{58P}
\figsetplot{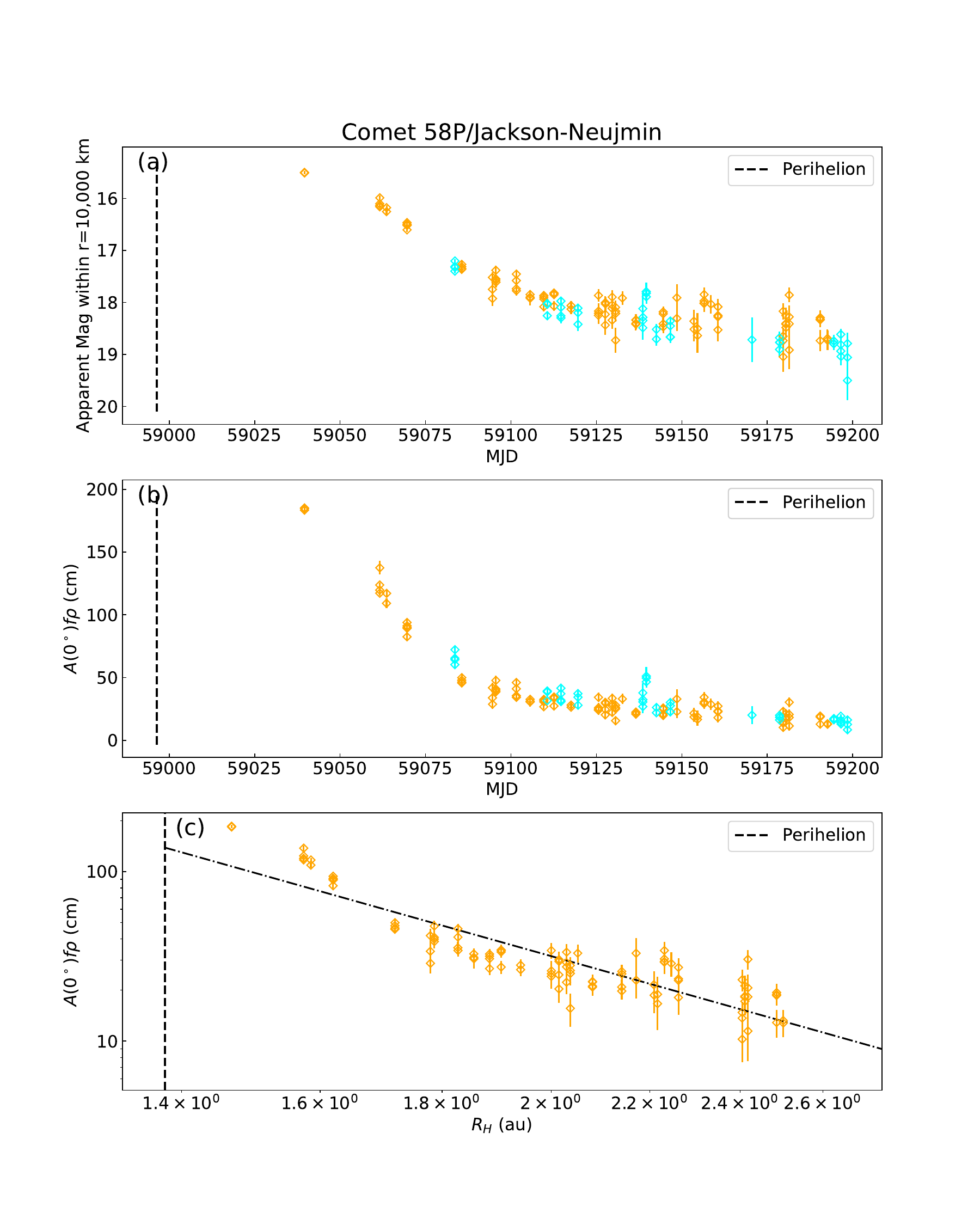}
\figsetgrpnote{58P/Jackson-Neujmin light curve is shown in the top panel, (a). The dust production parameter $A(0^\circ)f\rho$ vs time is displayed in plot (b), and (c) shows $A(0^\circ)f\rho$ vs heliocentric distance with the solid line representing the rate of change of activity pre-perihelion and the dot-dashed line showing the rate of change of activity post-perihelion. Orange and cyan points on the plots represent the \emph{orange} filter and \emph{cyan} filter respectively. Filled circular points represent pre-perihelion measurements and open diamonds represent post-perihelion. Perihelion is marked with dashed line in each panel.}

\figsetgrpnum{1.7}
\figsetgrptitle{108P}
\figsetplot{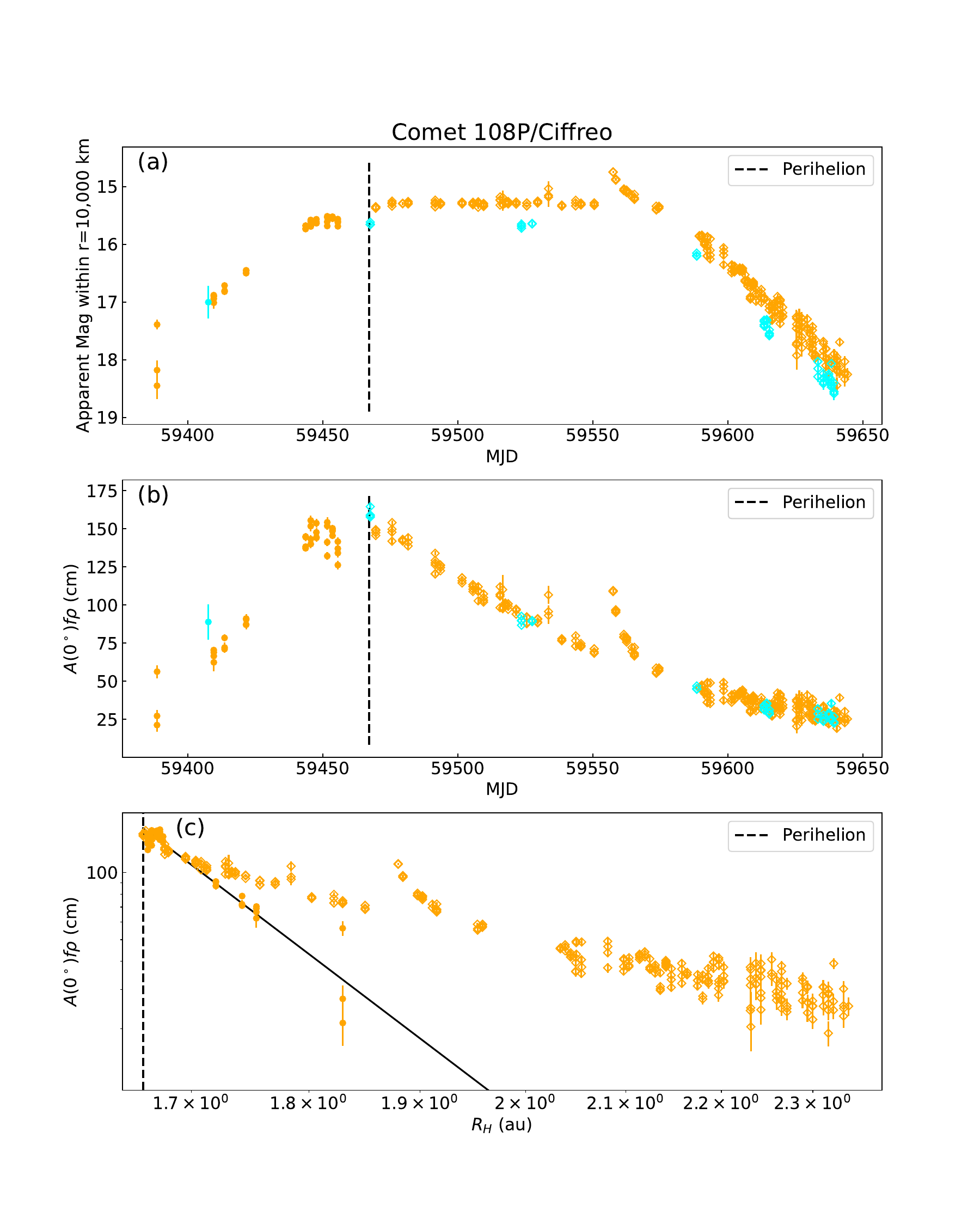}
\figsetgrpnote{\textbf{(Including outburst)} 108P/Ciffreo light curve is shown in the top panel, (a). The dust production parameter $A(0^\circ)f\rho$ vs time is displayed in plot (b), and (c) shows $A(0^\circ)f\rho$ vs heliocentric distance with the solid line representing the rate of change of activity pre-perihelion and the dot-dashed line showing the rate of change of activity post-perihelion. Orange and cyan points on the plots represent the \emph{orange} filter and \emph{cyan} filter respectively. Filled circular points represent pre-perihelion measurements and open diamonds represent post-perihelion. Perihelion is marked with dashed line in each panel.}

\figsetgrpnum{1.8}
\figsetgrptitle{108P}
\figsetplot{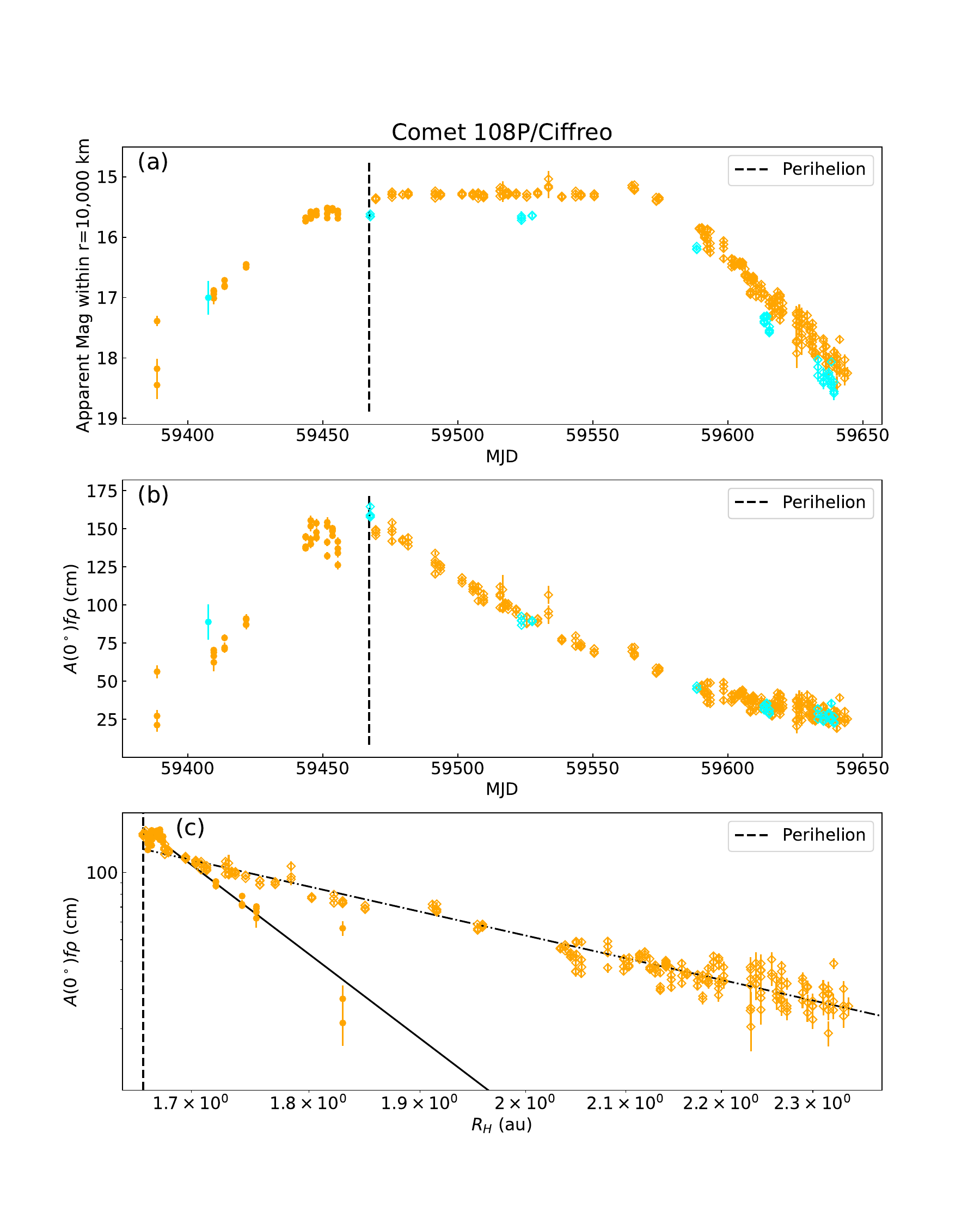}
\figsetgrpnote{\textbf{(Excluding outburst)} 108P/Ciffreo light curve is shown in the top panel, (a). The dust production parameter $A(0^\circ)f\rho$ vs time is displayed in plot (b), and (c) shows $A(0^\circ)f\rho$ vs heliocentric distance with the solid line representing the rate of change of activity pre-perihelion and the dot-dashed line showing the rate of change of activity post-perihelion. Orange and cyan points on the plots represent the \emph{orange} filter and \emph{cyan} filter respectively. Filled circular points represent pre-perihelion measurements and open diamonds represent post-perihelion. Perihelion is marked with dashed line in each panel.}

\figsetgrpnum{1.9}
\figsetgrptitle{114P}
\figsetplot{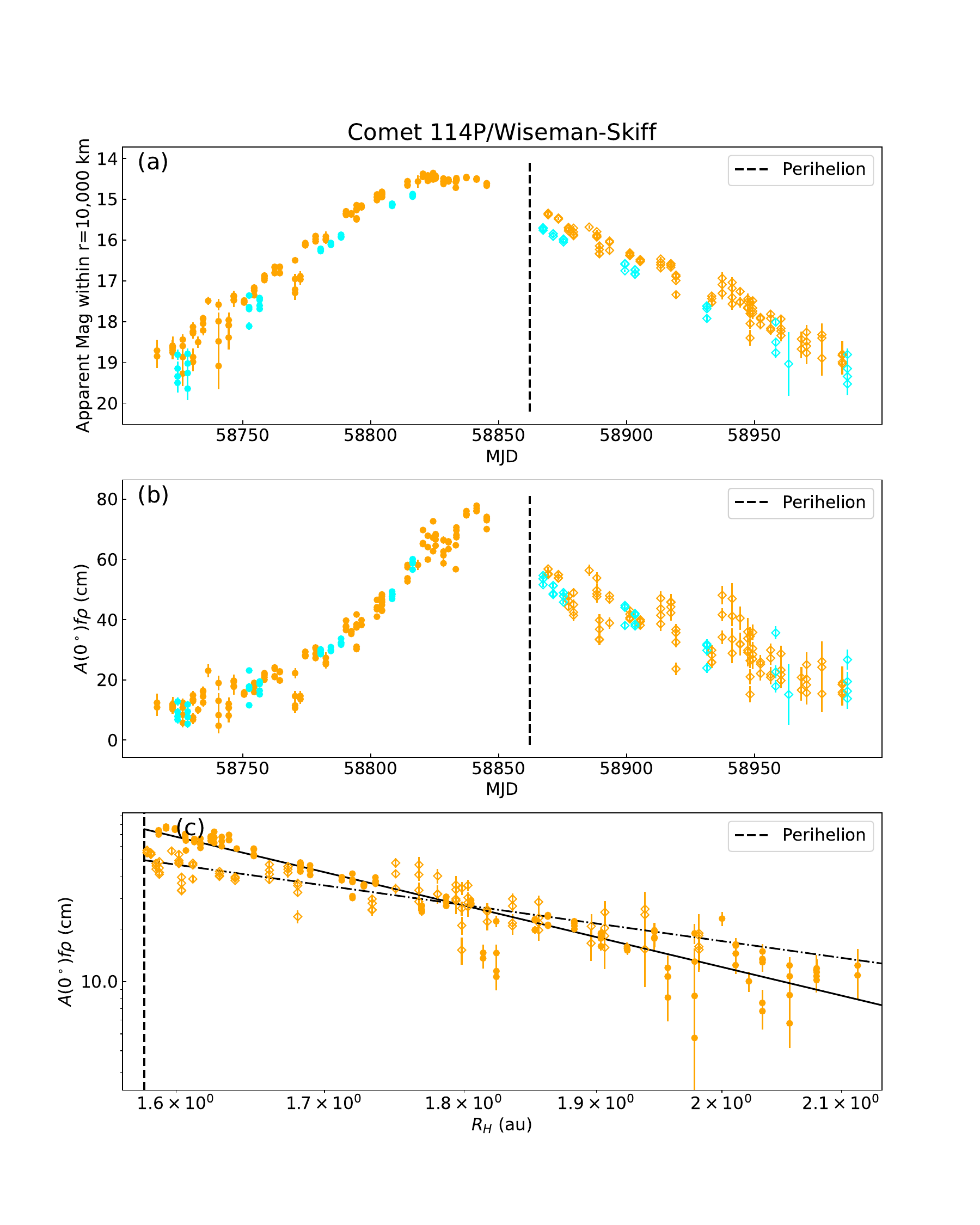}
\figsetgrpnote{114P/Wiseman-Skiff light curve is shown in the top panel, (a). The dust production parameter $A(0^\circ)f\rho$ vs time is displayed in plot (b), and (c) shows $A(0^\circ)f\rho$ vs heliocentric distance with the solid line representing the rate of change of activity pre-perihelion and the dot-dashed line showing the rate of change of activity post-perihelion. Orange and cyan points on the plots represent the \emph{orange} filter and \emph{cyan} filter respectively. Filled circular points represent pre-perihelion measurements and open diamonds represent post-perihelion. Perihelion is marked with dashed line in each panel.}

\figsetgrpnum{1.10}
\figsetgrptitle{132P}
\figsetplot{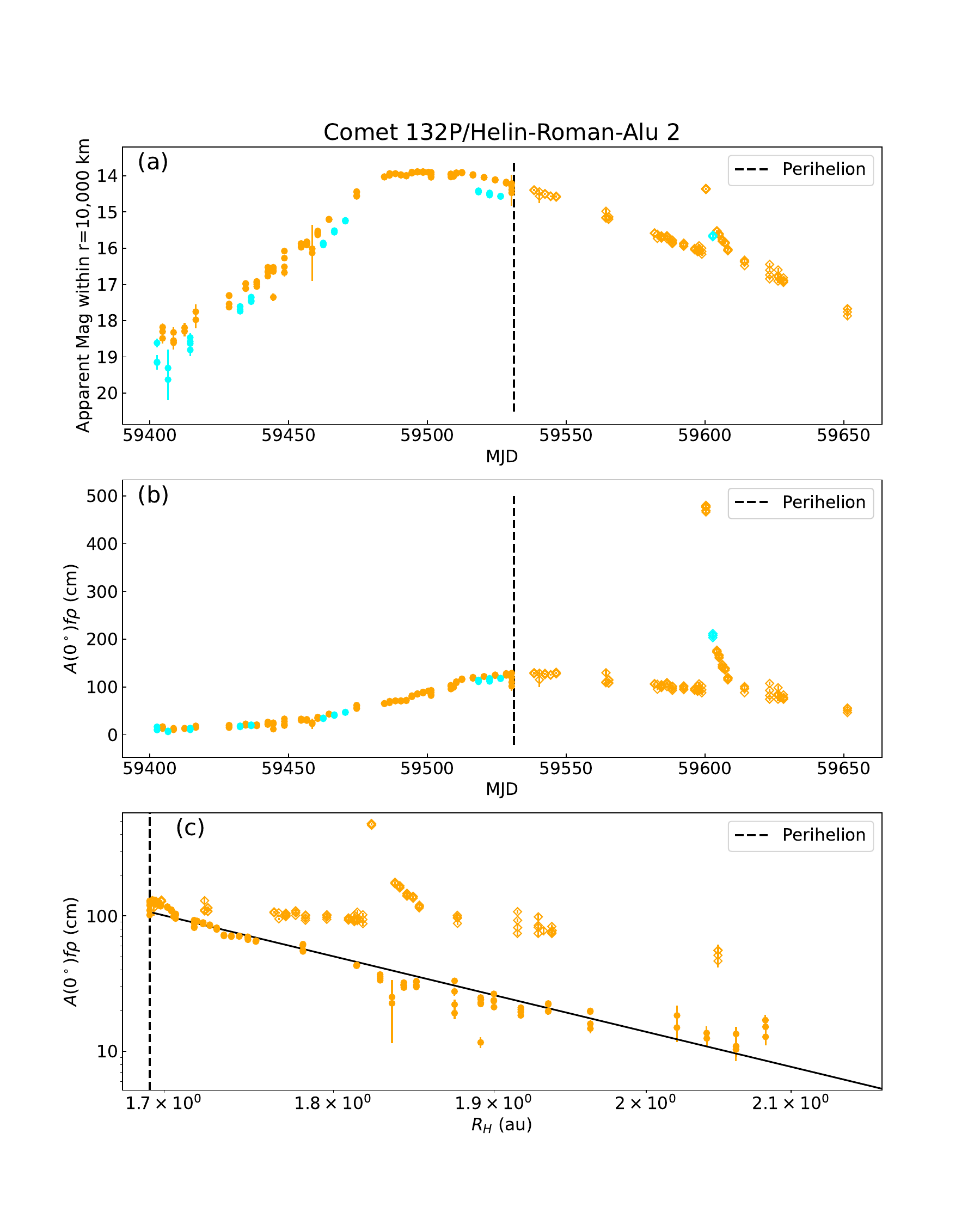}
\figsetgrpnote{\textbf{(Including outburst)} 132P/Helin-Roman-Alu 2 light curve is shown in the top panel, (a). The dust production parameter $A(0^\circ)f\rho$ vs time is displayed in plot (b), and (c) shows $A(0^\circ)f\rho$ vs heliocentric distance with the solid line representing the rate of change of activity pre-perihelion and the dot-dashed line showing the rate of change of activity post-perihelion. Orange and cyan points on the plots represent the \emph{orange} filter and \emph{cyan} filter respectively. Filled circular points represent pre-perihelion measurements and open diamonds represent post-perihelion. Perihelion is marked with dashed line in each panel.}

\figsetgrpnum{1.11}
\figsetgrptitle{132P}
\figsetplot{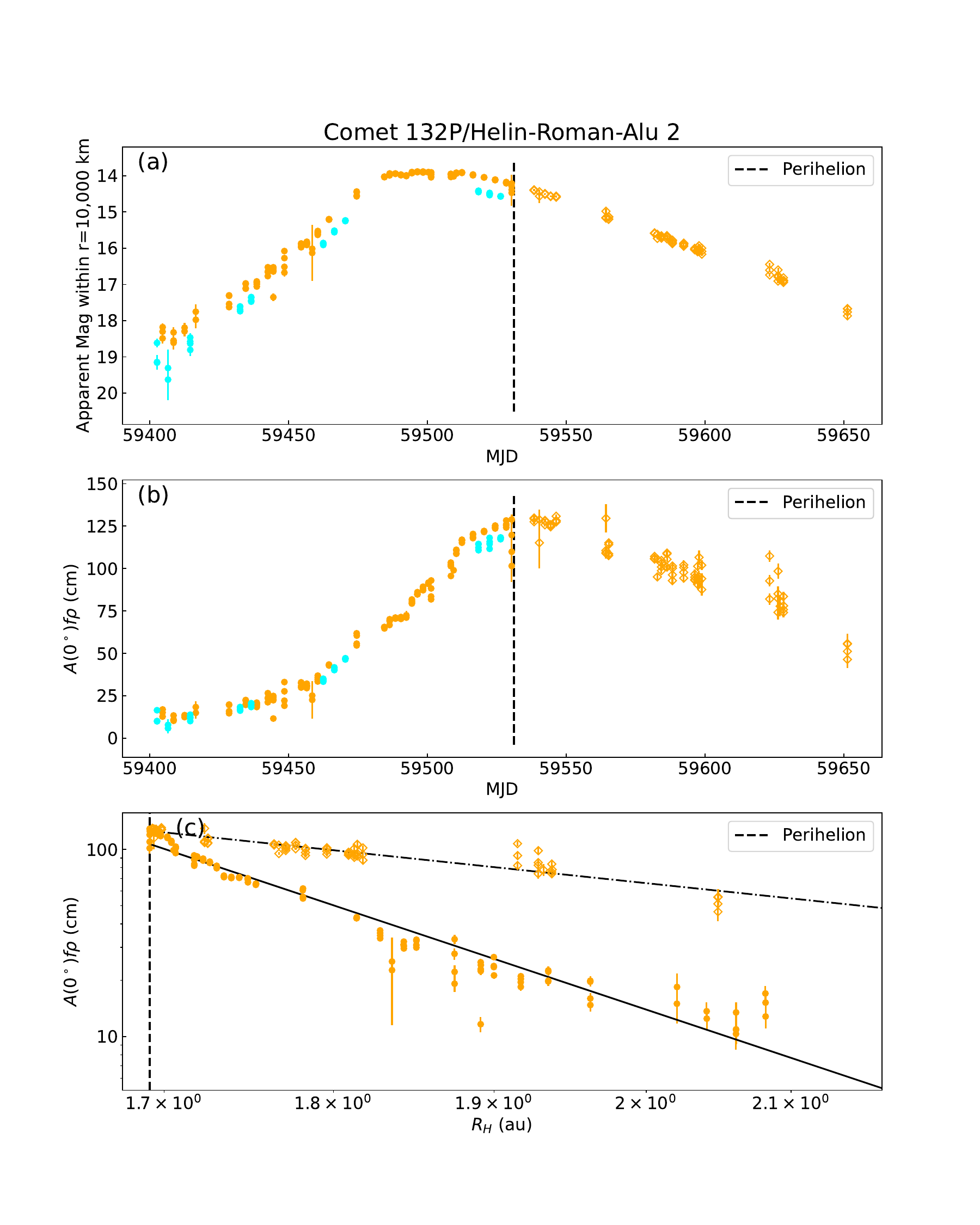}
\figsetgrpnote{\textbf{(Excluding outburst)} 132P/Helin-Roman-Alu 2 light curve is shown in the top panel, (a). The dust production parameter $A(0^\circ)f\rho$ vs time is displayed in plot (b), and (c) shows $A(0^\circ)f\rho$ vs heliocentric distance with the solid line representing the rate of change of activity pre-perihelion and the dot-dashed line showing the rate of change of activity post-perihelion. Orange and cyan points on the plots represent the \emph{orange} filter and \emph{cyan} filter respectively. Filled circular points represent pre-perihelion measurements and open diamonds represent post-perihelion. Perihelion is marked with dashed line in each panel.}

\figsetgrpnum{1.12}
\figsetgrptitle{141P}
\figsetplot{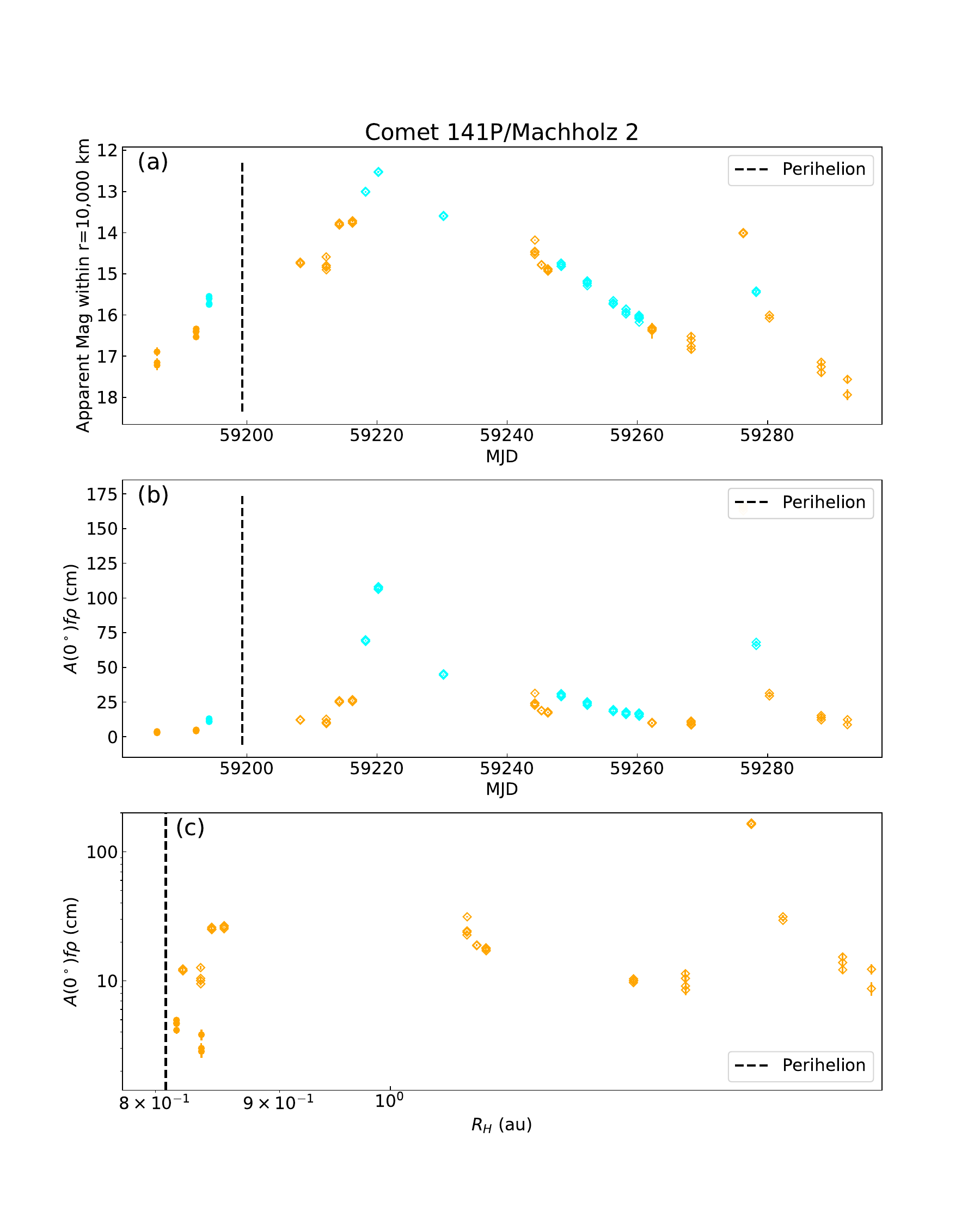}
\figsetgrpnote{\textbf{(Including outburst)} 141P/Machholz 2 light curve is shown in the top panel, (a). The dust production parameter $A(0^\circ)f\rho$ vs time is displayed in plot (b), and (c) shows $A(0^\circ)f\rho$ vs heliocentric distance with the solid line representing the rate of change of activity pre-perihelion and the dot-dashed line showing the rate of change of activity post-perihelion. Orange and cyan points on the plots represent the \emph{orange} filter and \emph{cyan} filter respectively. Filled circular points represent pre-perihelion measurements and open diamonds represent post-perihelion. Perihelion is marked with dashed line in each panel.}

\figsetgrpnum{1.13}
\figsetgrptitle{141P}
\figsetplot{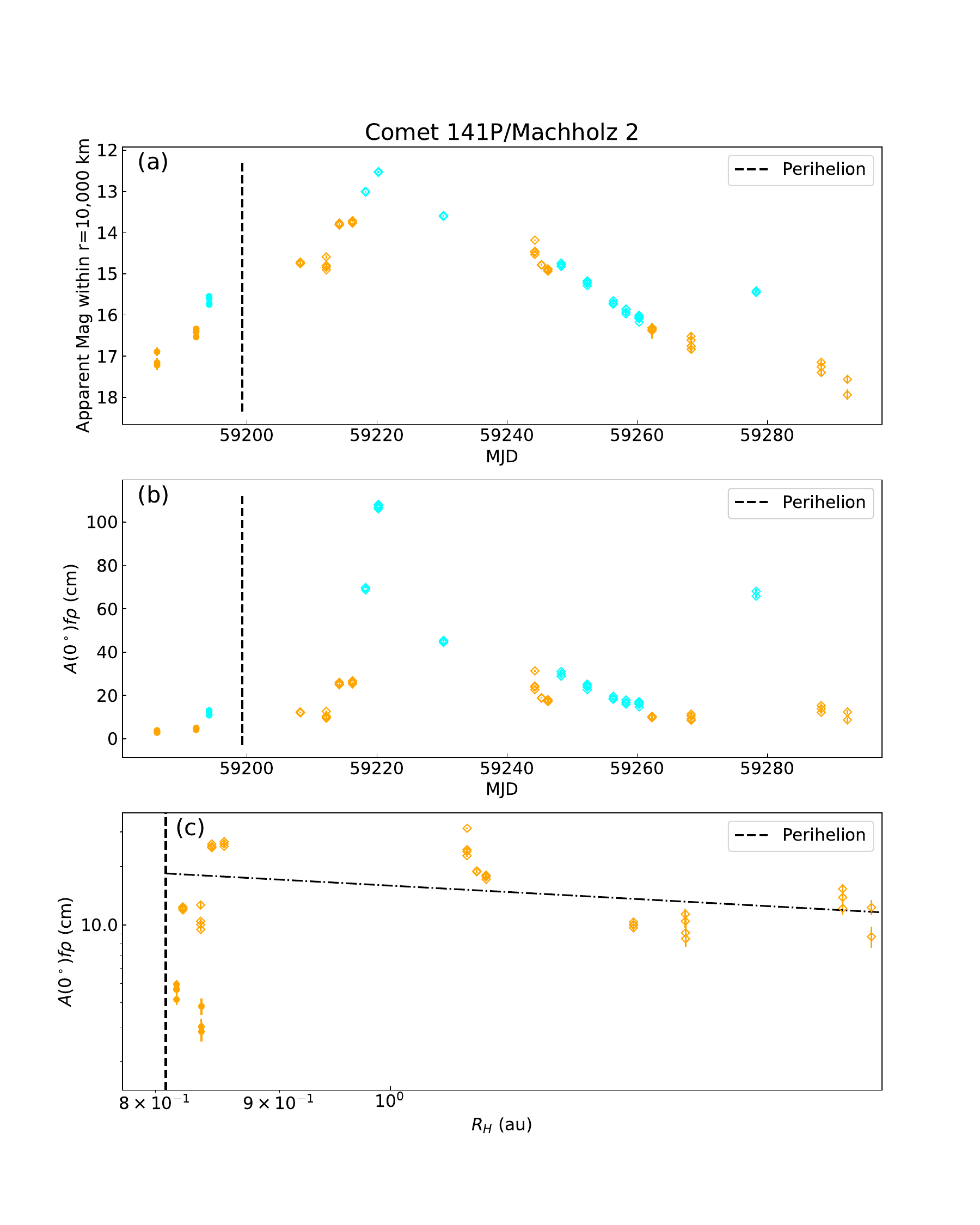}
\figsetgrpnote{\textbf{(Excluding outburst)} 141P/Machholz 2 light curve is shown in the top panel, (a). The dust production parameter $A(0^\circ)f\rho$ vs time is displayed in plot (b), and (c) shows $A(0^\circ)f\rho$ vs heliocentric distance with the solid line representing the rate of change of activity pre-perihelion and the dot-dashed line showing the rate of change of activity post-perihelion. Orange and cyan points on the plots represent the \emph{orange} filter and \emph{cyan} filter respectively. Filled circular points represent pre-perihelion measurements and open diamonds represent post-perihelion. Perihelion is marked with dashed line in each panel.}

\figsetgrpnum{1.14}
\figsetgrptitle{156P}
\figsetplot{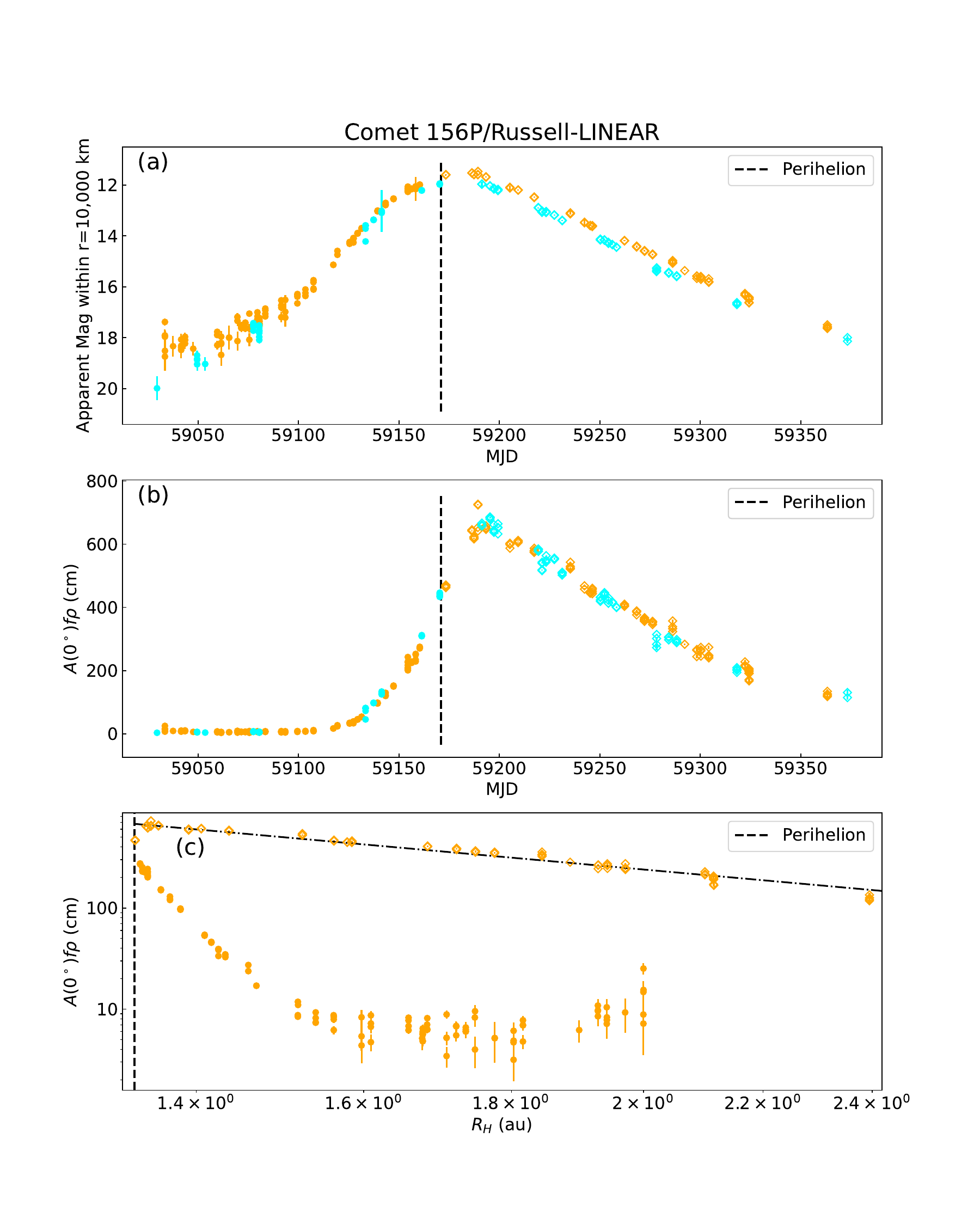}
\figsetgrpnote{156P/Russell-LINEAR light curve is shown in the top panel, (a). The dust production parameter $A(0^\circ)f\rho$ vs time is displayed in plot (b), and (c) shows $A(0^\circ)f\rho$ vs heliocentric distance with the solid line representing the rate of change of activity pre-perihelion and the dot-dashed line showing the rate of change of activity post-perihelion. Orange and cyan points on the plots represent the \emph{orange} filter and \emph{cyan} filter respectively. Filled circular points represent pre-perihelion measurements and open diamonds represent post-perihelion. Perihelion is marked with dashed line in each panel.}

\figsetgrpnum{1.15}
\figsetgrptitle{254P}
\figsetplot{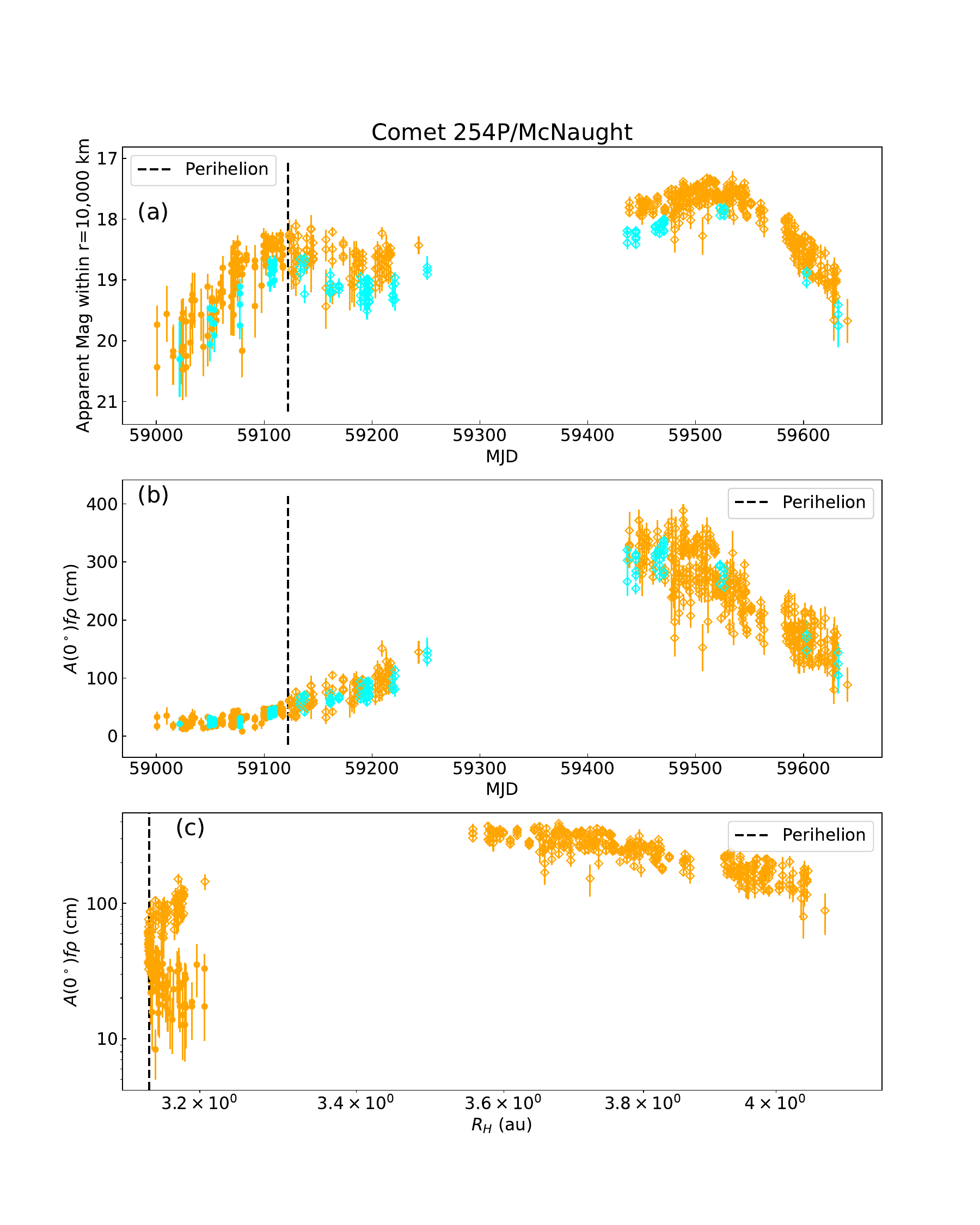}
\figsetgrpnote{254P/McNaught light curve is shown in the top panel, (a). The dust production parameter $A(0^\circ)f\rho$ vs time is displayed in plot (b), and (c) shows $A(0^\circ)f\rho$ vs heliocentric distance with the solid line representing the rate of change of activity pre-perihelion and the dot-dashed line showing the rate of change of activity post-perihelion. Orange and cyan points on the plots represent the \emph{orange} filter and \emph{cyan} filter respectively. Filled circular points represent pre-perihelion measurements and open diamonds represent post-perihelion. Perihelion is marked with dashed line in each panel.}

\figsetgrpnum{1.16}
\figsetgrptitle{312P}
\figsetplot{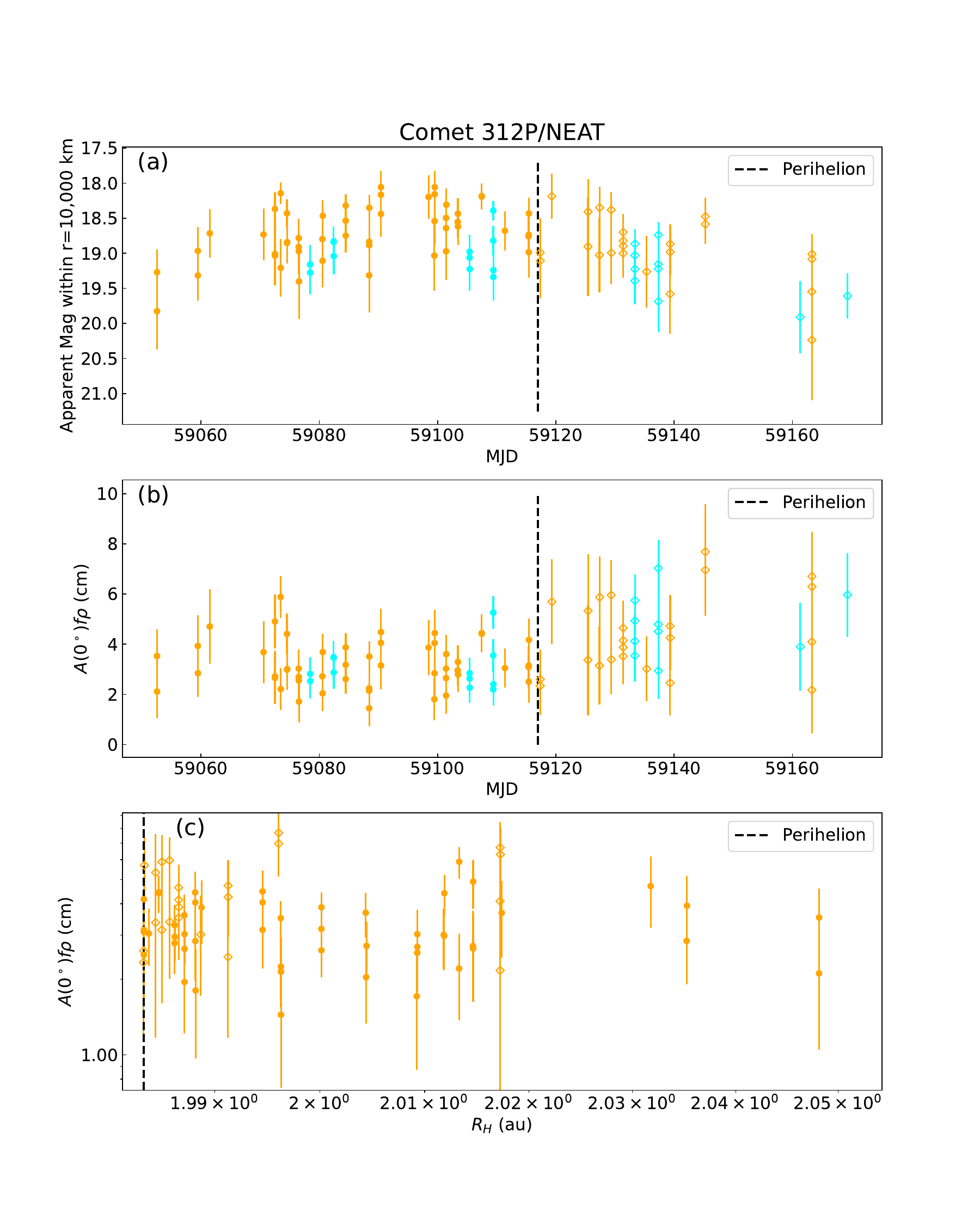}
\figsetgrpnote{312P/NEAT light curve is shown in the top panel, (a). The dust production parameter $A(0^\circ)f\rho$ vs time is displayed in plot (b), and (c) shows $A(0^\circ)f\rho$ vs heliocentric distance with the solid line representing the rate of change of activity pre-perihelion and the dot-dashed line showing the rate of change of activity post-perihelion. Orange and cyan points on the plots represent the \emph{orange} filter and \emph{cyan} filter respectively. Filled circular points represent pre-perihelion measurements and open diamonds represent post-perihelion. Perihelion is marked with dashed line in each panel.}

\figsetgrpnum{1.17}
\figsetgrptitle{377P}
\figsetplot{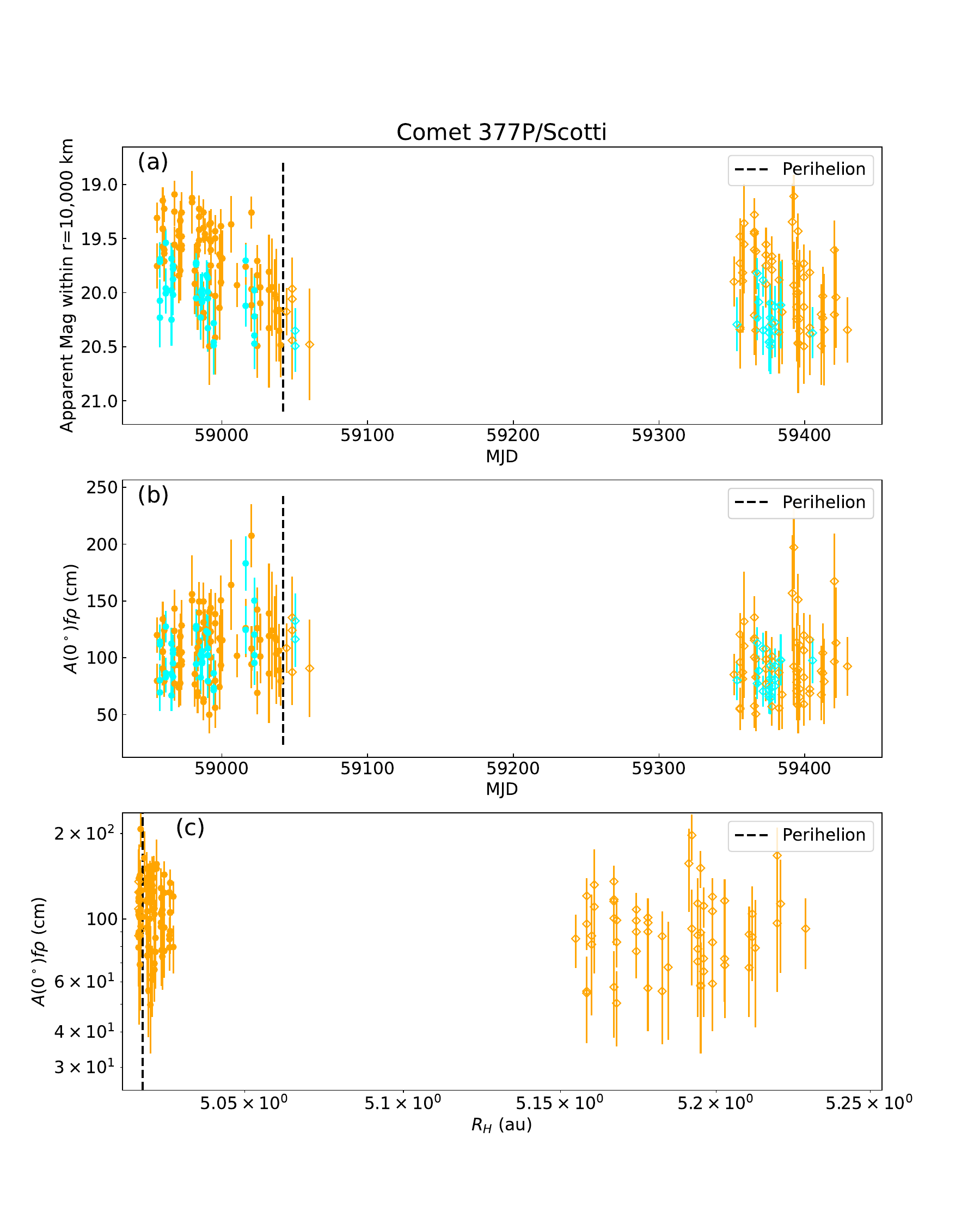}
\figsetgrpnote{377P/Scotti light curve is shown in the top panel, (a). The dust production parameter $A(0^\circ)f\rho$ vs time is displayed in plot (b), and (c) shows $A(0^\circ)f\rho$ vs heliocentric distance with the solid line representing the rate of change of activity pre-perihelion and the dot-dashed line showing the rate of change of activity post-perihelion. Orange and cyan points on the plots represent the \emph{orange} filter and \emph{cyan} filter respectively. Filled circular points represent pre-perihelion measurements and open diamonds represent post-perihelion. Perihelion is marked with dashed line in each panel.}

\figsetgrpnum{1.18}
\figsetgrptitle{378P}
\figsetplot{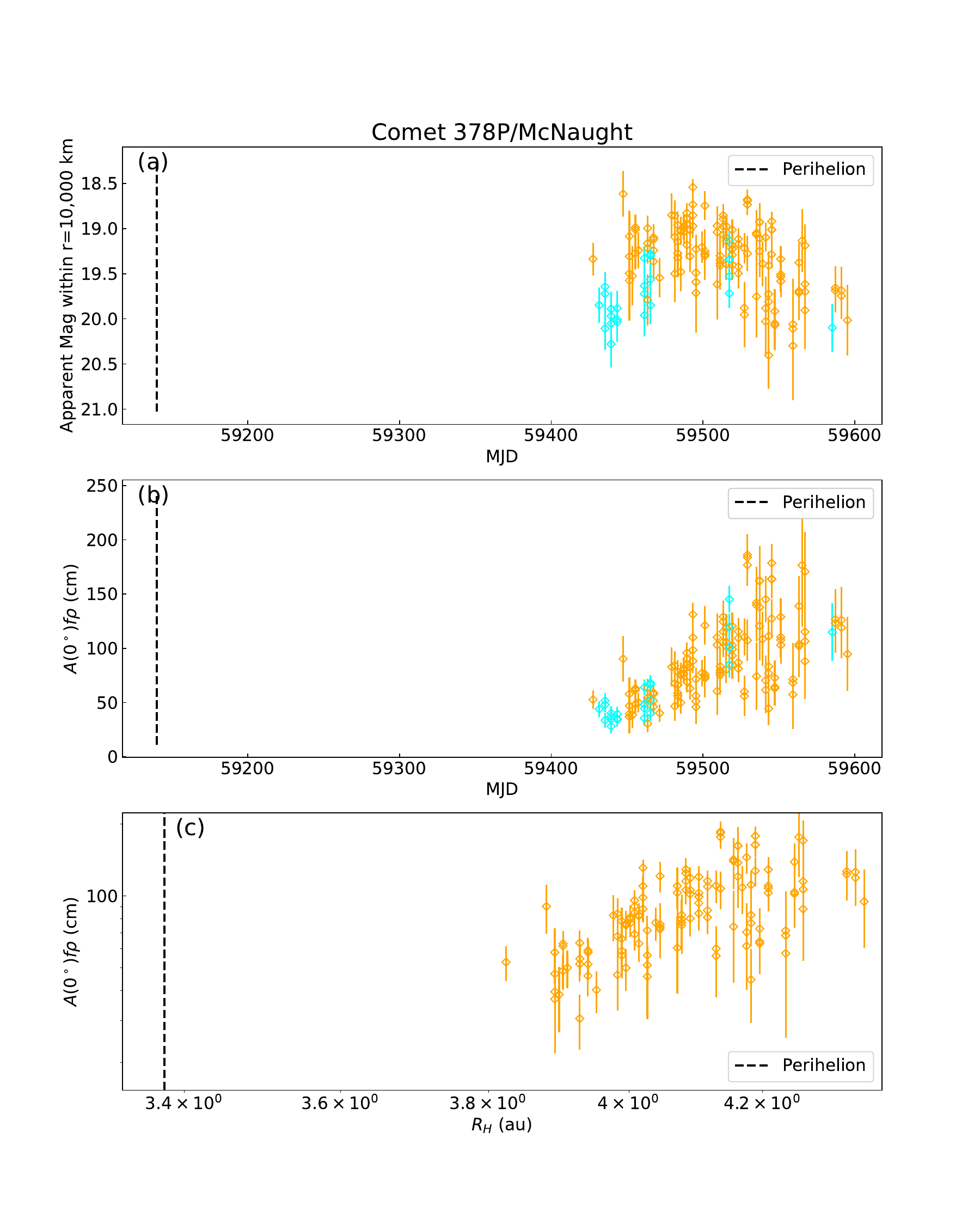}
\figsetgrpnote{378P/McNaught light curve is shown in the top panel, (a). The dust production parameter $A(0^\circ)f\rho$ vs time is displayed in plot (b), and (c) shows $A(0^\circ)f\rho$ vs heliocentric distance with the solid line representing the rate of change of activity pre-perihelion and the dot-dashed line showing the rate of change of activity post-perihelion. Orange and cyan points on the plots represent the \emph{orange} filter and \emph{cyan} filter respectively. Filled circular points represent pre-perihelion measurements and open diamonds represent post-perihelion. Perihelion is marked with dashed line in each panel.}

\figsetgrpnum{1.19}
\figsetgrptitle{390P}
\figsetplot{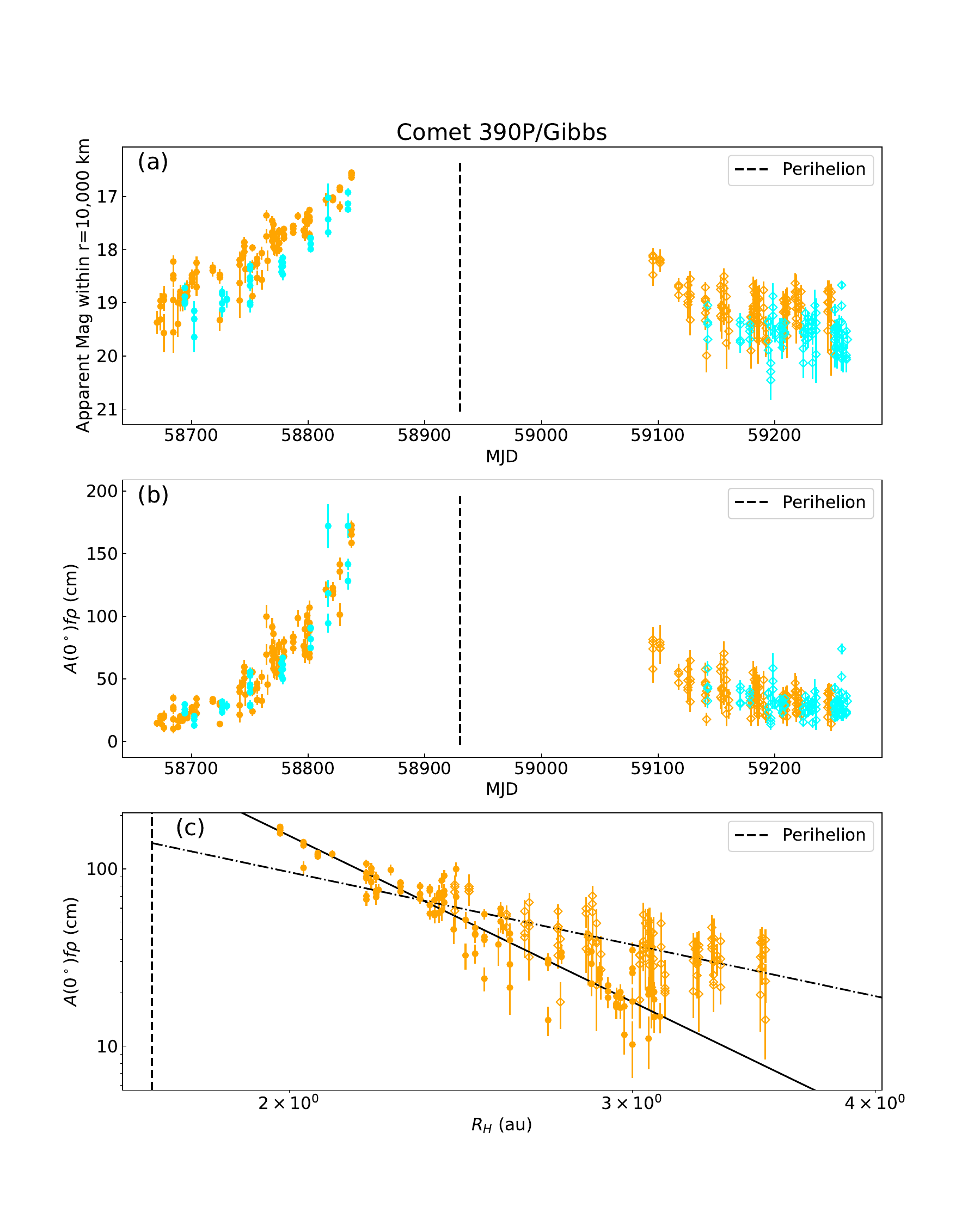}
\figsetgrpnote{390P/Gibbs light curve is shown in the top panel, (a). The dust production parameter $A(0^\circ)f\rho$ vs time is displayed in plot (b), and (c) shows $A(0^\circ)f\rho$ vs heliocentric distance with the solid line representing the rate of change of activity pre-perihelion and the dot-dashed line showing the rate of change of activity post-perihelion. Orange and cyan points on the plots represent the \emph{orange} filter and \emph{cyan} filter respectively. Filled circular points represent pre-perihelion measurements and open diamonds represent post-perihelion. Perihelion is marked with dashed line in each panel.}

\figsetgrpnum{1.20}
\figsetgrptitle{397P}
\figsetplot{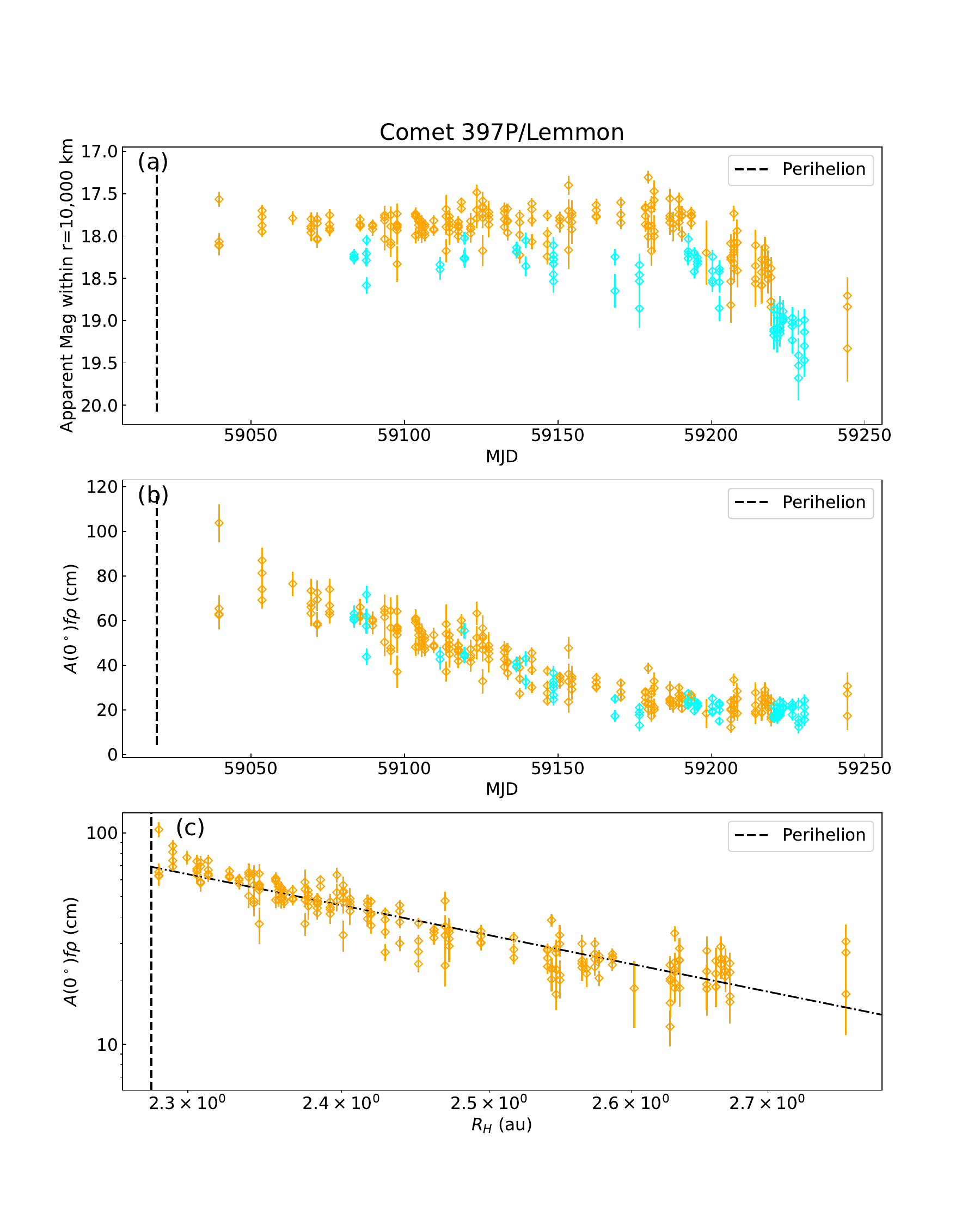}
\figsetgrpnote{397P/Lemmon light curve is shown in the top panel, (a). The dust production parameter $A(0^\circ)f\rho$ vs time is displayed in plot (b), and (c) shows $A(0^\circ)f\rho$ vs heliocentric distance with the solid line representing the rate of change of activity pre-perihelion and the dot-dashed line showing the rate of change of activity post-perihelion. Orange and cyan points on the plots represent the \emph{orange} filter and \emph{cyan} filter respectively. Filled circular points represent pre-perihelion measurements and open diamonds represent post-perihelion. Perihelion is marked with dashed line in each panel.}

\figsetgrpnum{1.21}
\figsetgrptitle{398P}
\figsetplot{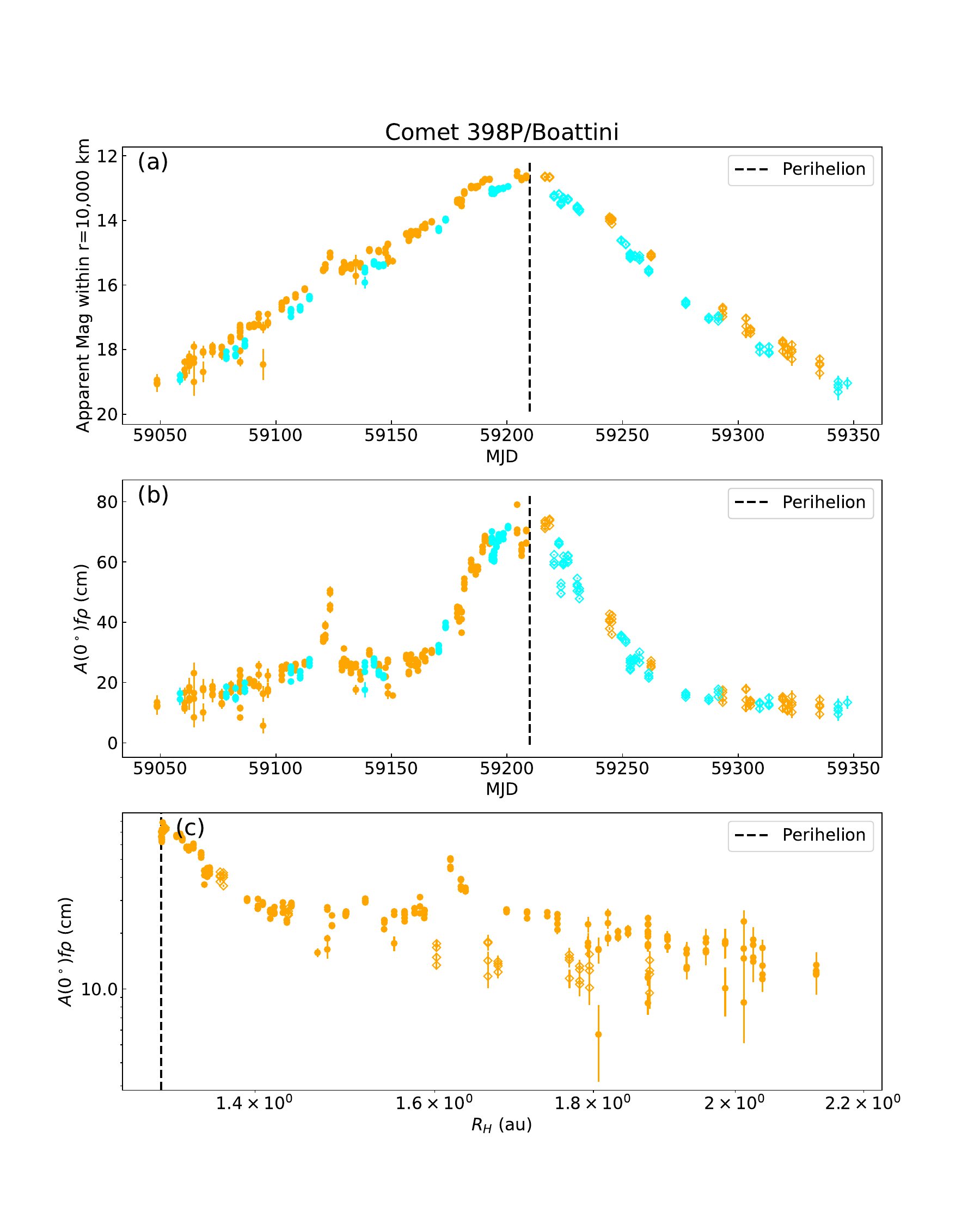}
\figsetgrpnote{\textbf{(Including outburst)} 398P/Boattini light curve is shown in the top panel, (a). The dust production parameter $A(0^\circ)f\rho$ vs time is displayed in plot (b), and (c) shows $A(0^\circ)f\rho$ vs heliocentric distance with the solid line representing the rate of change of activity pre-perihelion and the dot-dashed line showing the rate of change of activity post-perihelion. Orange and cyan points on the plots represent the \emph{orange} filter and \emph{cyan} filter respectively. Filled circular points represent pre-perihelion measurements and open diamonds represent post-perihelion. Perihelion is marked with dashed line in each panel.}

\figsetgrpnum{1.22}
\figsetgrptitle{398P}
\figsetplot{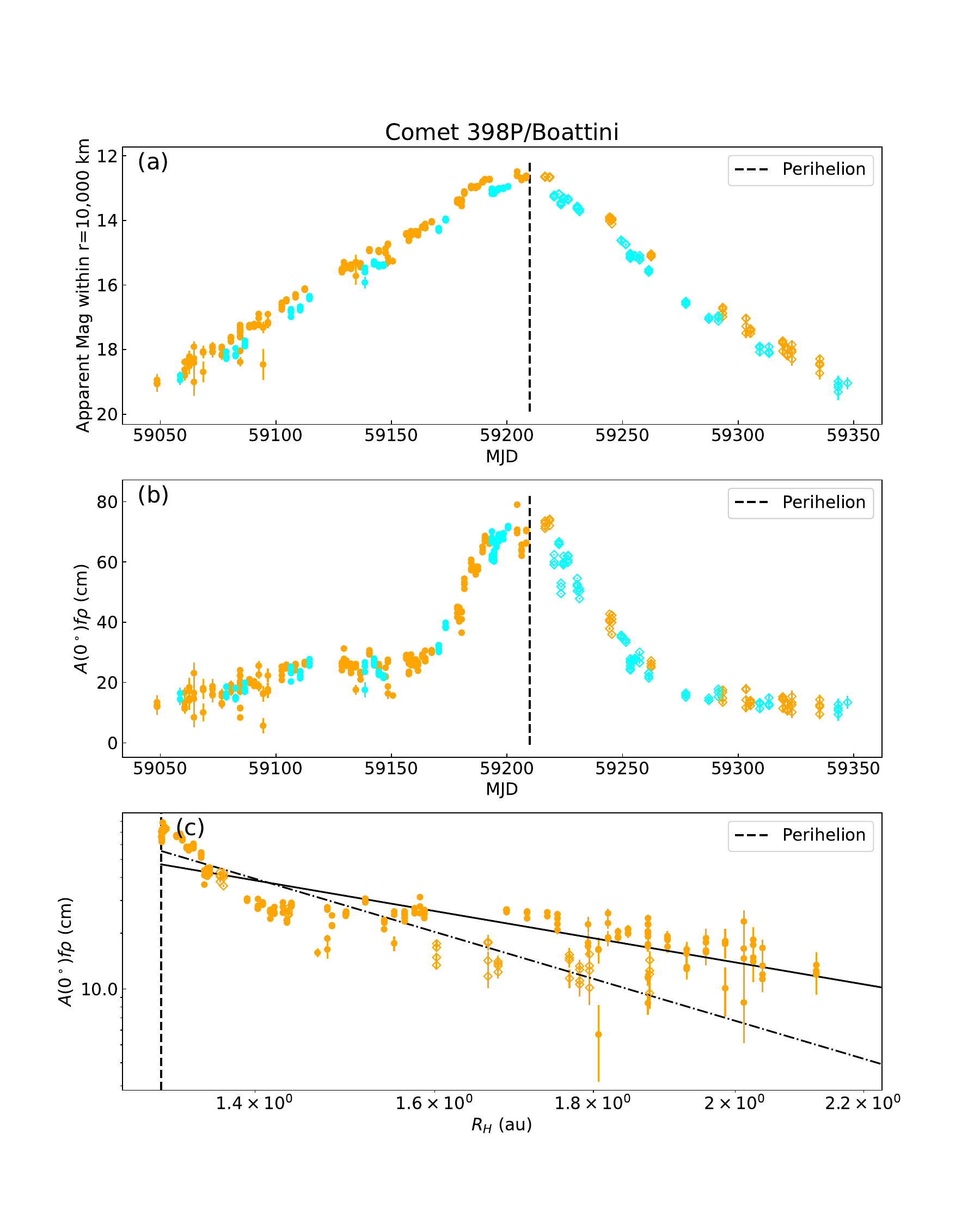}
\figsetgrpnote{\textbf{(Excluding outburst)} 398P/Boattini light curve is shown in the top panel, (a). The dust production parameter $A(0^\circ)f\rho$ vs time is displayed in plot (b), and (c) shows $A(0^\circ)f\rho$ vs heliocentric distance with the solid line representing the rate of change of activity pre-perihelion and the dot-dashed line showing the rate of change of activity post-perihelion. Orange and cyan points on the plots represent the \emph{orange} filter and \emph{cyan} filter respectively. Filled circular points represent pre-perihelion measurements and open diamonds represent post-perihelion. Perihelion is marked with dashed line in each panel.}

\figsetgrpnum{1.23}
\figsetgrptitle{399P}
\figsetplot{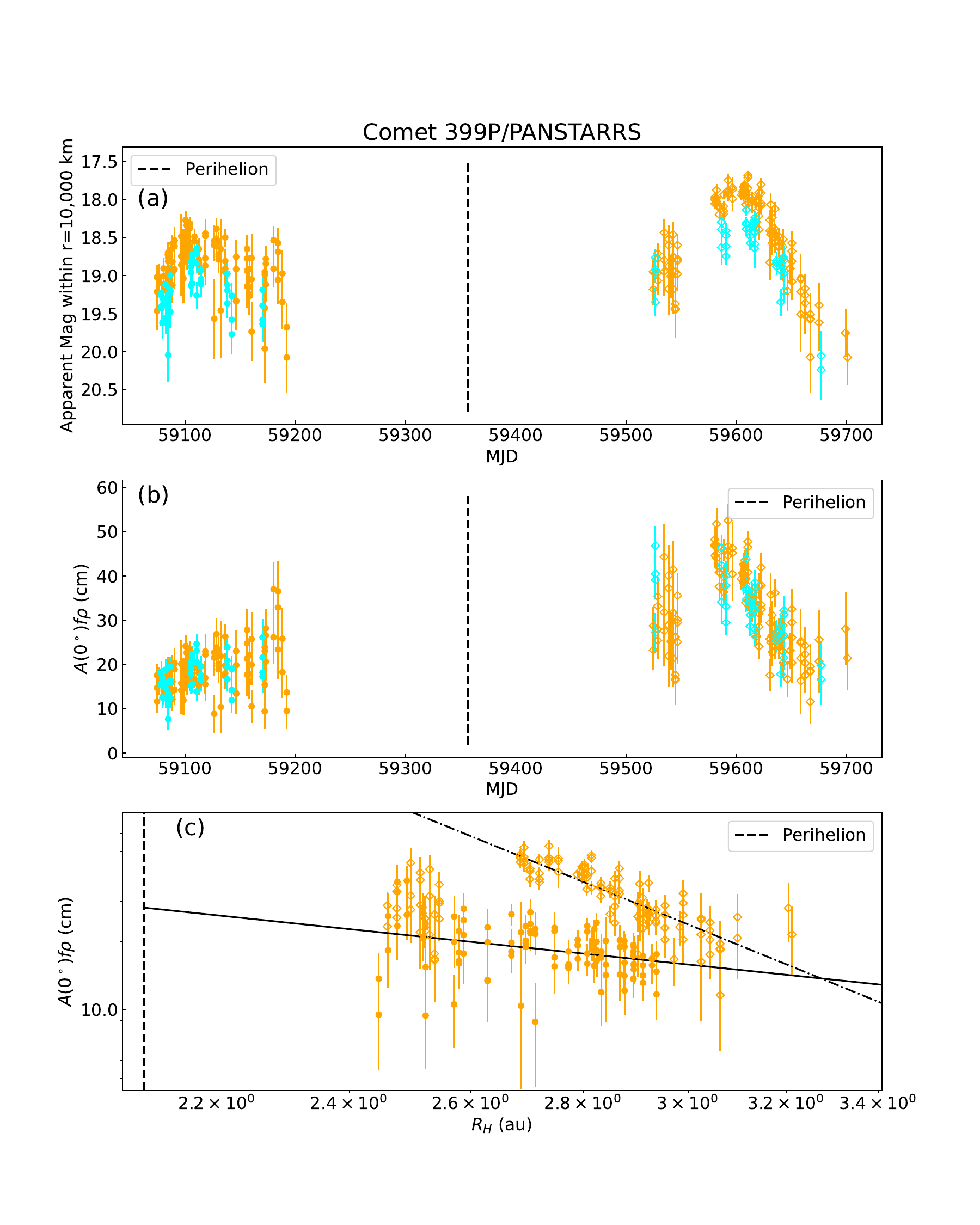}
\figsetgrpnote{399P/PANSTARRS light curve is shown in the top panel, (a). The dust production parameter $A(0^\circ)f\rho$ vs time is displayed in plot (b), and (c) shows $A(0^\circ)f\rho$ vs heliocentric distance with the solid line representing the rate of change of activity pre-perihelion and the dot-dashed line showing the rate of change of activity post-perihelion. Orange and cyan points on the plots represent the \emph{orange} filter and \emph{cyan} filter respectively. Filled circular points represent pre-perihelion measurements and open diamonds represent post-perihelion. Perihelion is marked with dashed line in each panel.}

\figsetgrpnum{1.24}
\figsetgrptitle{400P}
\figsetplot{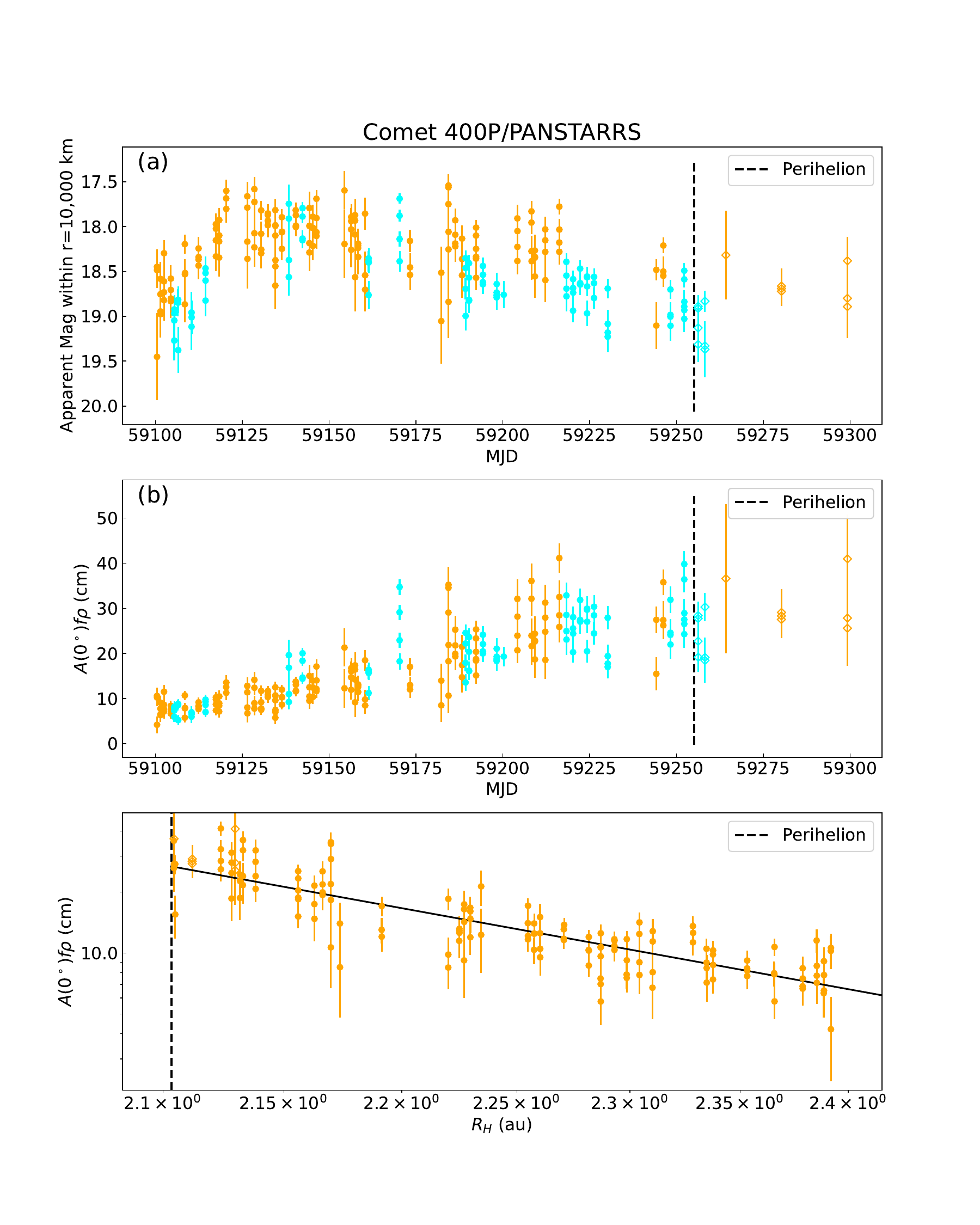}
\figsetgrpnote{400P/PANSTARRS light curve is shown in the top panel, (a). The dust production parameter $A(0^\circ)f\rho$ vs time is displayed in plot (b), and (c) shows $A(0^\circ)f\rho$ vs heliocentric distance with the solid line representing the rate of change of activity pre-perihelion and the dot-dashed line showing the rate of change of activity post-perihelion. Orange and cyan points on the plots represent the \emph{orange} filter and \emph{cyan} filter respectively. Filled circular points represent pre-perihelion measurements and open diamonds represent post-perihelion. Perihelion is marked with dashed line in each panel.}

\figsetgrpnum{1.25}
\figsetgrptitle{403P}
\figsetplot{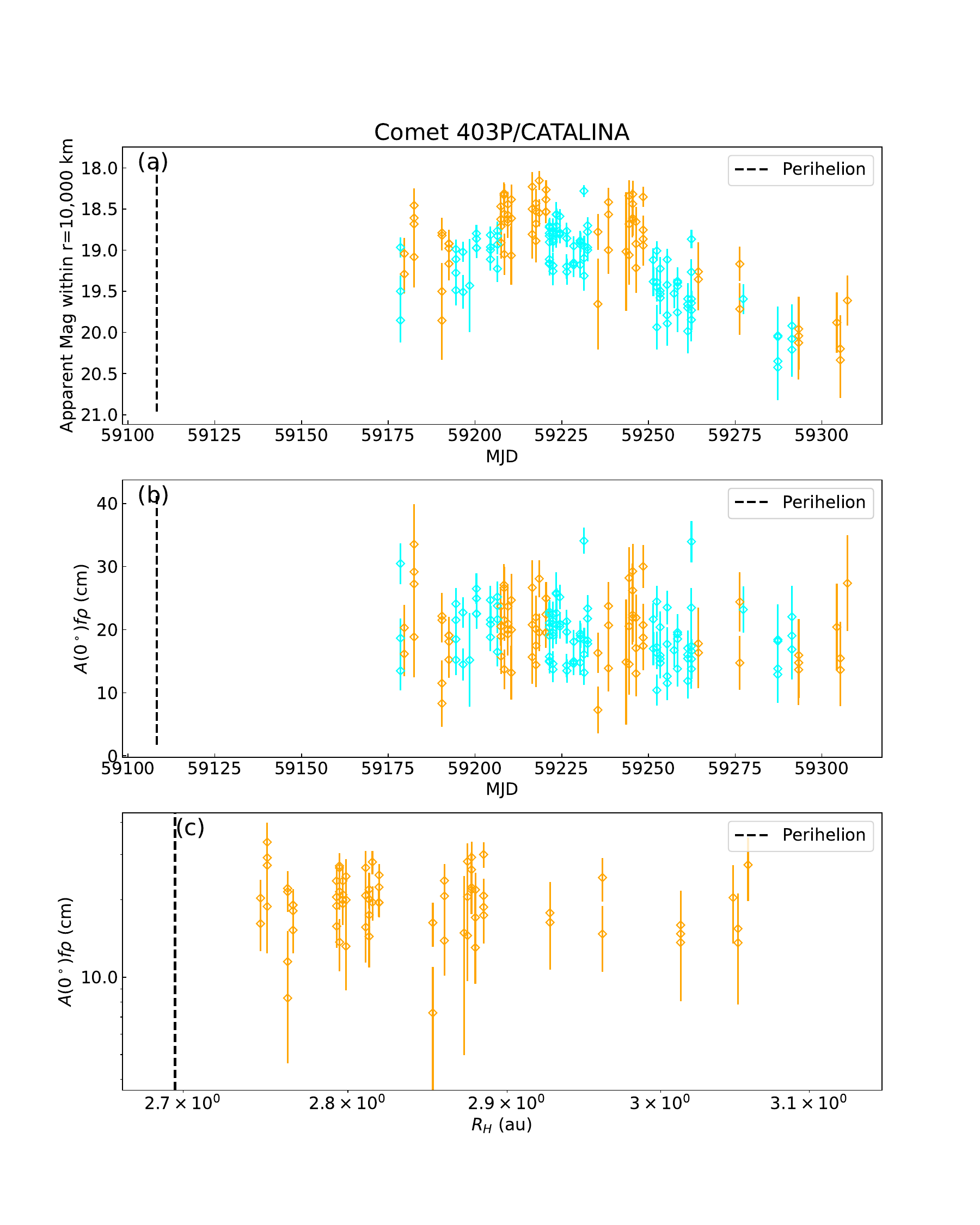}
\figsetgrpnote{403P/CATALINA light curve is shown in the top panel, (a). The dust production parameter $A(0^\circ)f\rho$ vs time is displayed in plot (b), and (c) shows $A(0^\circ)f\rho$ vs heliocentric distance with the solid line representing the rate of change of activity pre-perihelion and the dot-dashed line showing the rate of change of activity post-perihelion. Orange and cyan points on the plots represent the \emph{orange} filter and \emph{cyan} filter respectively. Filled circular points represent pre-perihelion measurements and open diamonds represent post-perihelion. Perihelion is marked with dashed line in each panel.}

\figsetgrpnum{1.26}
\figsetgrptitle{405P}
\figsetplot{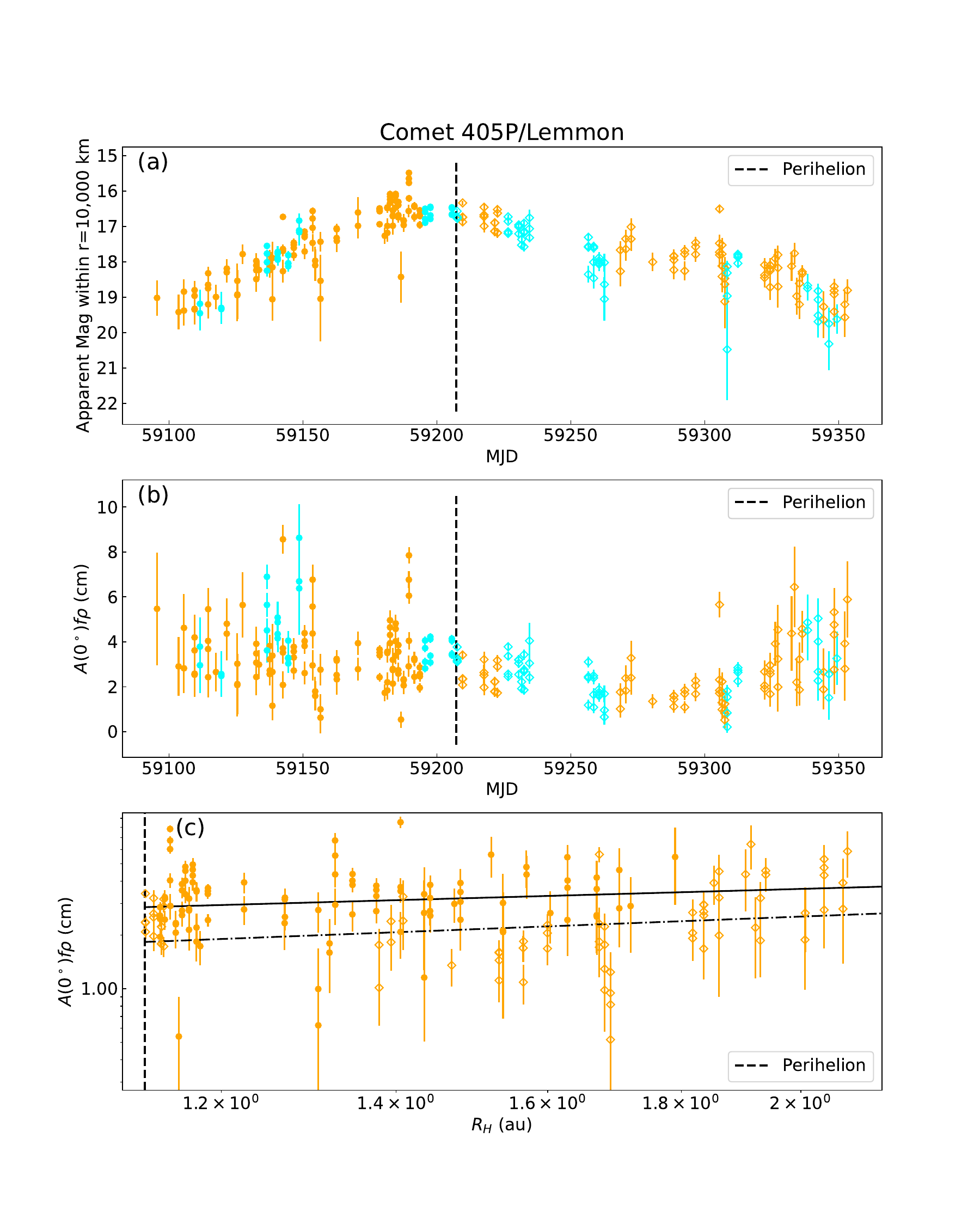}
\figsetgrpnote{405P/Lemmon light curve is shown in the top panel, (a). The dust production parameter $A(0^\circ)f\rho$ vs time is displayed in plot (b), and (c) shows $A(0^\circ)f\rho$ vs heliocentric distance with the solid line representing the rate of change of activity pre-perihelion and the dot-dashed line showing the rate of change of activity post-perihelion. Orange and cyan points on the plots represent the \emph{orange} filter and \emph{cyan} filter respectively. Filled circular points represent pre-perihelion measurements and open diamonds represent post-perihelion. Perihelion is marked with dashed line in each panel.}

\figsetgrpnum{1.27}
\figsetgrptitle{409P}
\figsetplot{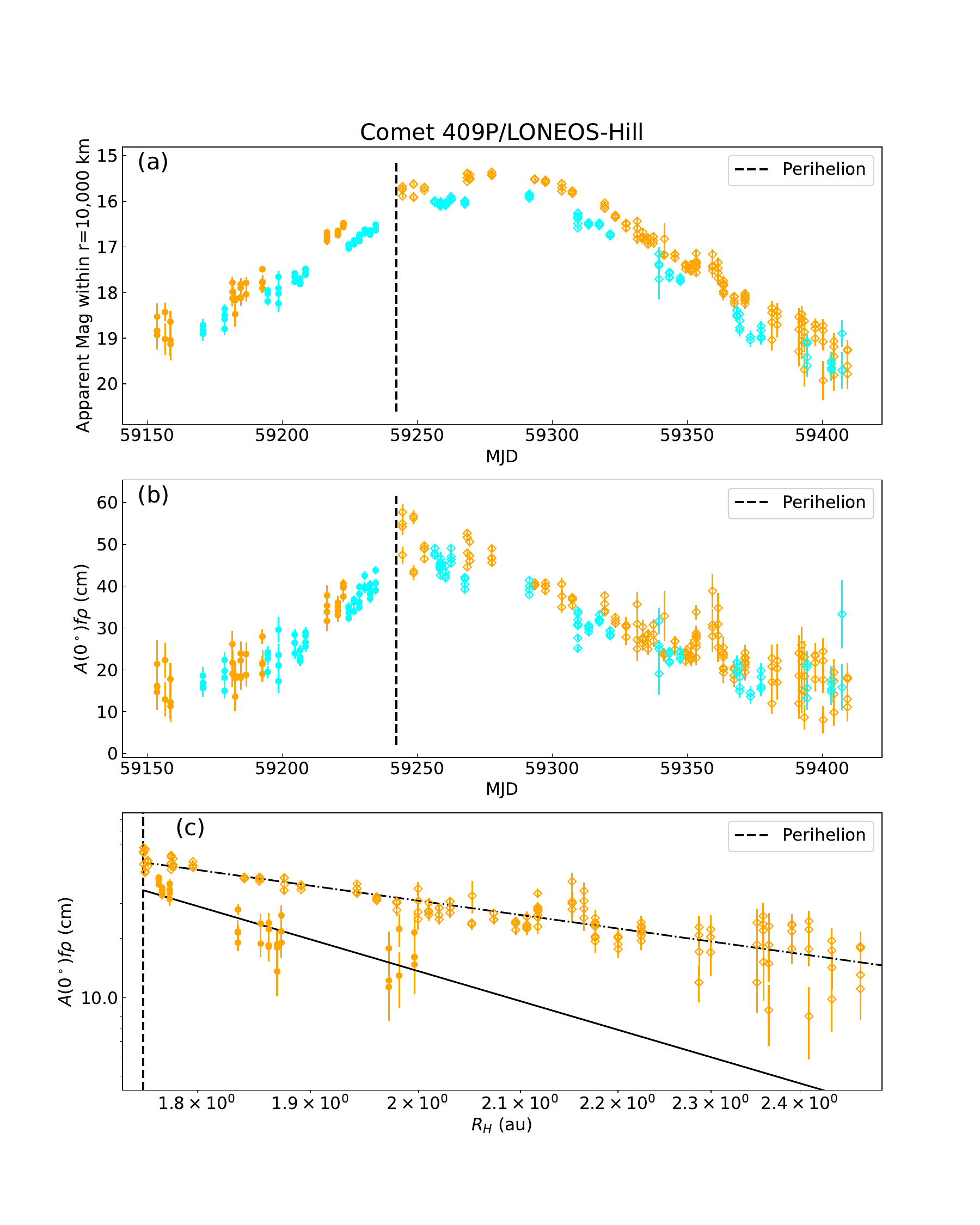}
\figsetgrpnote{409P/LONEOS-Hill light curve is shown in the top panel, (a). The dust production parameter $A(0^\circ)f\rho$ vs time is displayed in plot (b), and (c) shows $A(0^\circ)f\rho$ vs heliocentric distance with the solid line representing the rate of change of activity pre-perihelion and the dot-dashed line showing the rate of change of activity post-perihelion. Orange and cyan points on the plots represent the \emph{orange} filter and \emph{cyan} filter respectively. Filled circular points represent pre-perihelion measurements and open diamonds represent post-perihelion. Perihelion is marked with dashed line in each panel.}

\figsetgrpnum{1.28}
\figsetgrptitle{410P}
\figsetplot{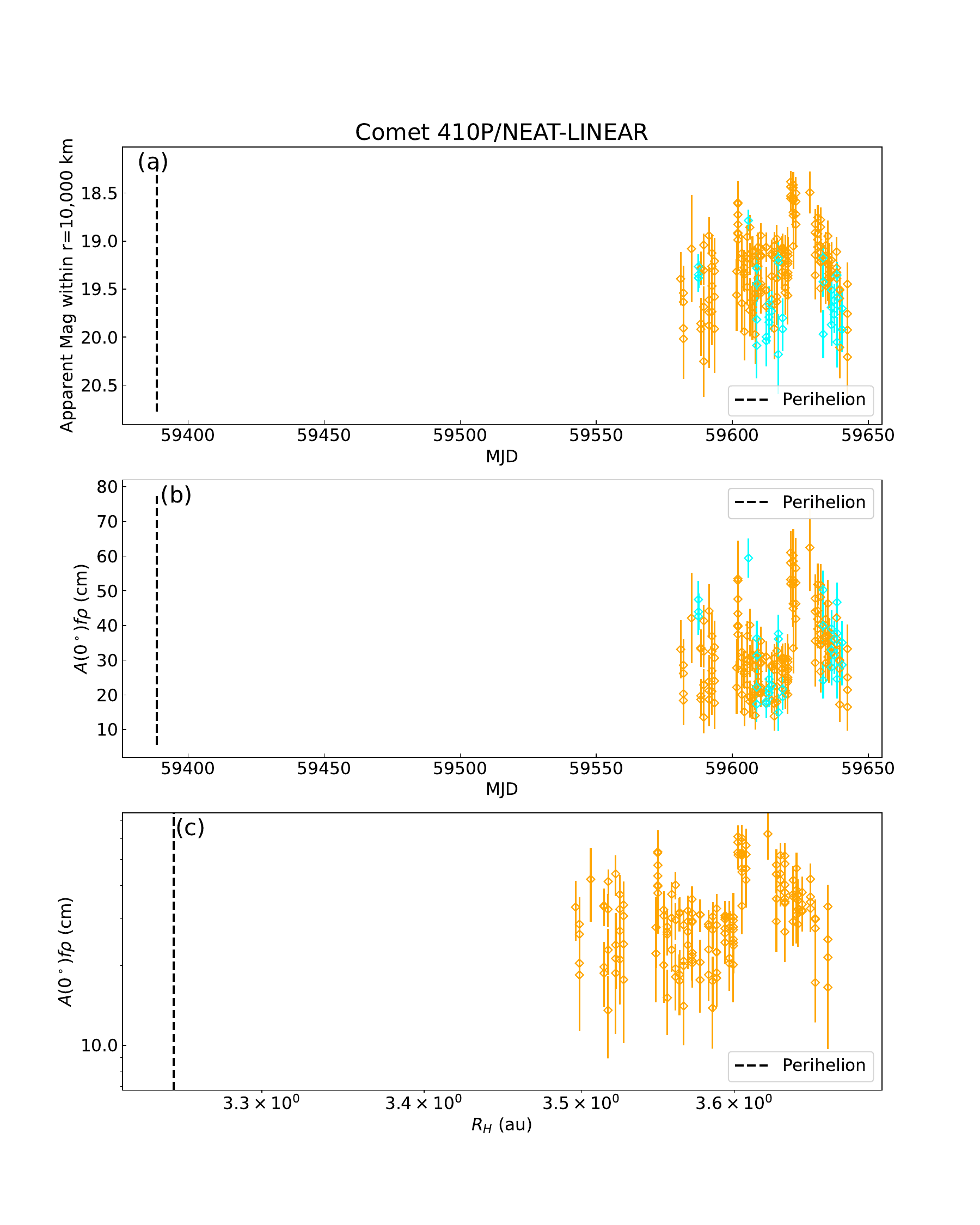}
\figsetgrpnote{410P/NEAT-LINEAR light curve is shown in the top panel, (a). The dust production parameter $A(0^\circ)f\rho$ vs time is displayed in plot (b), and (c) shows $A(0^\circ)f\rho$ vs heliocentric distance with the solid line representing the rate of change of activity pre-perihelion and the dot-dashed line showing the rate of change of activity post-perihelion. Orange and cyan points on the plots represent the \emph{orange} filter and \emph{cyan} filter respectively. Filled circular points represent pre-perihelion measurements and open diamonds represent post-perihelion. Perihelion is marked with dashed line in each panel.}

\figsetgrpnum{1.29}
\figsetgrptitle{414P}
\figsetplot{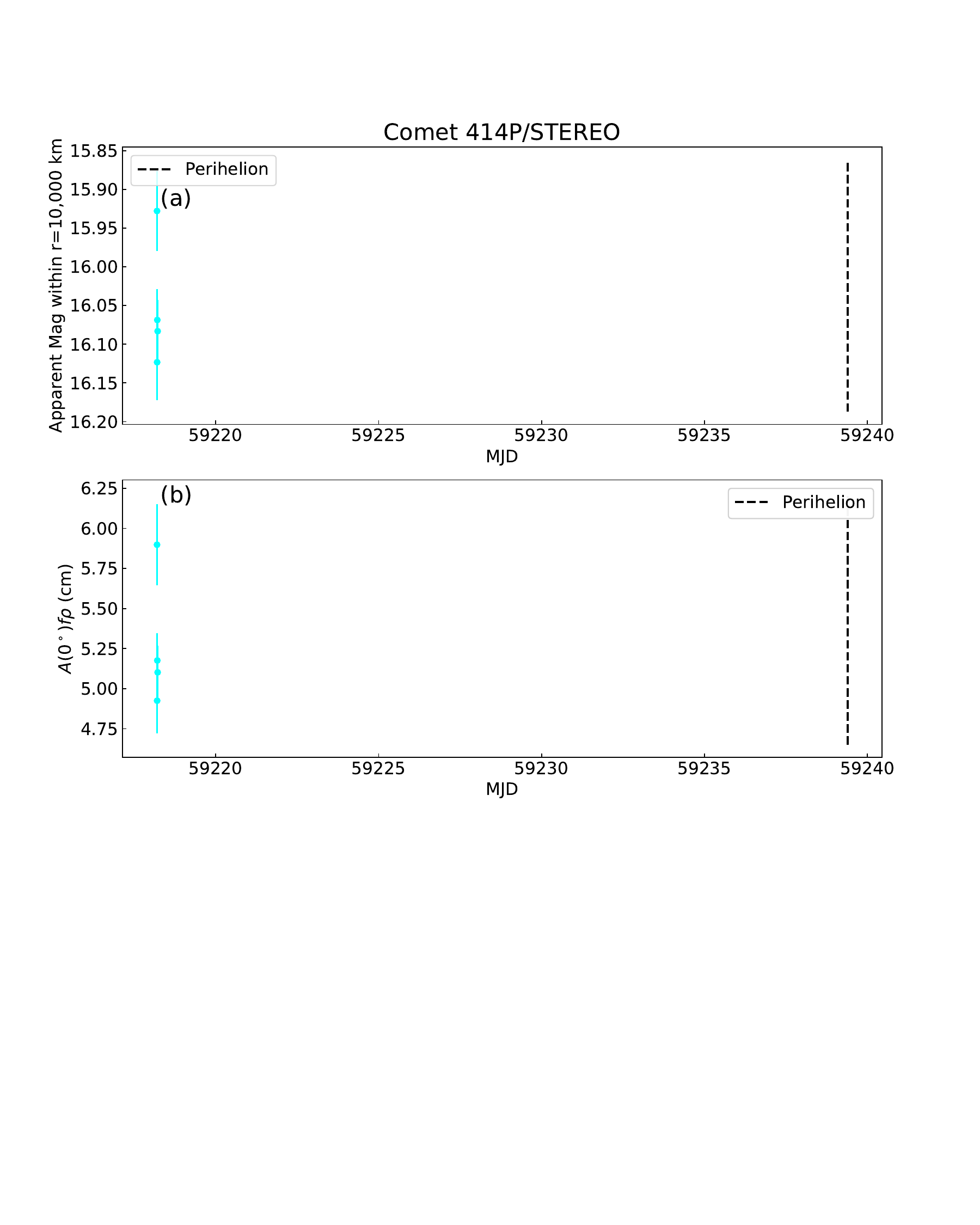}
\figsetgrpnote{414P/STEREO light curve is shown in the top panel, (a). The dust production parameter $A(0^\circ)f\rho$ vs time is displayed in plot (b). The cyan points on the plots represent the \emph{cyan} filter. Filled circular points represent pre-perihelion measurements. Perihelion is marked with dashed line in each panel.}

\figsetgrpnum{1.30}
\figsetgrptitle{415P}
\figsetplot{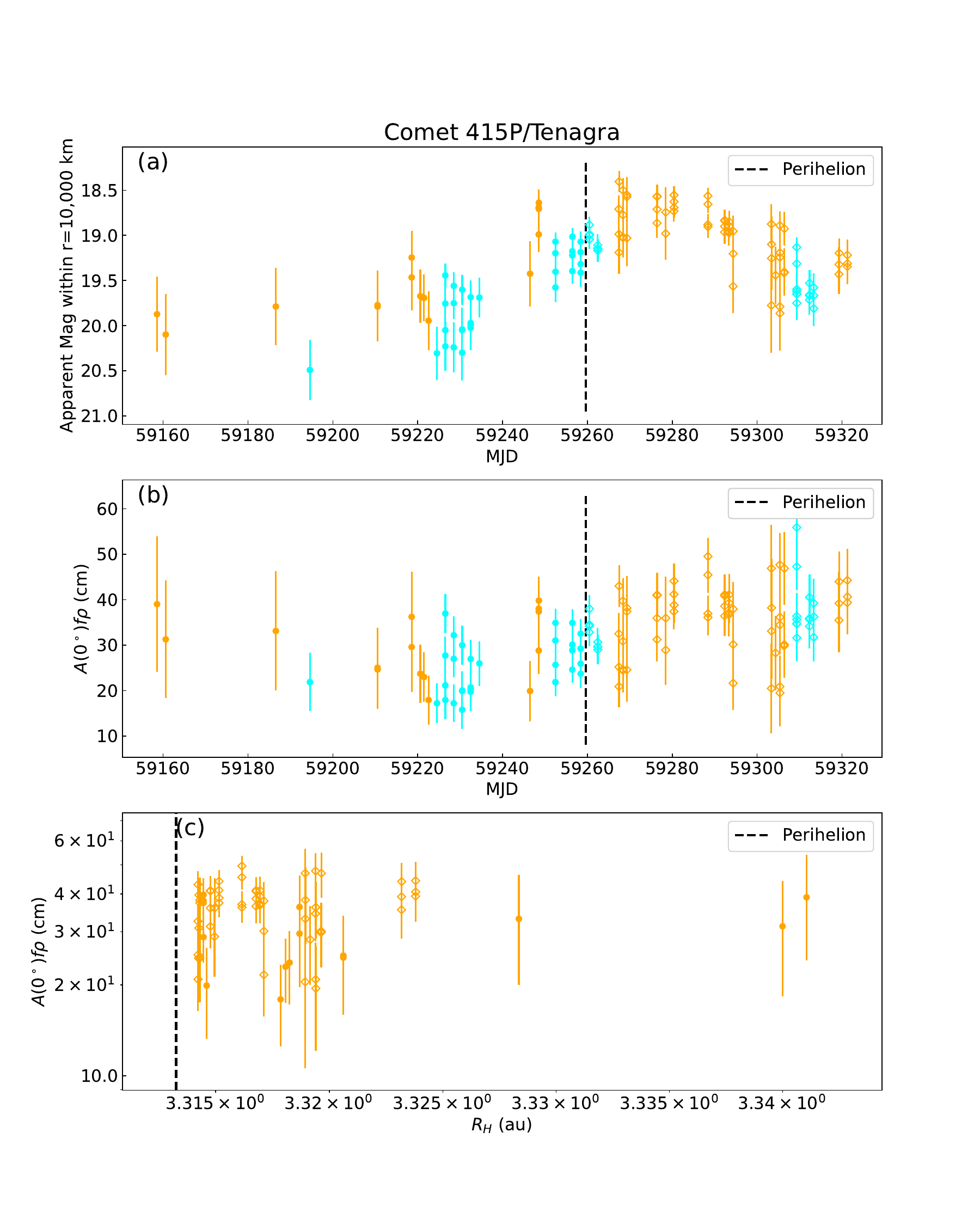}
\figsetgrpnote{415P/Tenagra light curve is shown in the top panel, (a). The dust production parameter $A(0^\circ)f\rho$ vs time is displayed in plot (b), and (c) shows $A(0^\circ)f\rho$ vs heliocentric distance with the solid line representing the rate of change of activity pre-perihelion and the dot-dashed line showing the rate of change of activity post-perihelion. Orange and cyan points on the plots represent the \emph{orange} filter and \emph{cyan} filter respectively. Filled circular points represent pre-perihelion measurements and open diamonds represent post-perihelion. Perihelion is marked with dashed line in each panel.}

\figsetgrpnum{1.31}
\figsetgrptitle{417P}
\figsetplot{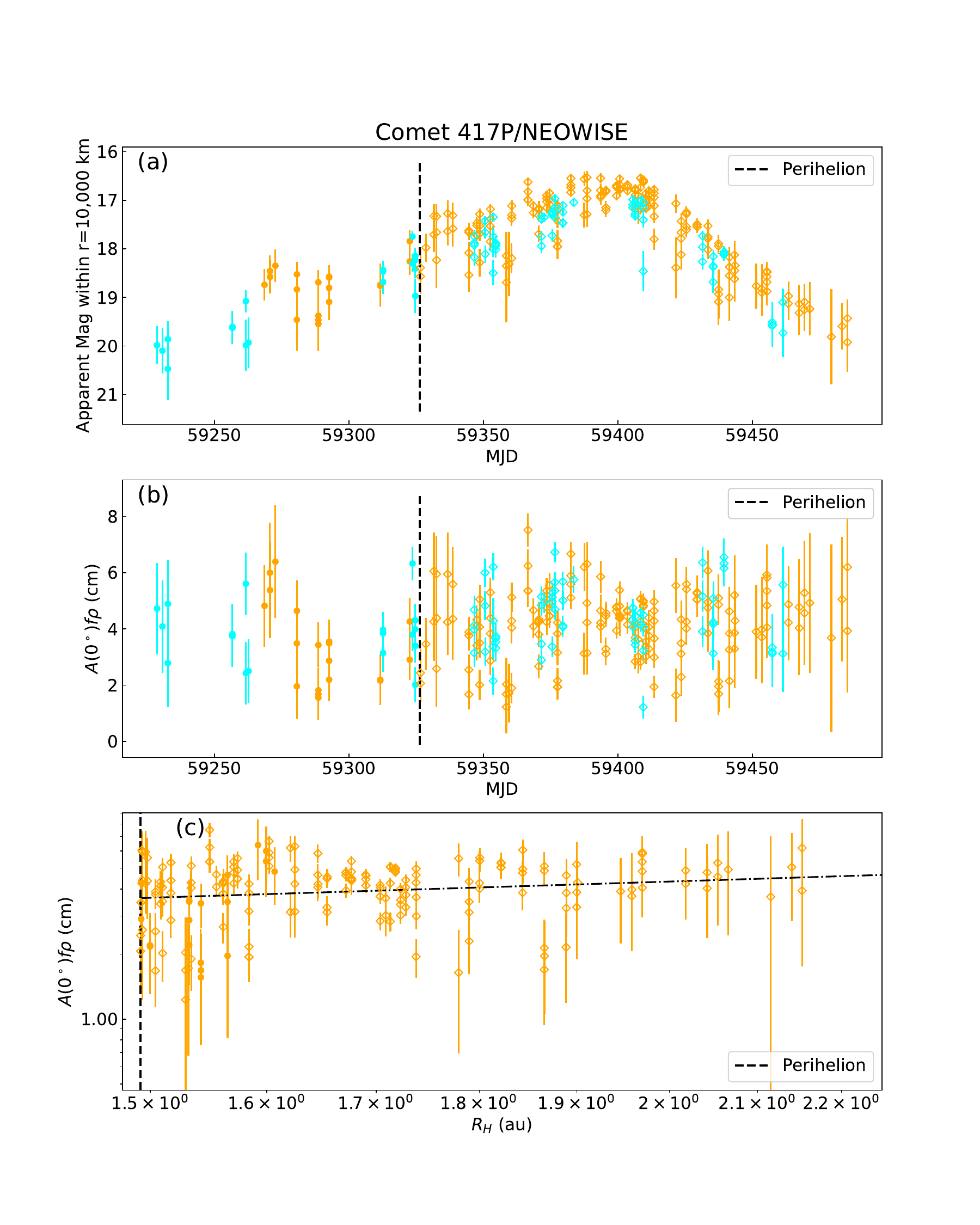}
\figsetgrpnote{417P/NEOWISE light curve is shown in the top panel, (a). The dust production parameter $A(0^\circ)f\rho$ vs time is displayed in plot (b), and (c) shows $A(0^\circ)f\rho$ vs heliocentric distance with the solid line representing the rate of change of activity pre-perihelion and the dot-dashed line showing the rate of change of activity post-perihelion. Orange and cyan points on the plots represent the \emph{orange} filter and \emph{cyan} filter respectively. Filled circular points represent pre-perihelion measurements and open diamonds represent post-perihelion. Perihelion is marked with dashed line in each panel.}

\figsetgrpnum{1.32}
\figsetgrptitle{425P}
\figsetplot{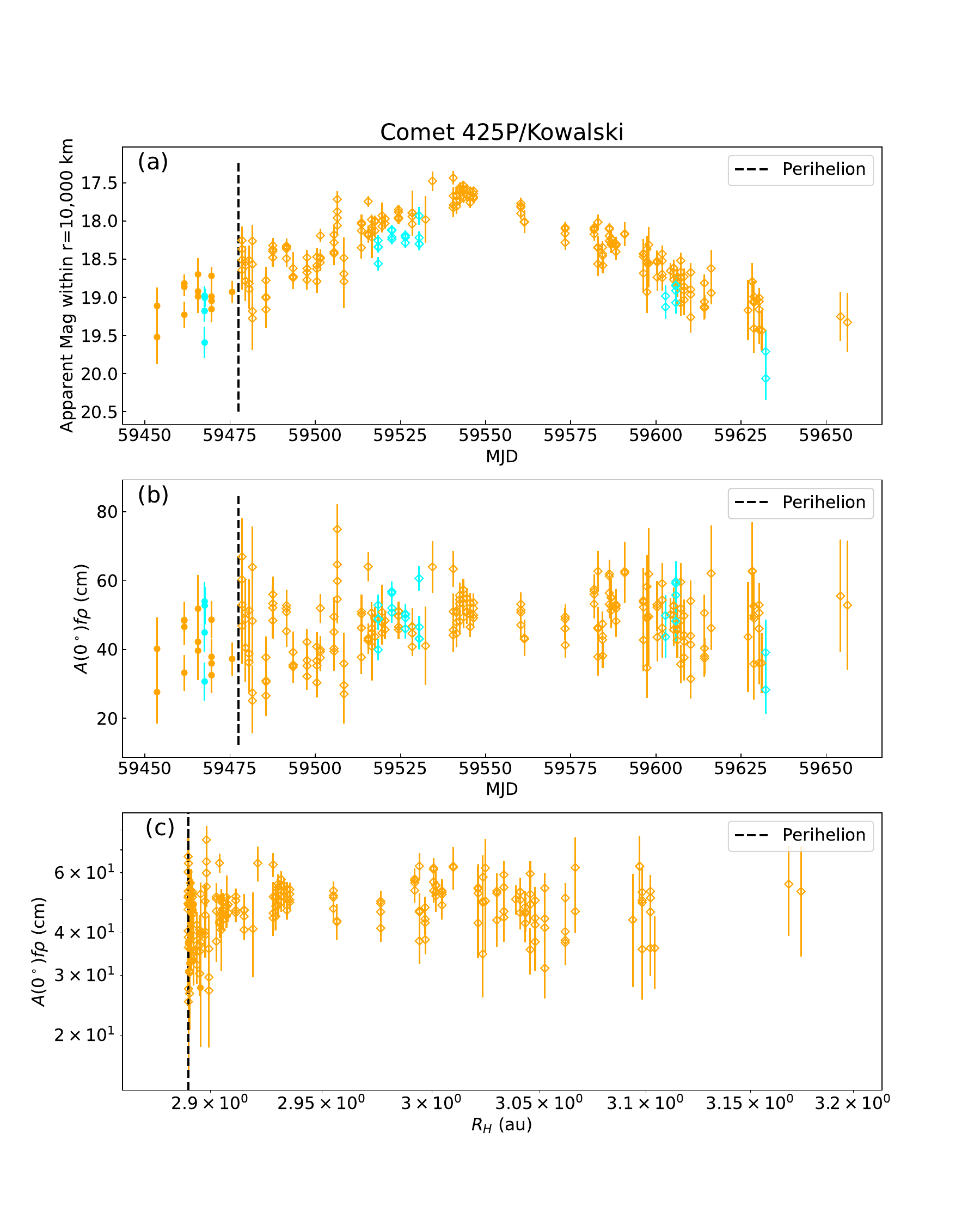}
\figsetgrpnote{425P/Kowalski light curve is shown in the top panel, (a). The dust production parameter $A(0^\circ)f\rho$ vs time is displayed in plot (b), and (c) shows $A(0^\circ)f\rho$ vs heliocentric distance with the solid line representing the rate of change of activity pre-perihelion and the dot-dashed line showing the rate of change of activity post-perihelion. Orange and cyan points on the plots represent the \emph{orange} filter and \emph{cyan} filter respectively. Filled circular points represent pre-perihelion measurements and open diamonds represent post-perihelion. Perihelion is marked with dashed line in each panel.}

\figsetgrpnum{1.33}
\figsetgrptitle{LeonardP2020S6}
\figsetplot{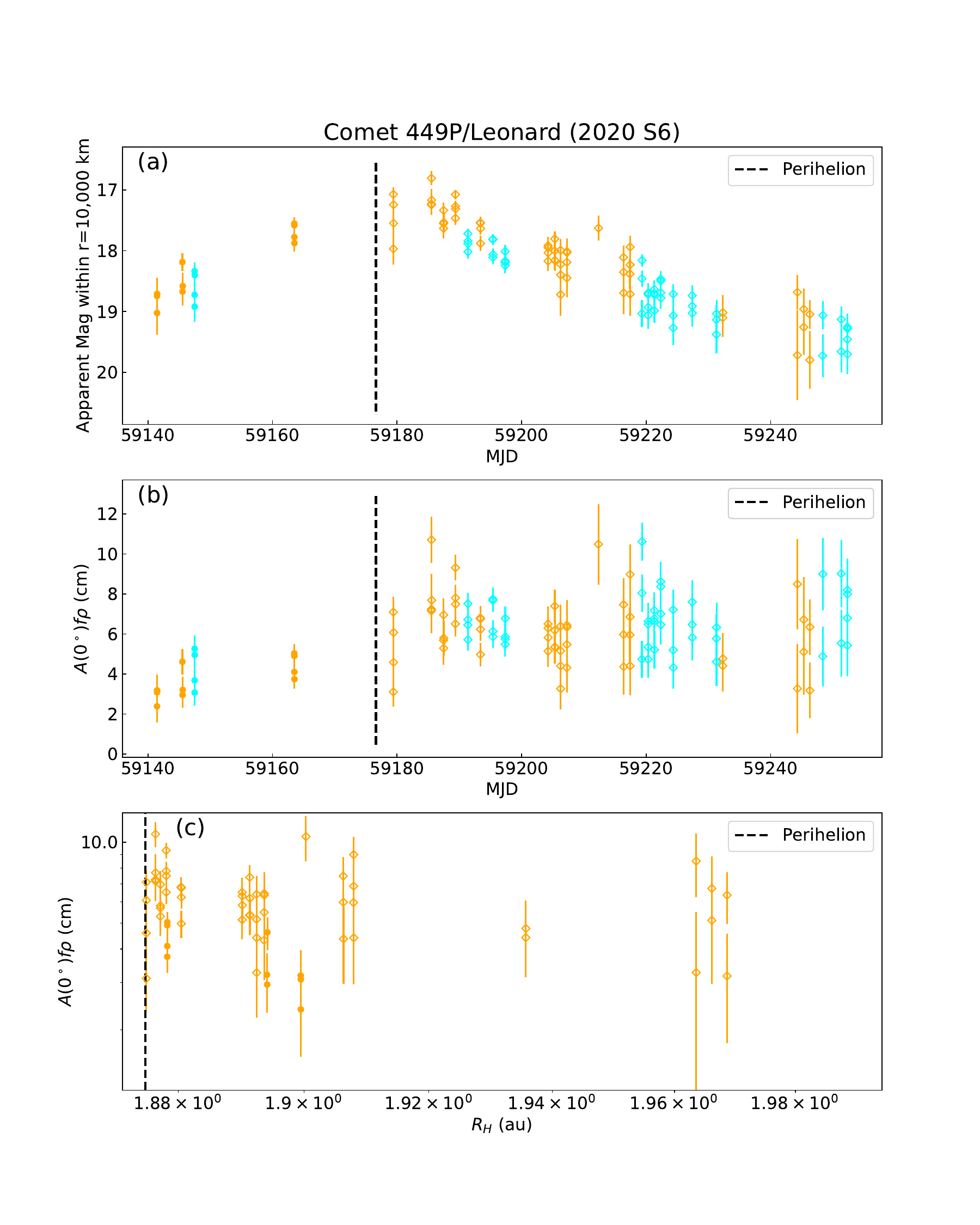}
\figsetgrpnote{449P/Leonard (2020 S6) light curve is shown in the top panel, (a). The dust production parameter $A(0^\circ)f\rho$ vs time is displayed in plot (b), and (c) shows $A(0^\circ)f\rho$ vs heliocentric distance with the solid line representing the rate of change of activity pre-perihelion and the dot-dashed line showing the rate of change of activity post-perihelion. Orange and cyan points on the plots represent the \emph{orange} filter and \emph{cyan} filter respectively. Filled circular points represent pre-perihelion measurements and open diamonds represent post-perihelion. Perihelion is marked with dashed line in each panel.}

\figsetgrpnum{1.34}
\figsetgrptitle{ATLASP2019LD2}
\figsetplot{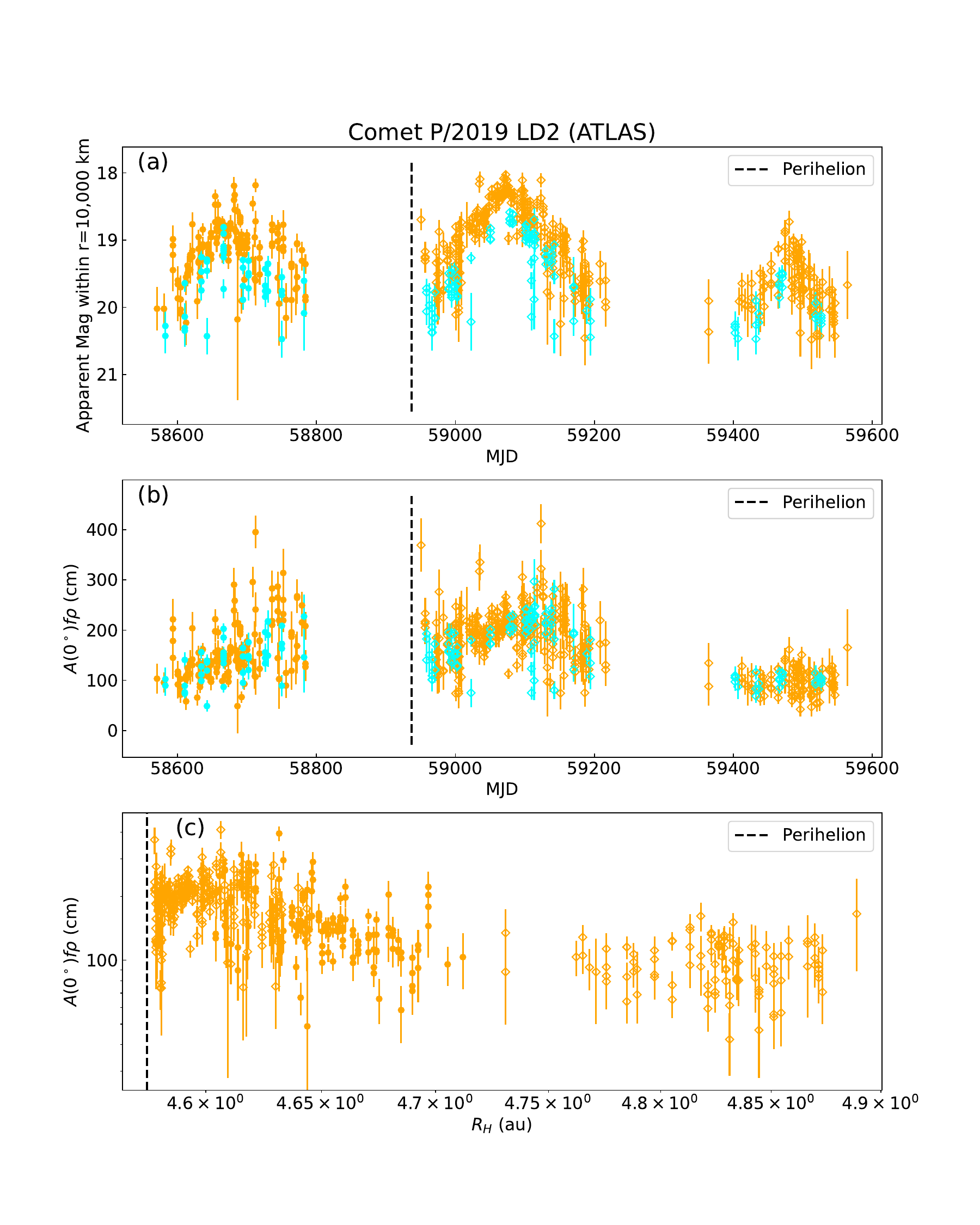}
\figsetgrpnote{P/2019 LD2 (ATLAS) light curve is shown in the top panel, (a). The dust production parameter $A(0^\circ)f\rho$ vs time is displayed in plot (b), and (c) shows $A(0^\circ)f\rho$ vs heliocentric distance with the solid line representing the rate of change of activity pre-perihelion and the dot-dashed line showing the rate of change of activity post-perihelion. Orange and cyan points on the plots represent the \emph{orange} filter and \emph{cyan} filter respectively. Filled circular points represent pre-perihelion measurements and open diamonds represent post-perihelion. Perihelion is marked with dashed line in each panel.}

\figsetgrpnum{1.35}
\figsetgrptitle{PimentelP2020G1}
\figsetplot{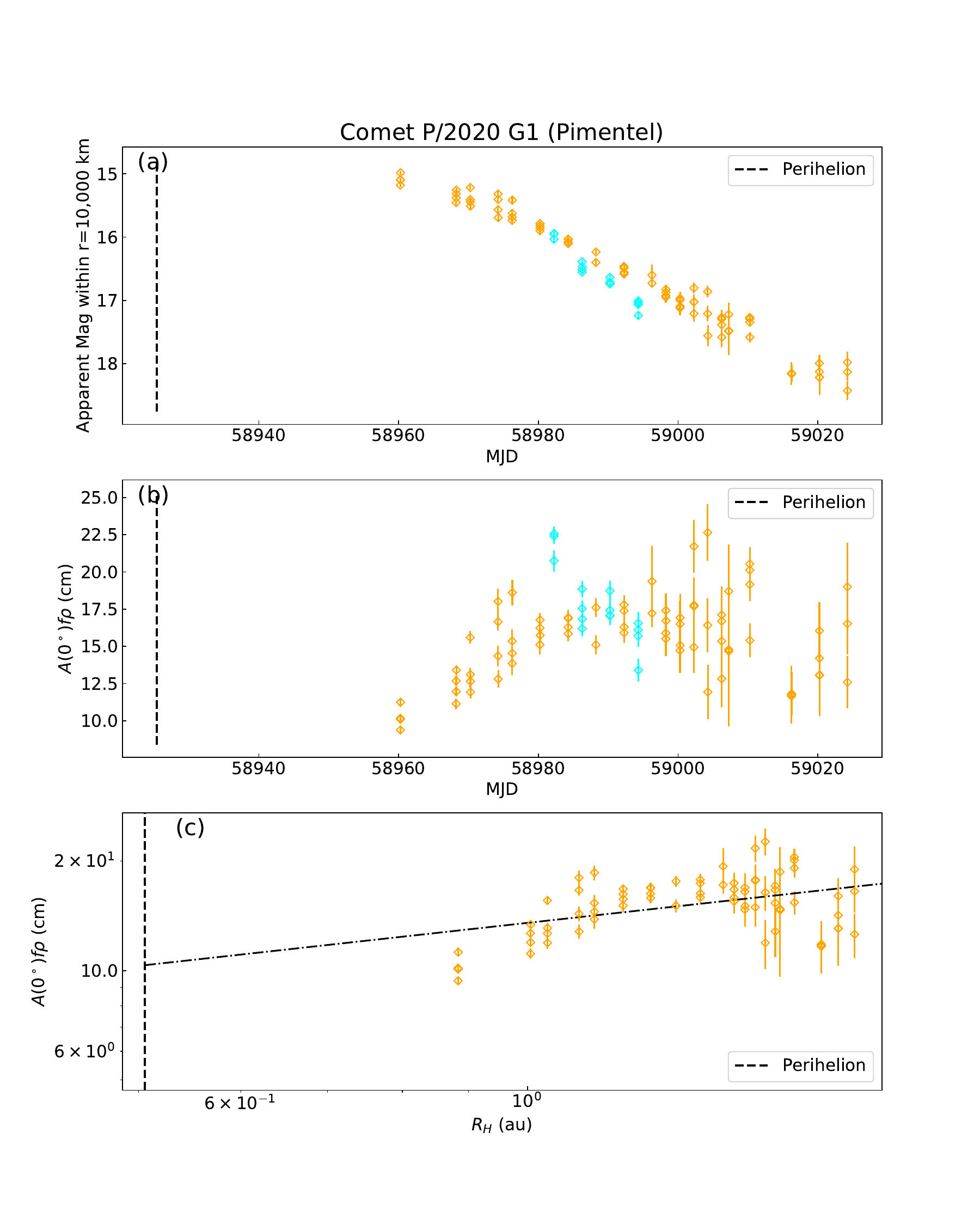}
\figsetgrpnote{P/2020 G1 (Pimentel) light curve is shown in the top panel, (a). The dust production parameter $A(0^\circ)f\rho$ vs time is displayed in plot (b), and (c) shows $A(0^\circ)f\rho$ vs heliocentric distance with the solid line representing the rate of change of activity pre-perihelion and the dot-dashed line showing the rate of change of activity post-perihelion. Orange and cyan points on the plots represent the \emph{orange} filter and \emph{cyan} filter respectively. Filled circular points represent pre-perihelion measurements and open diamonds represent post-perihelion. Perihelion is marked with dashed line in each panel.}

\figsetgrpnum{1.36}
\figsetgrptitle{LEMMONPANSTARRSP2020K9}
\figsetplot{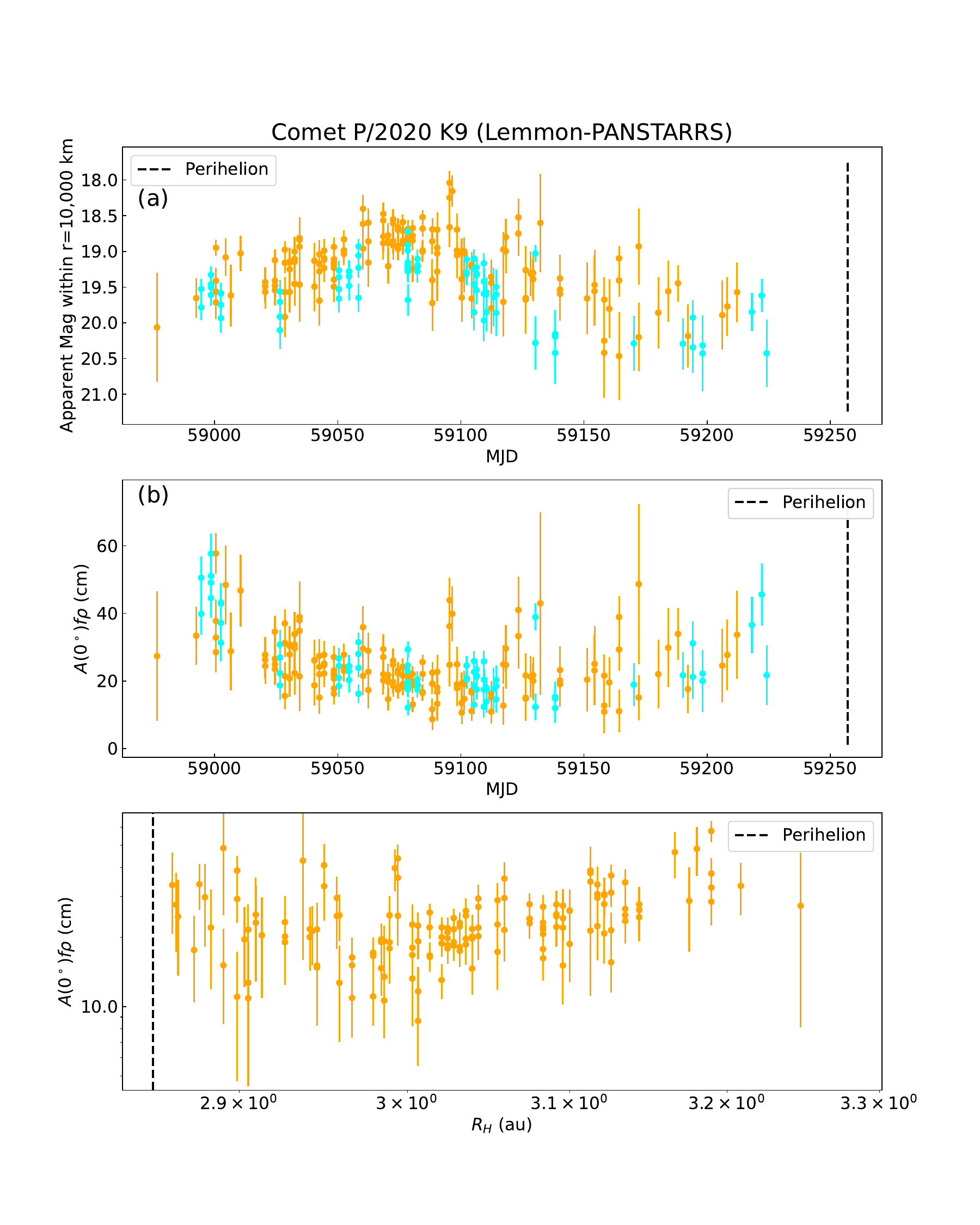}
\figsetgrpnote{P/2020 K9 (Lemmon-PANSTARRS) light curve is shown in the top panel, (a). The dust production parameter $A(0^\circ)f\rho$ vs time is displayed in plot (b), and (c) shows $A(0^\circ)f\rho$ vs heliocentric distance with the solid line representing the rate of change of activity pre-perihelion and the dot-dashed line showing the rate of change of activity post-perihelion. Orange and cyan points on the plots represent the \emph{orange} filter and \emph{cyan} filter respectively. Filled circular points represent pre-perihelion measurements and open diamonds represent post-perihelion. Perihelion is marked with dashed line in each panel.}

\figsetgrpnum{1.37}
\figsetgrptitle{PANSTARRSP2020S5}
\figsetplot{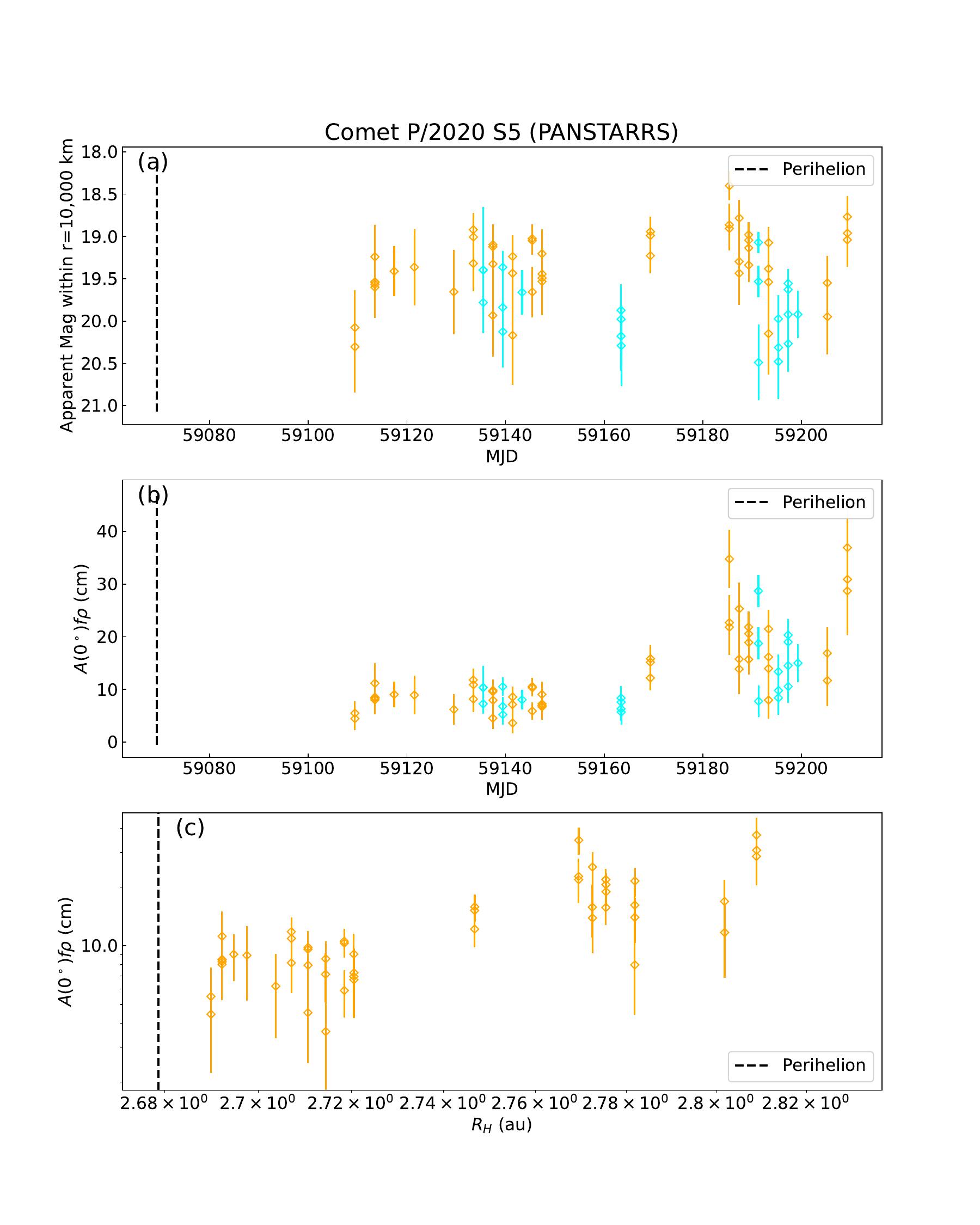}
\figsetgrpnote{P/2020 S5 (PANSTARRS) light curve is shown in the top panel, (a). The dust production parameter $A(0^\circ)f\rho$ vs time is displayed in plot (b), and (c) shows $A(0^\circ)f\rho$ vs heliocentric distance with the solid line representing the rate of change of activity pre-perihelion and the dot-dashed line showing the rate of change of activity post-perihelion. Orange and cyan points on the plots represent the \emph{orange} filter and \emph{cyan} filter respectively. Filled circular points represent pre-perihelion measurements and open diamonds represent post-perihelion. Perihelion is marked with dashed line in each panel.}

\figsetgrpnum{1.38}
\figsetgrptitle{PANSTARRSP2020T3}
\figsetplot{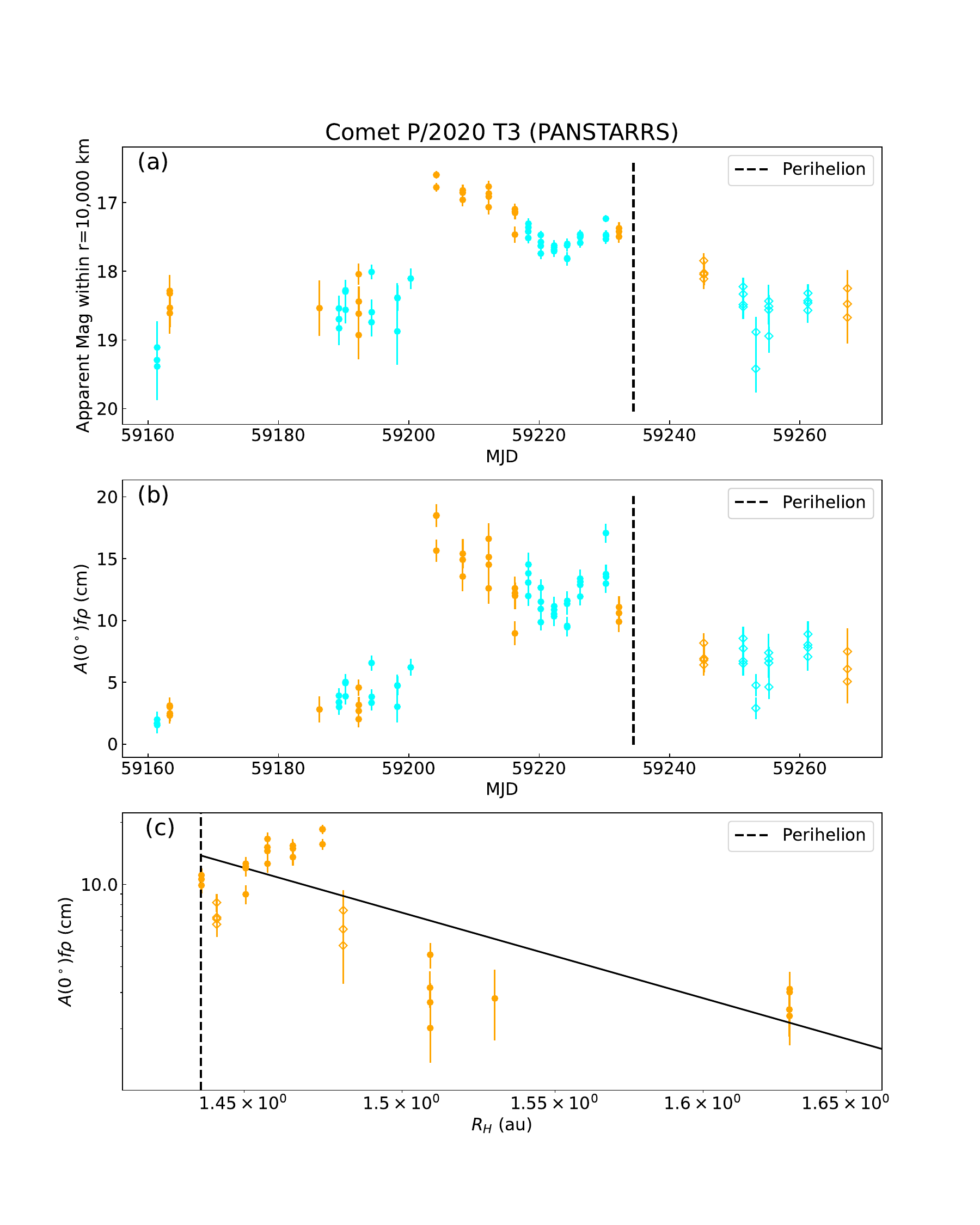}
\figsetgrpnote{P/2020 T3 (PANSTARRS) light curve is shown in the top panel, (a). The dust production parameter $A(0^\circ)f\rho$ vs time is displayed in plot (b), and (c) shows $A(0^\circ)f\rho$ vs heliocentric distance with the solid line representing the rate of change of activity pre-perihelion and the dot-dashed line showing the rate of change of activity post-perihelion. Orange and cyan points on the plots represent the \emph{orange} filter and \emph{cyan} filter respectively. Filled circular points represent pre-perihelion measurements and open diamonds represent post-perihelion. Perihelion is marked with dashed line in each panel.}

\figsetgrpnum{1.39}
\figsetgrptitle{PANSTARRSP2020U2}
\figsetplot{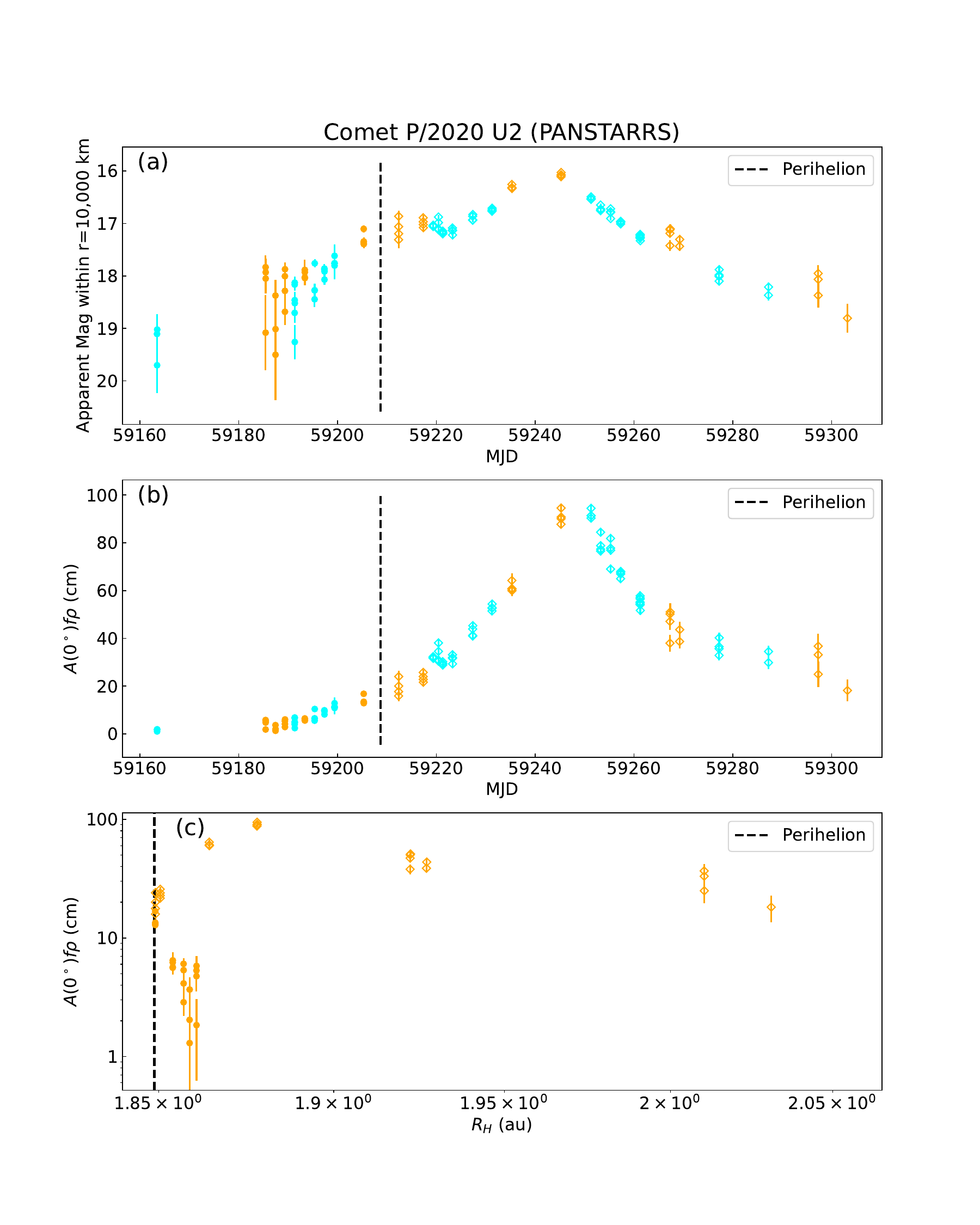}
\figsetgrpnote{P/2020 U2 (PANSTARRS) light curve is shown in the top panel, (a). The dust production parameter $A(0^\circ)f\rho$ vs time is displayed in plot (b), and (c) shows $A(0^\circ)f\rho$ vs heliocentric distance with the solid line representing the rate of change of activity pre-perihelion and the dot-dashed line showing the rate of change of activity post-perihelion. Orange and cyan points on the plots represent the \emph{orange} filter and \emph{cyan} filter respectively. Filled circular points represent pre-perihelion measurements and open diamonds represent post-perihelion. Perihelion is marked with dashed line in each panel.}

\figsetgrpnum{1.40}
\figsetgrptitle{LemmonP2020WJ5}
\figsetplot{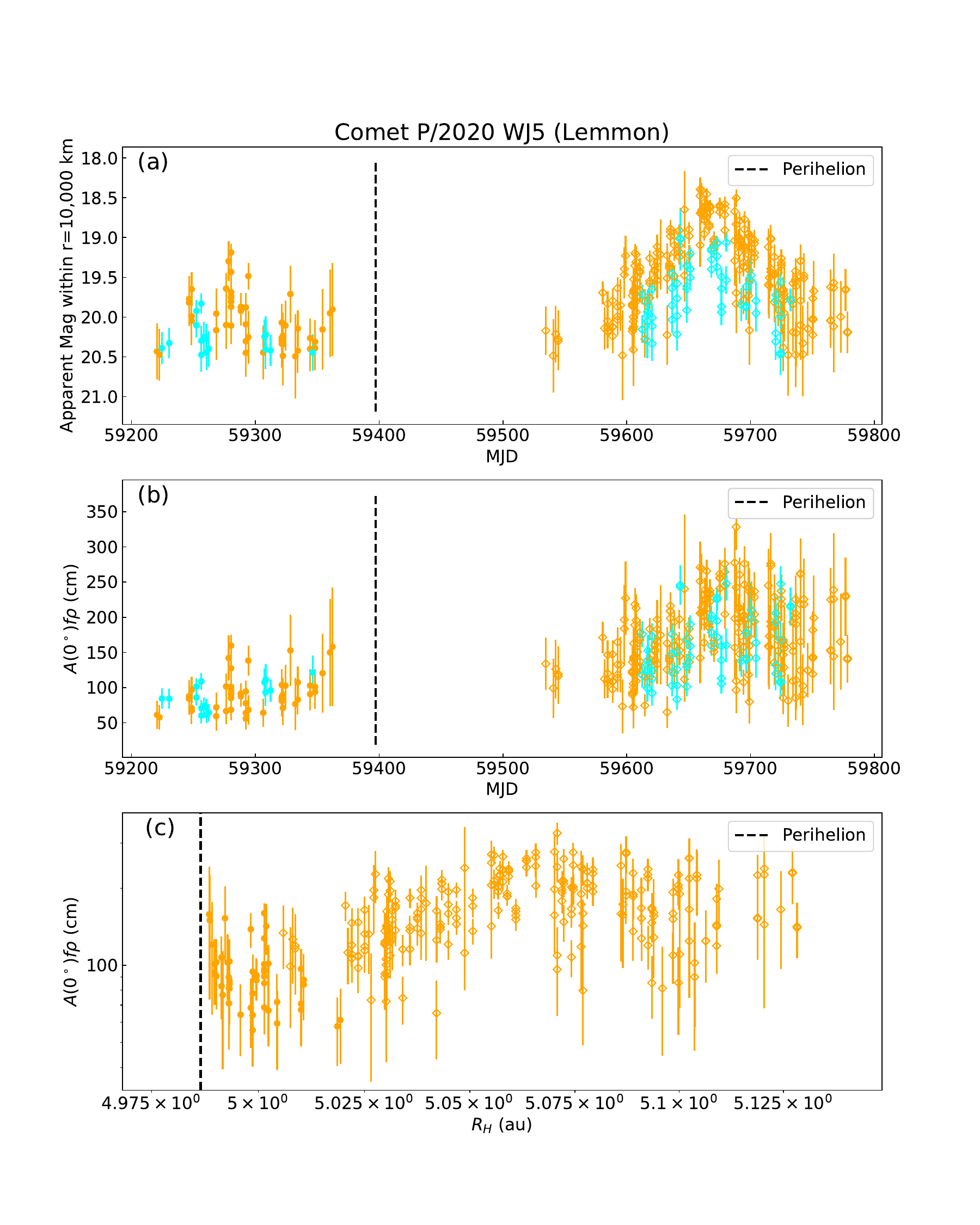}
\figsetgrpnote{P/2020 WJ5 (Lemmon) light curve is shown in the top panel, (a). The dust production parameter $A(0^\circ)f\rho$ vs time is displayed in plot (b), and (c) shows $A(0^\circ)f\rho$ vs heliocentric distance with the solid line representing the rate of change of activity pre-perihelion and the dot-dashed line showing the rate of change of activity post-perihelion. Orange and cyan points on the plots represent the \emph{orange} filter and \emph{cyan} filter respectively. Filled circular points represent pre-perihelion measurements and open diamonds represent post-perihelion. Perihelion is marked with dashed line in each panel.}

\figsetgrpnum{1.41}
\figsetgrptitle{ATLASP2020X1}
\figsetplot{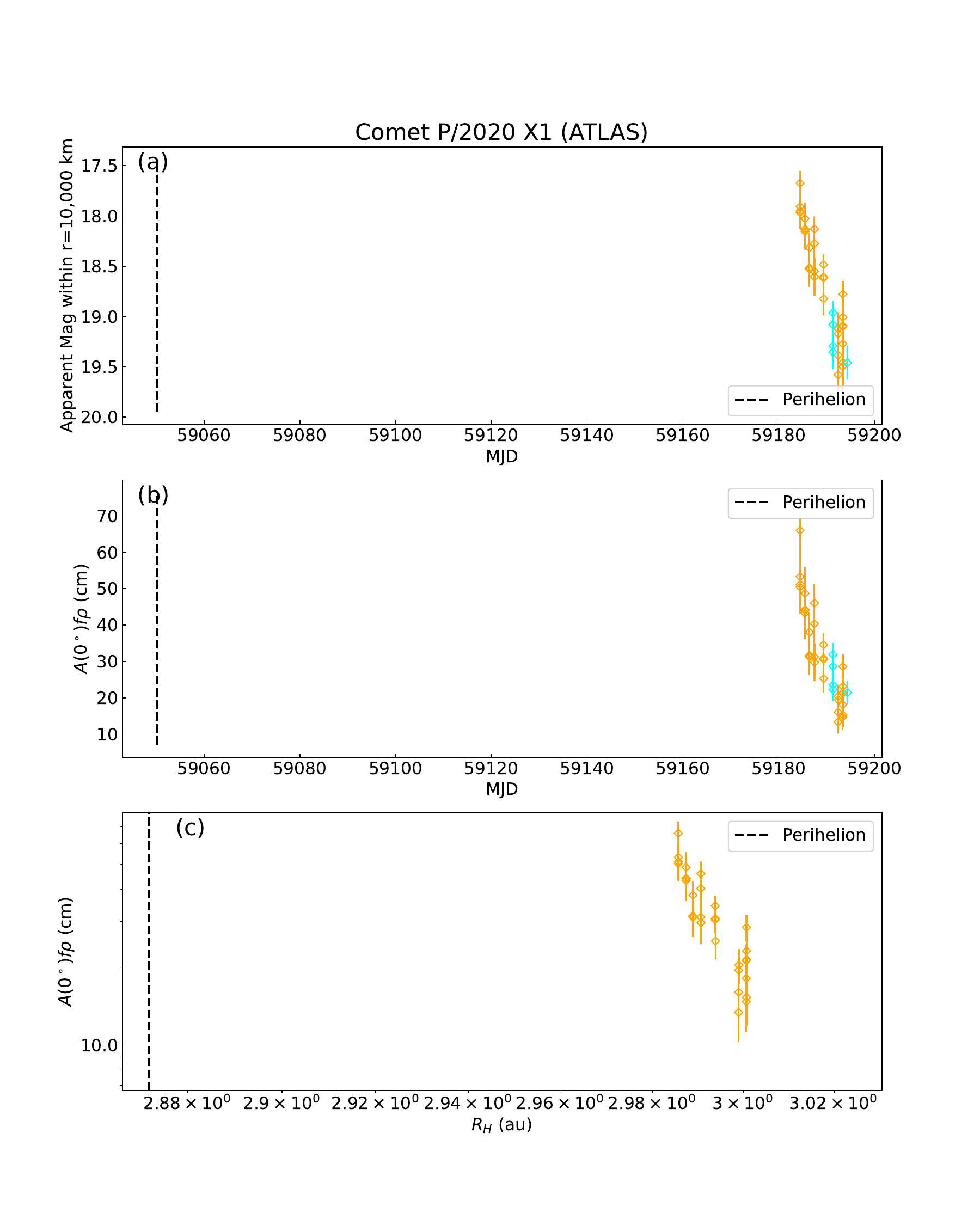}
\figsetgrpnote{P/2020 X1 (ATLAS) light curve is shown in the top panel, (a). The dust production parameter $A(0^\circ)f\rho$ vs time is displayed in plot (b), and (c) shows $A(0^\circ)f\rho$ vs heliocentric distance with the solid line representing the rate of change of activity pre-perihelion and the dot-dashed line showing the rate of change of activity post-perihelion. Orange and cyan points on the plots represent the \emph{orange} filter and \emph{cyan} filter respectively. Filled circular points represent pre-perihelion measurements and open diamonds represent post-perihelion. Perihelion is marked with dashed line in each panel.}

\figsetgrpnum{1.42}
\figsetgrptitle{LeonardP2021L2}
\figsetplot{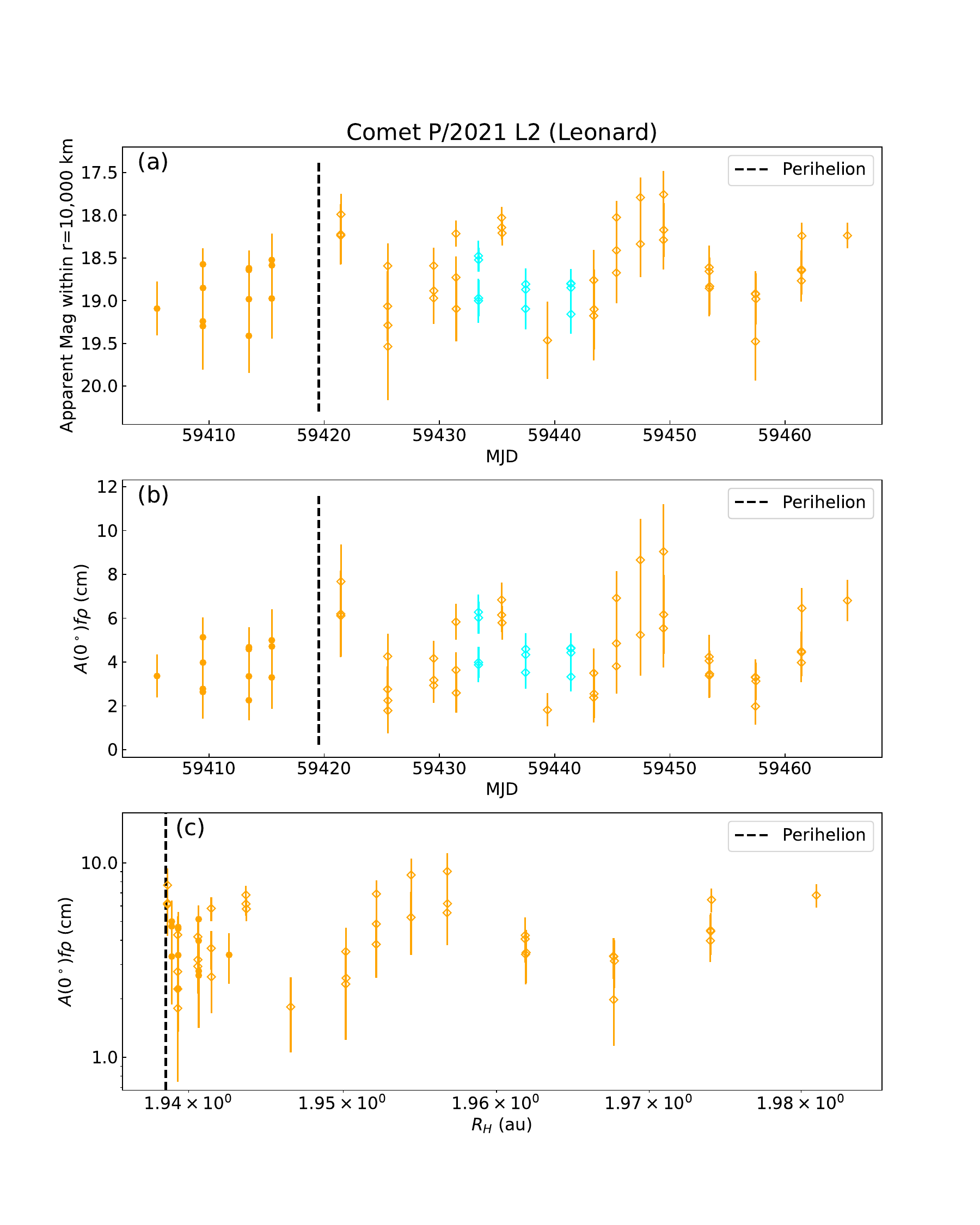}
\figsetgrpnote{P/2021 L2 (Leonard) light curve is shown in the top panel, (a). The dust production parameter $A(0^\circ)f\rho$ vs time is displayed in plot (b), and (c) shows $A(0^\circ)f\rho$ vs heliocentric distance with the solid line representing the rate of change of activity pre-perihelion and the dot-dashed line showing the rate of change of activity post-perihelion. Orange and cyan points on the plots represent the \emph{orange} filter and \emph{cyan} filter respectively. Filled circular points represent pre-perihelion measurements and open diamonds represent post-perihelion. Perihelion is marked with dashed line in each panel.}

\figsetgrpnum{1.43}
\figsetgrptitle{ZTFP2021N1}
\figsetplot{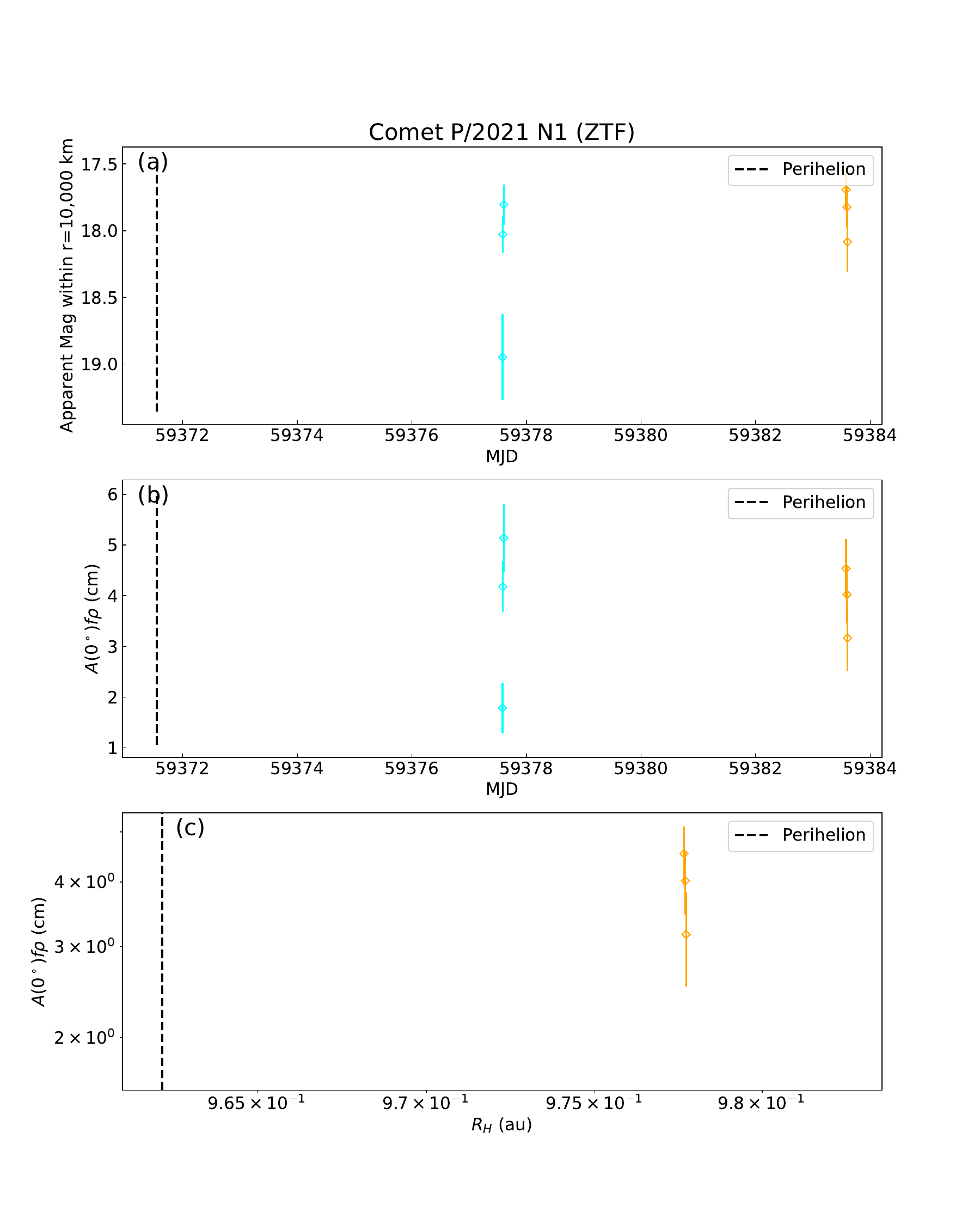}
\figsetgrpnote{P/2021 N1 (ZTF) light curve is shown in the top panel, (a). The dust production parameter $A(0^\circ)f\rho$ vs time is displayed in plot (b), and (c) shows $A(0^\circ)f\rho$ vs heliocentric distance with the solid line representing the rate of change of activity pre-perihelion and the dot-dashed line showing the rate of change of activity post-perihelion. Orange and cyan points on the plots represent the \emph{orange} filter and \emph{cyan} filter respectively. Filled circular points represent pre-perihelion measurements and open diamonds represent post-perihelion. Perihelion is marked with dashed line in each panel.}

\figsetgrpnum{1.44}
\figsetgrptitle{FulsP2021N2}
\figsetplot{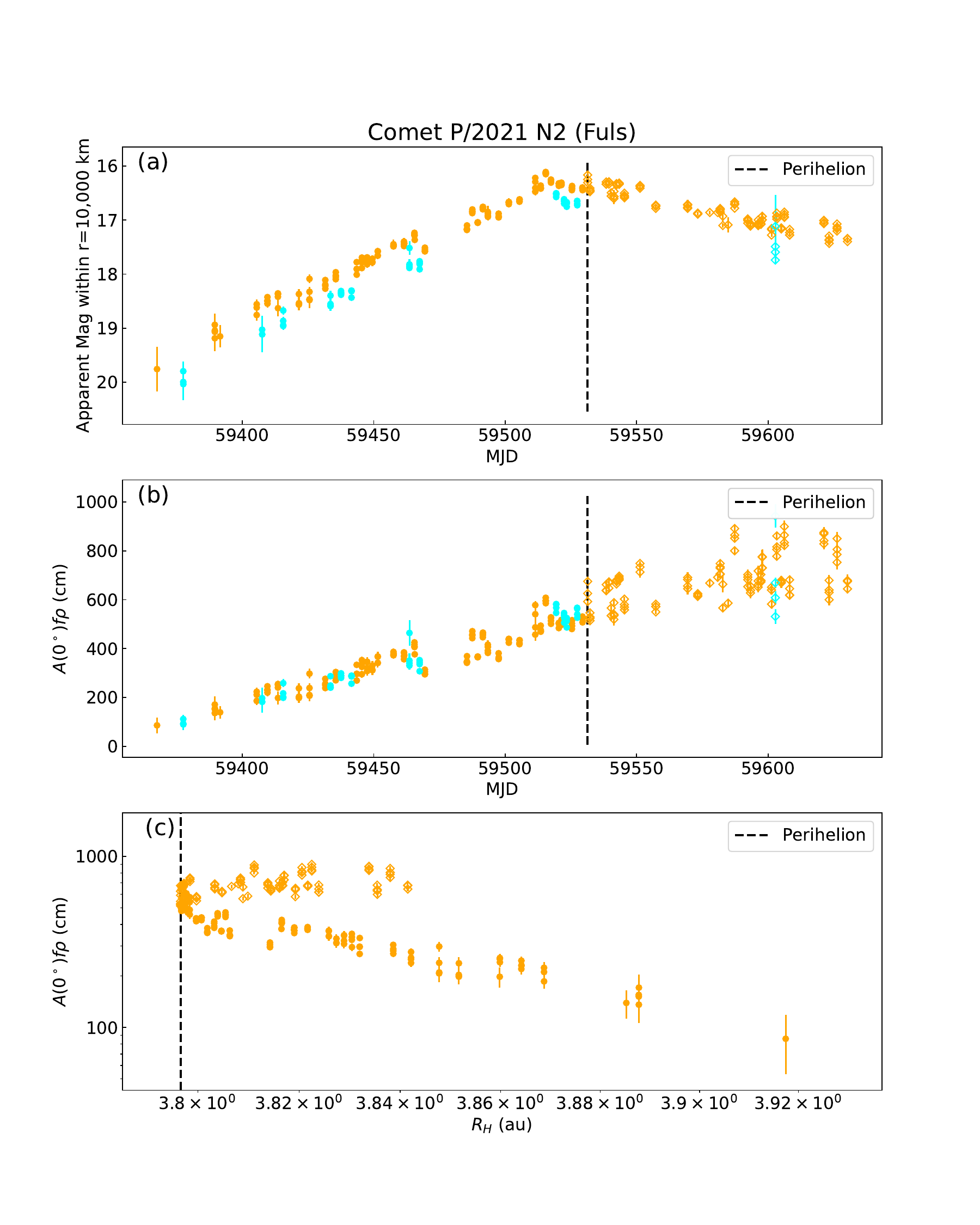}
\figsetgrpnote{P/2021 N2 (Fuls) light curve is shown in the top panel, (a). The dust production parameter $A(0^\circ)f\rho$ vs time is displayed in plot (b), and (c) shows $A(0^\circ)f\rho$ vs heliocentric distance with the solid line representing the rate of change of activity pre-perihelion and the dot-dashed line showing the rate of change of activity post-perihelion. Orange and cyan points on the plots represent the \emph{orange} filter and \emph{cyan} filter respectively. Filled circular points represent pre-perihelion measurements and open diamonds represent post-perihelion. Perihelion is marked with dashed line in each panel.}

\figsetgrpnum{1.45}
\figsetgrptitle{PANSTARRSP2021P3}
\figsetplot{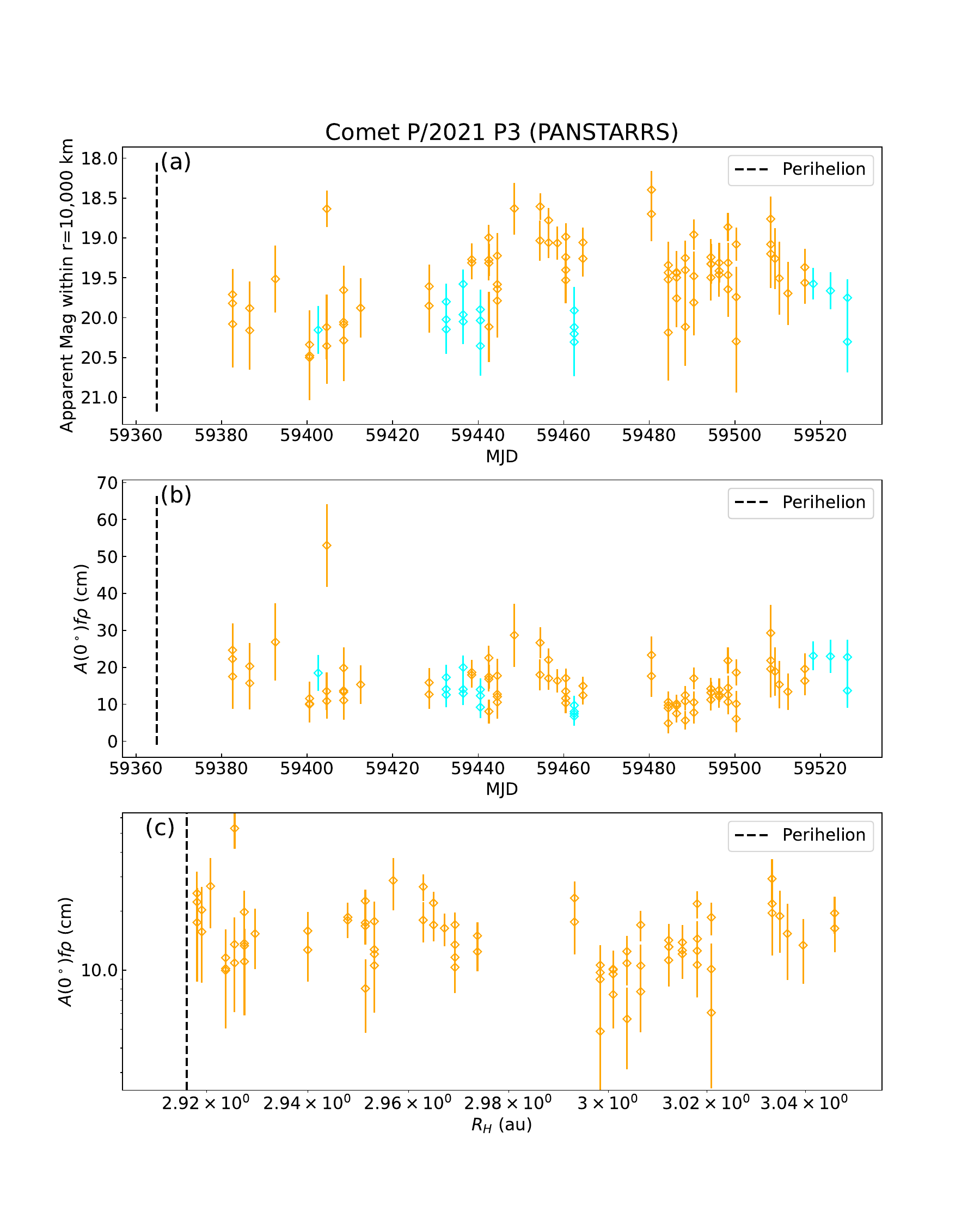}
\figsetgrpnote{P/2021 P3 (PANSTARRS) light curve is shown in the top panel, (a). The dust production parameter $A(0^\circ)f\rho$ vs time is displayed in plot (b), and (c) shows $A(0^\circ)f\rho$ vs heliocentric distance with the solid line representing the rate of change of activity pre-perihelion and the dot-dashed line showing the rate of change of activity post-perihelion. Orange and cyan points on the plots represent the \emph{orange} filter and \emph{cyan} filter respectively. Filled circular points represent pre-perihelion measurements and open diamonds represent post-perihelion. Perihelion is marked with dashed line in each panel.}

\figsetgrpnum{1.46}
\figsetgrptitle{ATLASP2021Q5}
\figsetplot{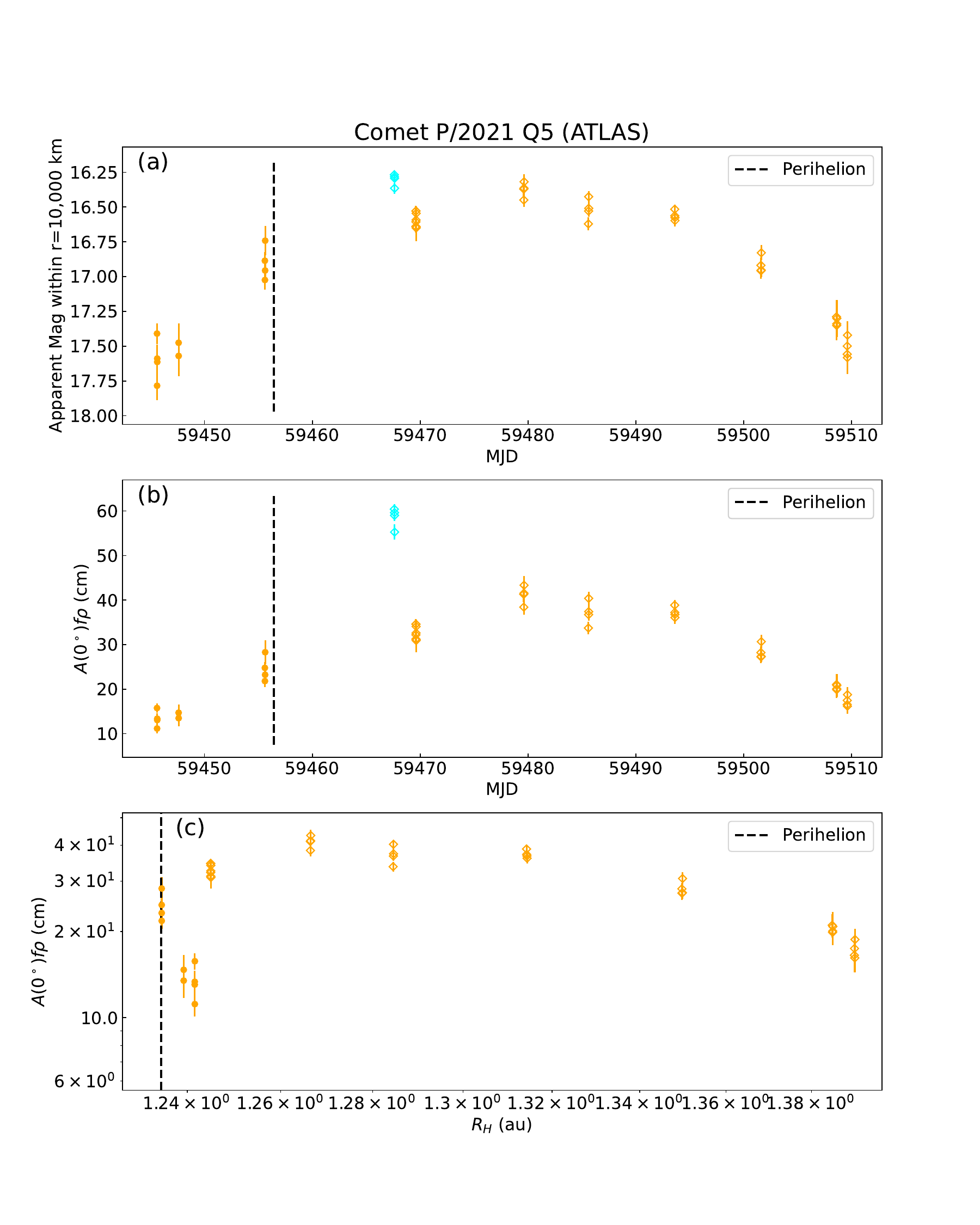}
\figsetgrpnote{P/2021 Q5 (ATLAS) light curve is shown in the top panel, (a). The dust production parameter $A(0^\circ)f\rho$ vs time is displayed in plot (b), and (c) shows $A(0^\circ)f\rho$ vs heliocentric distance with the solid line representing the rate of change of activity pre-perihelion and the dot-dashed line showing the rate of change of activity post-perihelion. Orange and cyan points on the plots represent the \emph{orange} filter and \emph{cyan} filter respectively. Filled circular points represent pre-perihelion measurements and open diamonds represent post-perihelion. Perihelion is marked with dashed line in each panel.}

\figsetgrpnum{1.47}
\figsetgrptitle{GroellerP2021R6}
\figsetplot{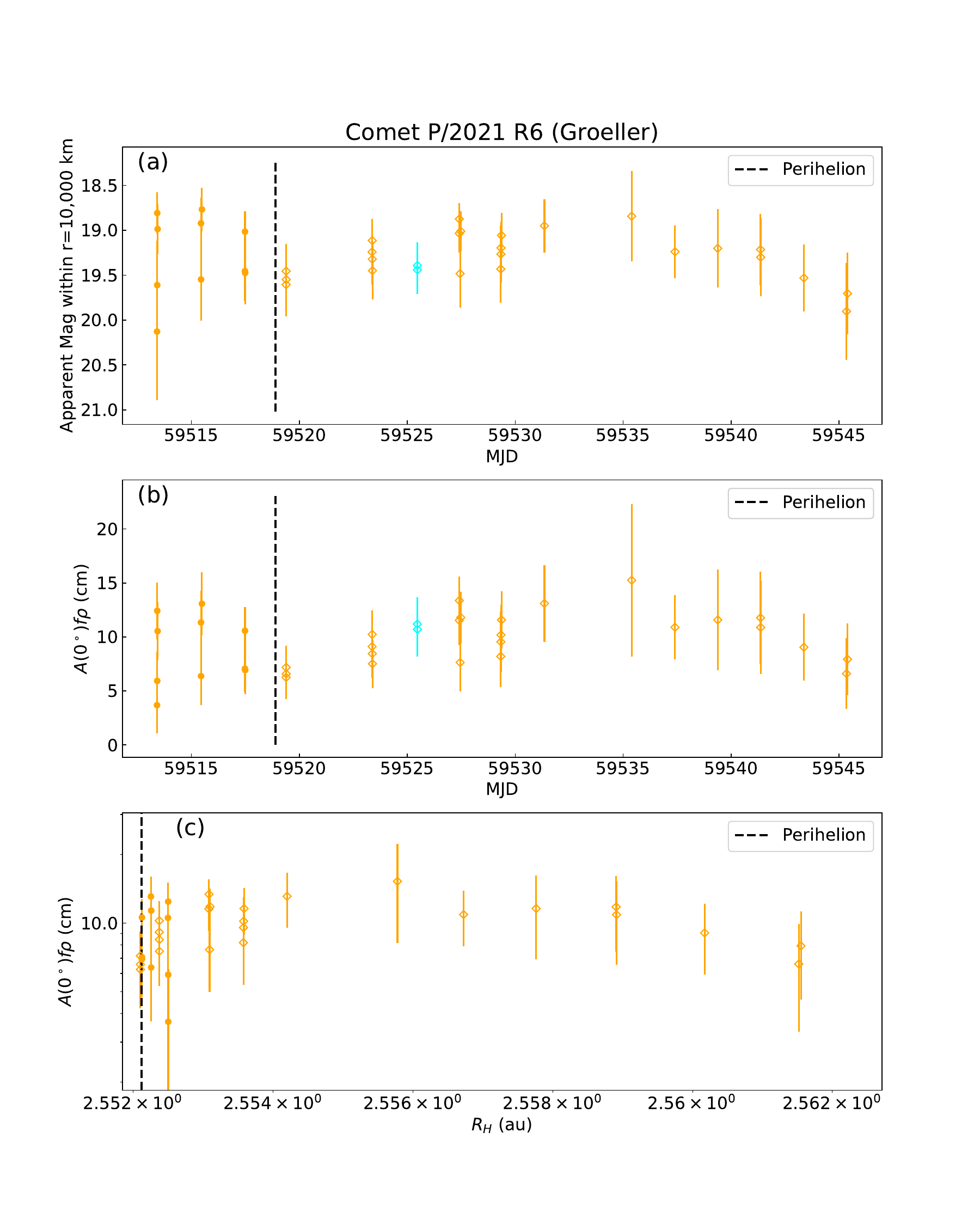}
\figsetgrpnote{P/2021 R6 (Groeller) light curve is shown in the top panel, (a). The dust production parameter $A(0^\circ)f\rho$ vs time is displayed in plot (b), and (c) shows $A(0^\circ)f\rho$ vs heliocentric distance with the solid line representing the rate of change of activity pre-perihelion and the dot-dashed line showing the rate of change of activity post-perihelion. Orange and cyan points on the plots represent the \emph{orange} filter and \emph{cyan} filter respectively. Filled circular points represent pre-perihelion measurements and open diamonds represent post-perihelion. Perihelion is marked with dashed line in each panel.}

\figsetgrpend
\figsetend

\end{document}